\documentclass[a4paper,fleqn]{cas-sc}
\usepackage{amsmath}
\usepackage{amssymb}
\usepackage{graphicx}
\usepackage{mathrsfs}
\usepackage{amsfonts}
\usepackage{amsthm}
\usepackage{color}
\usepackage{titletoc}

\titlecontents{section}
              [1.1cm]
              { \normalsize}%
              {\contentslabel{2em}}%
              {}%
              {\titlerule*[0.3pc]{$\cdot$}\normalsize\contentspage\hspace*{0cm}}%
\titlecontents{subsection}
              [2cm]
              {\normalsize}%
              {\contentslabel{2em}}%
              {}%
              {\titlerule*[0.3pc]{$\cdot$}\normalsize\contentspage\hspace*{0cm}}%
\titlecontents{subsubsection}
              [2.9cm]
              {\normalsize}%
              {\contentslabel{2.5em}}%
              {}%
              {\titlerule*[0.3pc]{$\cdot$}\normalsize\contentspage\hspace*{0cm}}%

\usepackage{lipsum}
\usepackage{gensymb}
\usepackage[numbers,sort&compress]{natbib}
\usepackage[numbers]{natbib} 
\def\tsc#1{\csdef{#1}{\textsc{\lowercase{#1}}\xspace}}
\tsc{WGM}
\tsc{QE}
\tsc{EP}
\tsc{PMS}
\tsc{BEC}
\tsc{DE}

\begin{document}
\let\WriteBookmarks\relax
\def\floatpagepagefraction{1}
\def\textpagefraction{.001}
\shorttitle{H. Yang, L. Song, Y. Cao, and P. Yan}
\shortauthors{H. Yang et~al.}

\title [mode = title]{Circuit realization of topological physics}                      

\author[]{Huanhuan Yang}[orcid=0000-0003-0157-6872]

\address[]{School of Physics and State Key Laboratory of Electronic Thin Films and Integrated Devices, University of Electronic Science and Technology of China, Chengdu 610054, China}

\author[]{Lingling Song}[orcid=0000-0001-6961-0788]

\author[]{Yunshan Cao}[orcid=0000-0002-3409-2578]

\author[]{Peng Yan}[orcid=0000-0001-6369-2882]
\cormark[1]
\ead{yan@uestc.edu.cn}

\cortext[cor1]{Corresponding author.}

\begin{abstract}
Recently, topolectrical circuits (TECs) boom in studying the topological states of matter. The resemblance between circuit Laplacians and tight-binding models in condensed matter physics allows for the exploration of exotic topological phases on the circuit platform. In this review, we begin by presenting the basic equations for the circuit elements and units, along with the fundamentals and experimental methods for TECs. Subsequently, we retrospect the main literature in this field, encompassing the circuit realization of (higher-order) topological insulators and semimetals. Due to the abundant electrical elements and flexible connections, many unconventional topological states like the non-Hermitian, nonlinear, non-Abelian, non-periodic, non-Euclidean, and higher-dimensional topological states that are challenging to observe in conventional condensed matter physics, have been observed in circuits and summarized in this review. Furthermore, we show the capability of electrical circuits for exploring the physical phenomena in other systems, such as photonic and magnetic ones. Importantly, we highlight TEC systems are convenient for manufacture and miniaturization because of their compatibility with the traditional integrated circuits. Finally, we prospect the future directions in this exciting field, and connect the emerging TECs with the development of topology physics, (meta)material designs, and device applications.
\end{abstract}


\maketitle

\tableofcontents
              
\section{Introduction}
Electrical circuits, conventionally, are used to manipulate the flow of electrical charge. An electrical circuit is a network consisting of closed loops with basic electrical components, e.g., sources, wires, switches, resistors (R), inductors (L), and capacitors (C), giving return paths for the current \cite{Nilssonbook}. In the lumped-element model, the operating rules of a circuit are determined by the structures of circuits [described by Kirchhoff's current law (KCL), Kirchhoff's voltage law (KVL)] and the characteristics of the components [characterized by voltage-current relations (VCRs)], respectively. These three laws constitute the foundation of the circuit theorem. With electrical circuits, one may explore many fascinating physical phenomena. For example, with the nonlinear circuit platforms, one can study chaos \cite{Chenbook}, fractals \cite{Lazareck}, solitons \cite{Nagashima1979,Muroya1981,Muroya1982,Kuusela1987}, and other nonlinear problems \cite{Chua69,Muthuswamybook}.
In non-Hermitian circuits, the (anti-)parity-time [$\mathcal{(A)PT}$] symmetrical behavior can be well rendered in RLC resonant units \cite{Schindler040101,Choi2182,Chen297,Sahoo023508,Cao262}. In such cases, the (imaginary)real eigenvalues and complex conjugate pairs emerge for the exact and broken $\mathcal{(A)PT}$ symmetry phases, respectively, separated by the so-called exceptional point (EP). Near the EP, the ultra-sensitive sensors can be designed \cite{Chen297,Sahoo023508}. 

Topological phenomena have attracted considerable long-term attention because of their fundamental interest and application prospects \cite{Hasan2010,Qi2011,Chiu2016,BXie2021}, such as the quantized information transport and robust wave propagation. Although topological orders originate from condensed matter physics, researchers from other fields, like photonics \cite{Ozawa015006,Lu821}, phononics \cite{Susstrunk2015,Serra342,Peterson346,Liu1904784}, mechanics \cite{Fan204301,Huber621,Zheng1987}, spintronics \cite{Li164,McClarty171}, and electrical circuits, are also interested in this topic. Among these platforms, electric circuits stand out as a competitive one for studying topological physics with simple circuit networks. A new terminology, topolectrical circuit (TEC), emerged recently for describing this prosperous field \cite{Jia2015,Albert2015,Lee2018}. 

The key point of TEC is that one can fully map the tight-binding Hamiltonian in condensed matter physics to circuit Laplacians (admittance matrices). The large variety of electrical elements and variable connection modes allow the researchers to realize a wide variety of topological states, some of which are extremely challenging to observe in condensed matter systems. In particular, the hoppings and on-site potentials host generous freedoms in circuits, such as the strength, direction, and dimension, which enable us to introduce interactions between two arbitrary nodes and control the on-site energy for each node. Then, one can study the topological physics in the systems with various interactions (e.g. long-range interactions), and investigate the topological states in higher-dimensional, non-periodic, and non-Euclidean lattices. Drawing upon the nonlinear and non-reciprocal circuit components, one can explore the nonlinear and non-Hermitian topological phenomena in the circuit platform as well. Besides, it is convenient to manufacture and miniaturize the TEC device due to its compatibility with the integrated circuit. With these superiorities, TECs can be used to explore exotic topological physics to deepen the understanding of topological phases and demonstrate practical applications.

\begin{figure}[htbp!]
  \centering
  \includegraphics[width=0.9\textwidth]{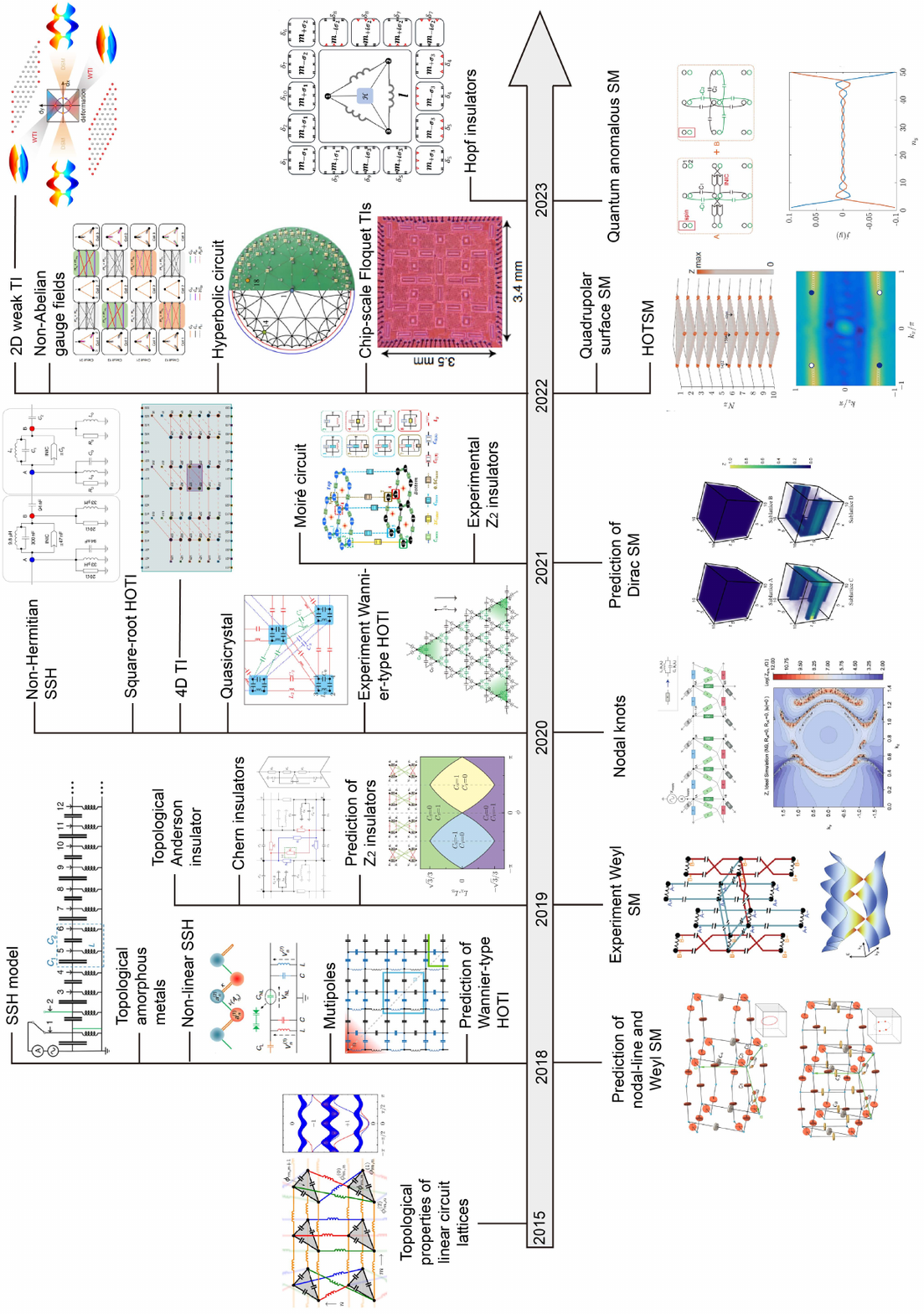}
  \caption{Timeline of developments of topological phases in electrical circuits. SSH, Su-Schrieffer-Heeger; HOTI, higher-order topological insulators; 4D TI, four-dimensional topological insulators; 2D weak TI, two-dimensional weak topological insulators; SM, semimetal; HOTSM, higher-order topological semimetal. }\label{Developement}
\end{figure}

\begin{figure}
  \centering
  \includegraphics[width=0.75\textwidth]{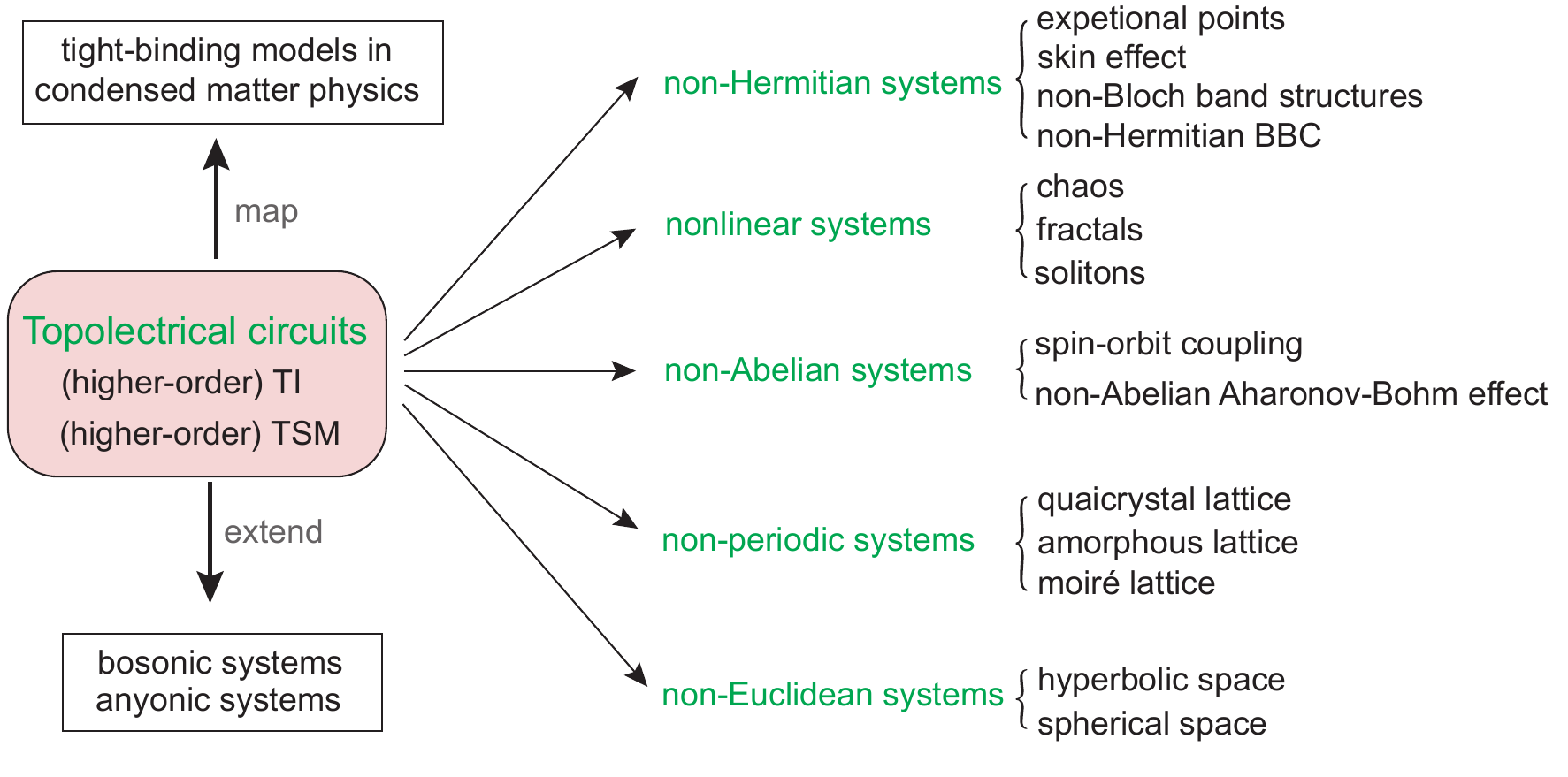}
  \caption{Framework of topolectrical circuits. The main contents of TECs consist of the (higher-order) TIs and TSMs, originating from the mapping for tight-binding models of topological states to circuit Laplacians. TECs can also intersect with other physical systems, such as nonlinear, non-Hermitian, non-Abelian, non-periodic, and non-Euclidean systems, producing novel phenomena. With the superiorities of TECs, the topological states can be extended to bosonic and anyonic systems. }\label{Overview}
\end{figure}

In this article, we review the recent development of TECs, as outlined in Fig. \ref{Developement}, making the growing field of TECs accessible to a broad community of researchers. This review is organized by the framework in Fig. \ref{Overview}. In Section \ref{S2}, we introduce the fundamentals of TECs, including the basic circuit equations and the construction methods. In Section \ref{S3}, we review the topological insulators (TIs) containing both the first-order and higher-order topological (HOT) states in the TEC platform. The circuit realization of (higher-order) topological semimetals (TSMs) is summarized in Section \ref{S4}, covering the Dirac, Weyl, and nodal-line ones. In Section \ref{S5}, unconventional topological states in circuit are presented, including non-Hermitian, nonlinear, non-Abelian, non-periodic, non-Euclidean, higher-dimensional, and other exotic topological states. Section \ref{S6} summarizes the circuit simulation of the physical phenomena in other systems, like photonic and magnetic ones. Conclusion and outlook are drawn in Section \ref{S7}, accompanied by the discussion on the integration of TECs with complementary metal-oxide semiconductors (CMOS) technique.

\section{Fundamentals of topolectrical circuits} \label{S2}
In this section, we show the basic circuit equations for electrical elements, including resistors, inductors, varactors, and negative impedance converters with current inversion (INIC). We use two examples to show the diversity of the circuit systems, i.e., the nonlinear circuits with attractors or solitons and the $\mathcal{(A)PT}$ symmetrical circuits with EPs. Subsequently, we elaborate on the details of the construction method for TECs and show the map from the circuit Laplacian to tight-binding Hamiltonian and from the circuit equation to Schr\"{o}dinger equation. Finally, we introduce how to observe the topological states in circuit experiments.

\subsection{Basic circuit elements and equations}

In electrical circuits, there are three fundamental laws \cite{Nilssonbook}.

\textbf{KCL} The sum of the currents into any circuit node is zero.

\textbf{KVL} The sum of the voltage drops around any loop is zero.

\textbf{VCR} The relationship between the current and voltage when the electric current goes through a component.

Based on KCL and KVL, two methods are developed for circuit analysis, that is, \emph{nodal analysis} and \emph{mesh analysis}. One can use these two methods to analyze any linear circuits by obtaining a set of simultaneous equations that are solved to obtain the node voltage and loop current.

\begin{figure}
  \centering
  \includegraphics[width=0.9\textwidth]{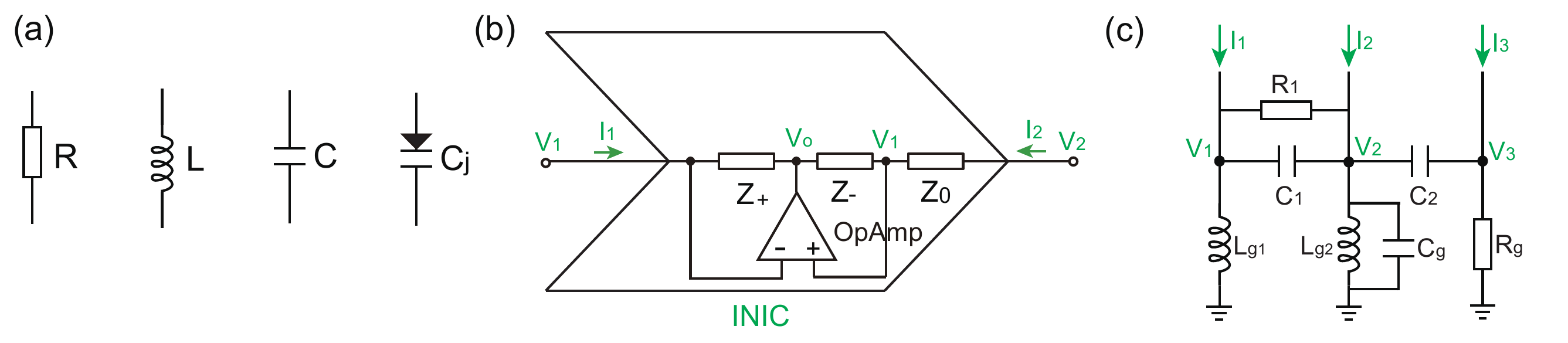}
  \caption{(a) Basic electrical elements used in circuits, including the resistor, inductor, capacitor, and varactor diode, respectively. (b) Set up of the negative impedance converter with current inversion. (c) A paradigmatic circuit diagram with three nodes.}\label{elements}
\end{figure}

In Fig. \ref{elements}(a), we present basic elements used in electrical circuits, containing resistors, inductors, capacitors, and varactors. The VCR equations for these four components are written as 
\begin{equation}
\begin{aligned}
i&=\frac{1}{R}v,\\
i&=C\frac{dv}{dt}, \\
i&=\frac{1}{L}\int_{-\infty}^tvdt,\\
i&=C_j\frac{dv}{dt},
\end{aligned}
\end{equation}
where $C_j=\frac{C_0}{(V_b-V)^m}$ is diode capacitance, $C_0$ is diode capacitance without voltage across it, $V$ is the applied voltage, $V_b$ is barrier voltage at the junction, and $m$ is constant depending upon the materials (range from a few decimal points to several hundred). These RLC elements constitute the basic linear circuits, while varactors are used in nonlinear systems. It is noted that the memory effects can also be taken into consideration in circuit with the usage of memristors, memcapacitors, and meminductors \cite{Pershin2010}.

Figure \ref{elements}(b) shows the circuit unit named INIC. According to the properties of operational amplifiers, one can write the VCR of this unit as 
\begin{equation}\label{INIC}
\left(
  \begin{array}{c}
    I_1 \\
    I_2 \\
  \end{array}
\right)=\frac{1}{Z_0}\left(
          \begin{array}{cc}
            -\nu & \nu \\
            -1 & 1 \\
          \end{array}
        \right)\left(
  \begin{array}{c}
    V_1 \\
    V_2 \\
  \end{array}
\right)
\end{equation} 
with $\nu=Z_-/Z_+$. Here, $Z_-$, $Z_+$, and $Z_0$ are impedances of the components shown in Fig. \ref{elements}(b). Governed by Eq. \eqref{INIC}, the unit acts as a positive (negative) impedance from right to left (left to right),  which can be used to realize non-reciprocal interactions and negative impedance elements.

To show how to write the circuit equations, we take a simple RLC circuit as an example, as displayed in Fig. \ref{elements}(c). By utilizing the nodal-analysis method, the responses of this circuit are described by the following equations 
\begin{equation}\label{Ex}
\begin{aligned}
    i_1&=\frac{v_1-v_2}{R}+C_1\frac{d(v_1-v_2)}{dt}+\frac{1}{L_{g1}}\int_{-\infty}^tv_1dt, \\
    i_2&=-\frac{v_1-v_2}{R}-C_1\frac{d(v_1-v_2)}{dt}+C_2\frac{d(v_2-v_3)}{dt}+C_g\frac{dv_2}{dt}+
    \frac{1}{L_{g2}}\int_{-\infty}^tv_2dt, \\
    i_3&=-C_2\frac{d(v_2-v_3)}{dt}+\frac{v_3}{R_g}. \\
\end{aligned}
\end{equation}

For a sinusoidal steady-state analysis, one can convert these equations to frequency space by supposing $v_n=V_n\exp(i\omega t)$ and $i_n=I_n\exp(i\omega t)$ ($n=1,2,3$) at frequency $\omega$. Equation\eqref{Ex} then can be written as
\begin{equation}\label{EQ4}
\left(
  \begin{array}{c}
    I_1 \\
    I_2 \\
    I_3 \\
  \end{array}
\right)=\left(
          \begin{array}{ccc}
            G+i\omega C_1+1(i\omega L_{g1}) & -G-i\omega C_1 & 0 \\
            -G-i\omega C_1 & G+i\omega C_1+i\omega C_2+i\omega C_g+1(i\omega L_{g2}) & -i\omega C_2 \\
            0 & -i\omega C_2 & i\omega C_2+G_g \\
          \end{array}
        \right)\left(
                 \begin{array}{c}
                   V_1 \\
                   V_2 \\
                   V_3 \\
                 \end{array}
               \right),
\end{equation} 
with $G=1/R$ and $G_g=1/R_g$. The $3\times 3$ matrix in Eq. \eqref{EQ4} is an admittance matrix, which is also referred to the circuit Laplacian. It is convenient to use this formula if we only concern the steady-state results. 

\begin{figure}
  \centering
  \includegraphics[width=0.9\textwidth]{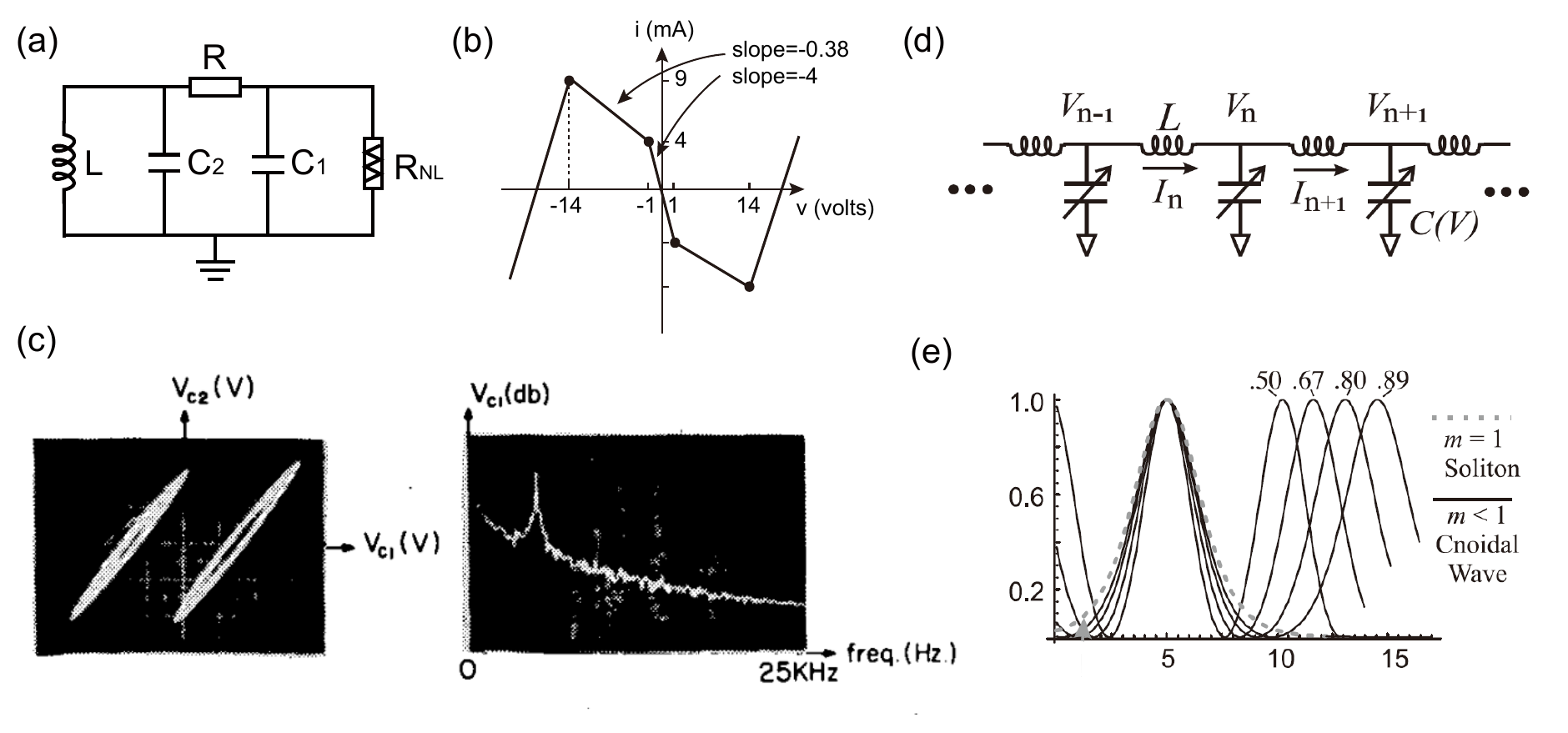}
  \caption{(a) Chua's chaotic circuit model. (b) The voltage-current relation of the nonlinear resistor $R_{\rm NL}$. (c) Chaotic attractors measured from Chua’s circuit. (d) Nonlinear transmission line with varactors. (e) The solutions of cnoidal waves (solid lines) and solitons (dashed line). \\
  {\emph{Source}:} The figures are adapted 
  from Refs. \cite{Zhong501,Ricketts}}\label{CC}
\end{figure}

Below we present two classical examples to show the capabilities of circuits on studying novel physical phenomena, one of which is the chaos and solitons in the nonlinear circuits and another one is the realization of the $\mathcal{(A)PT}$ symmetrical dimers in non-Hermitian circuits.

The study of nonlinear circuits can be traced back to 20th century and continues until now because of their abundant physical phenomena and valuable applications \cite{Chua69,Muthuswamybook}. Here, we show two typical systems in nonlinear circuits, i.e., Chua's circuit and nonlinear transmission line with solitons. Figure \ref{CC}(a) shows a typical Chua's circuit with two capacitors $C_{1,2}$, one inductor $L$, one linear resistor $R$, and one nonlinear resistor $R_{\rm NL}$. This circuit is described by the following equations
\begin{equation}\label{ChuaE}
\begin{aligned}
    &\frac{du_{C_1}}{dt}=\frac{G}{C_1}(u_{C_2}-u_{C_1})-\frac{1}{C_1}g_{\rm NL}(u_{C_1}), \\
    &\frac{du_{C_2}}{dt}=\frac{1}{C_2}i_L+\frac{g}{C_2}(u_{C_1}-u_{C_2}), \\
    &\frac{di_{L}}{dt}=-\frac{1}{L}u_{C_2},\\
\end{aligned}
\end{equation}
with $g=1/R$ the conductance and $g_{\rm NL}(u)=1/R_{\rm NL}$ the equivalent nonlinear conductance. Zhong et al. considered a nonlinear resistor $R_{\rm NL}$ with the voltage-current relation shown in Fig. \ref{CC}(b) \cite{Zhong501}, and they observed a pair of chaotic attractors, shown in Fig. \ref{CC}(c), which compare well with the numerical solutions of Eq. \eqref{ChuaE}. Chaotic phenomena are ubiquitous in circuits, which can constitute rich phase diagrams. With these circuits, one can study chaos phenomena and use them for application, such as secret communications \cite{Kocarev6}.

Figure \ref{CC}(d) displays a nonlinear transmission line with each node grounded by a varactor \cite{Ricketts}, the behavior of which is described by the circuit equation
\begin{equation} \label{KdV}
V_T+\frac{b}{LC_0}VV_S+\frac{1}{24\sqrt{LC_0}}V_{SSS}=0,
\end{equation}
where $S=\varepsilon^{1/2}(n-v_0t)$, $T=\varepsilon^{3/2}t$ with $\varepsilon=n-ct$ ($c$ is the velocity of the soliton and $n$ is the node number), and $b$ is a parameter for the nonlinear capacitor, typically ranging from 0.05 to 0.2.

Equation \eqref{KdV} is equivalent to the Korteweg–De Vries (KdV) equation
\begin{equation}\label{kdv}
u_t+6uu_x+u_{xxx}=0,
\end{equation}
with renormalized coefficients. Equation \eqref{kdv} has two kinds of solutions, i.e., cnoidal waves and solitons, displayed by the solid and dashed lines in Fig. \ref{CC}(e), respectively. 

\begin{figure}
  \centering
  \includegraphics[width=1\textwidth]{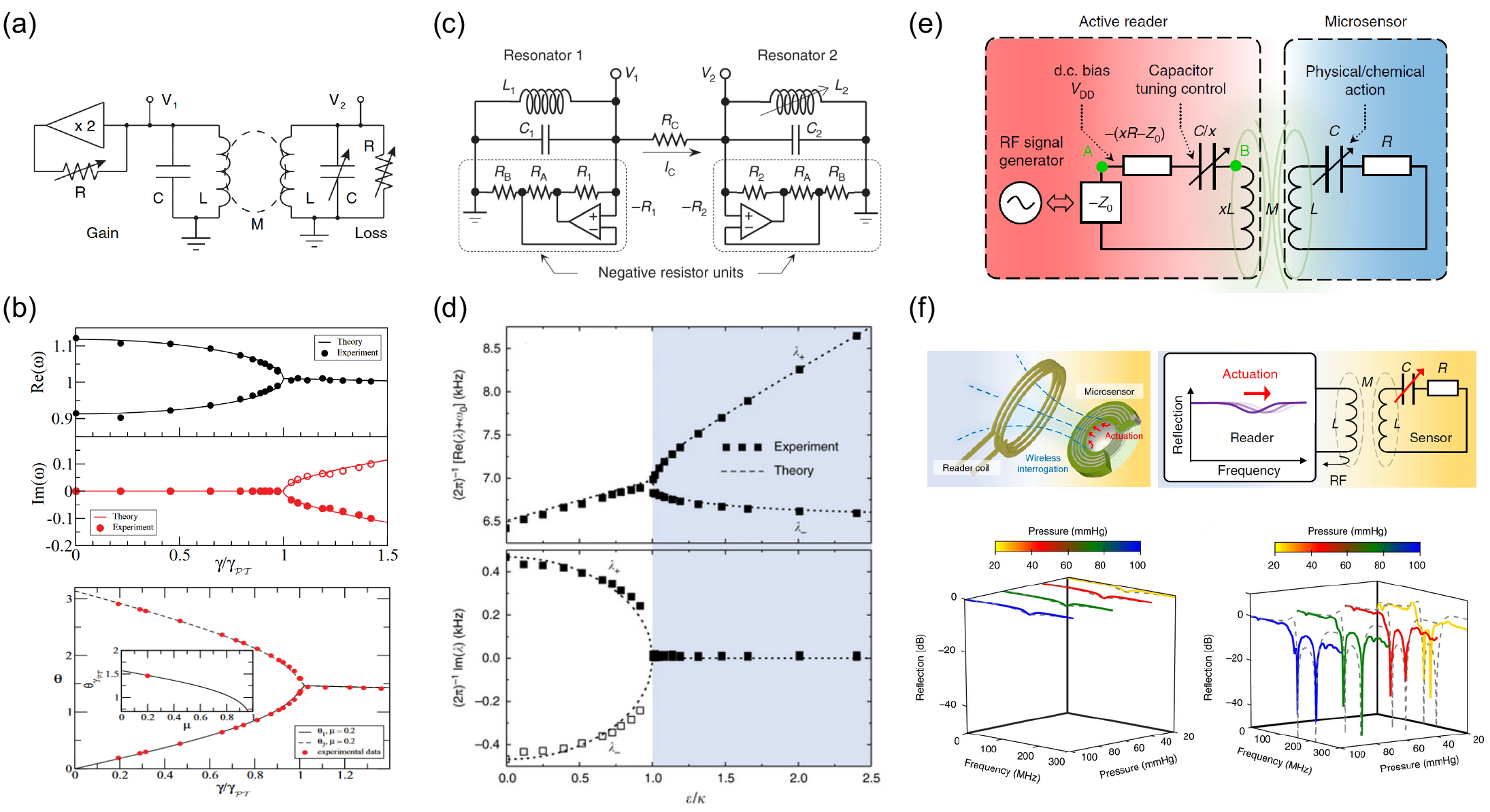}
  \caption{(a) Inductor-coupled two RLC units. (b) Measured frequency spectra as a function of the gain-loss parameter (top). The black and red lines represent the real and imaginary parts. The evolution of phase difference between two eigenmodes (bottom). (c) Resistor-coupled RLC units with two gains. (d) Measured eigenvalue splitting. $W_1=|F[V_1(t)]|$ is the Fourier-transformed intensity of $V_1$. (e) Sensors based on the spectra splitting of PTX symmetrical circuit. (f) Design of the sensor. Responses of the sensors in the passive (left) and PTX-symmetrical (right) systems.\\
   {\emph{Source}:} The figures are adapted from Refs. \cite{Schindler040101,Choi2182,Chen297} }\label{PTS}
\end{figure}

Next, we discuss three kinds of $\mathcal{(A)PT}$ circuits. As some special cases of the non-Hermitian systems, they can host pure (imaginary)real spectra in the $\mathcal{(A)PT}$ symmetrical conserved phase. Figure \ref{PTS}(a) displays a $\mathcal{PT}$-symmetric dimer with balanced gain and loss, coupled by the mutual inductance. This system is described by the following equations
\begin{equation} \label{PTE}
\begin{aligned}
\frac{d^2Q_1^C}{d\tau^2}&=-\frac{1}{1-\mu^2}Q_1^C+\frac{\mu}{1-\mu^2}Q_2^C+\gamma\frac{dQ_1^C}{d\tau},\\
\frac{d^2Q_2^C}{d\tau^2}&=\frac{\mu}{1-\mu^2}Q_1^C-\frac{1}{1-\mu^2}Q_2^C-\gamma\frac{dQ_2^C}{d\tau},\\
\end{aligned}
\end{equation}
with $\tau=\omega_0t$, $\mu=M/L$ ($M$ is the mutual inductance), $\gamma$ is the gain-loss parameter, and $Q^C_{1,2}$ are the capacitor charges for gain and loss units, respectively. The circuit respects $\mathcal{PT}$ symmetry because Eqs. \eqref{PTE} are invariant under the combined $\mathcal{P}$ ($1\leftrightarrow 2$) and $\mathcal{T}$ ($t\rightarrow -t$) operations. 

Recasting Eqs. \eqref{PTE} as $\frac{d\Phi}{dt}=\mathcal{H}\Phi$ with
\begin{equation}
\mathcal{H}=\left(
                   \begin{array}{cc}
                     0 & \sigma_0 \\
                     -\frac{1}{1-\mu^2}\sigma_0+\frac{\mu}{1-\mu^2}\sigma_x & \gamma\sigma_z \\
                   \end{array}
                 \right),
\end{equation}
and $\Phi=(Q_1^C,Q_2^C,\dot{Q}_1^C,\dot{Q}_2^C)$. Here, $\sigma_0$ is the identity matrix and $\sigma_{x,z}$ are Pauli matrices. One obtains the spectra by diagonalizing $\mathcal{H}$
\begin{equation}\label{omegaPT}
\omega=\pm\sqrt{-\frac{2+\gamma^2(\mu^2-1)\pm\sqrt{4(\mu^2-1)+[2+\gamma^2(\mu^2-1)]^2}}{2(\mu^2-1)}}.
\end{equation}

Figure \ref{PTS}(b) shows the spectra and phase difference dependence on $\gamma$ both in theory [solid line from Eq. \eqref{omegaPT}] and experiment (dots). Two real eigenvalues appear below the phase transition point ($\gamma_{PT}=1$, namely the EP), while a pair of conjugated complex roots are observed in $\mathcal{PT}$-symmetry-broken phase ($\gamma_{PT}>1$). These results manifest as the main feature of $\mathcal{PT}$-symmetry system.

Similar phenomenon can be observed in $\mathcal{APT}$ system. Figure \ref{PTS}(c) displays two gain circuit units coupled by a resistor, characterized by the Hamiltonian
\begin{equation} \label{EQAPT}
\mathcal{H}=\left(
              \begin{array}{cc}
                -\epsilon+i\gamma & i\kappa \\
                i\kappa & \epsilon+i\gamma \\
              \end{array}
            \right),
\end{equation}
with $\epsilon$ the energy level, $\gamma$ the amplifying ratio, and $\kappa$ being the coupling strength. If one executes the $\mathcal{P}$ and $\mathcal{T}$ operations on the Eq. \eqref{EQAPT}, $\mathcal{H}$ will become to $\mathcal{-H}$, which means the system respects $\mathcal{APT}$ symmetry. One can observe the splitting of spectra as the increasing of energy level, as shown in Fig. \ref{PTS}(d). Beyond the EPs ($\epsilon/\kappa>1$), this system hosts two real eigenvalues.

As an important application of $\mathcal{(A)PT}$ system, one can design the ultra-sensitive sensor. Near the EPs, a perturbation will lead to the splitting of the spectrum, which follows the rule of $\epsilon^{1/N}$ with $\epsilon$ the strength disturbance and $N$ the order of EPs. As shown in Fig. \ref{PTS}(e), Chen et al. designed a sensor based on $\mathcal{PTX}$ symmetrical dimer (here $\mathcal{X}$ means a scaling parameter). The pressure-sensitive varactor is used as the detector. The change of pressure will influence the resonant peak of passive sensor [top of Fig. \ref{PTS}(f)] and yield the frequency splitting of the reflection spectrum for $\mathcal{PT}$-symmetric sensors. Figure \ref{PTS}(f) shows the response of two different sensors, where the signal of $\mathcal{PT}$-symmetric sensor (right) is much stronger than the passive one (left). 

Apart from the two prominent instances in electrical circuits, recent studies found topological physics can be fasten to circuit platform. Next, we will introduce the key theories and construction methods of TECs.

\subsection{Circuit Laplacian and Schrödinger equation}

The major method used in TECs is the \emph{nodal analysis} because we often concern about the node voltages rather than mesh currents. Considering a general linear circuit consisting of $RLC$, one can use the $N$-component vectors {\textbf V} and {\textbf I} to label the voltages measured at the nodes of the circuit board against ground and the external input currents, respectively. The equations of motion of the circuit are given by
\begin{equation} \label{ME}
\frac{d}{dt}{\textbf I}(t)={\mathcal C}\frac{d^2}{dt^2}{\textbf V}(t)+\Sigma\frac{d}{dt}{\textbf V}(t)+{\mathcal L}{\textbf V}(t),
\end{equation}
where the ${\mathcal C}$, $\Sigma$, and ${\mathcal L}$ are the $N\times N$ real-valued matrices \cite{Hofmann2019}. The element in matrices ${\mathcal C}$, $\Sigma$, ${\mathcal L}$ represents the capacitance, the inverse of resistance, and the inverse of inductance between two nodes or node to ground.

The response of the circuit at frequency $\omega$ follows Kirchhoff's law in the frequency domain
\begin{equation}
I(\omega)=J(\omega)V(\omega),
\end{equation}
with
\begin{equation} \label{CL}
J(\omega)=i\omega {\mathcal C}+\Sigma+\frac{1}{i\omega}{\mathcal L}
\end{equation}
being the circuit Laplacian. $J$($\omega$) can be expressed as $i \mathcal{H}(\omega)$ with $\mathcal{H}(\omega)$ being analogous to the tight-binding Hamiltonian formalism in condensed matter physics. The diagonal and non-diagonal elements correspond to the on-site potentials and hoppings, respectively. By diagonalizing the circuit Hamiltonian $\mathcal{H}(\omega)$, one can obtain the admittance spectra $j_n$ and wave functions $\phi_n$ ($n$ indicates the state number with $n=1,2,3,...,N$).

Analogy to Schrödinger's equation, the homogeneous equations of Eq. \eqref{ME} ($\textbf{I}=0$) can be rewritten as $2N$ differential equations
\begin{equation}
-i\frac{d}{dt}\psi(t)=\mathcal{H}_S\psi(t),
\end{equation}
with $\psi=[\dot{\textbf{V}}(t),\textbf{V}(t)]^T$ and the block Hamiltonian matrix being
\begin{equation} \label{CS}
\mathcal{H}_S=i\left(
               \begin{array}{cc}
                 {\mathcal C}^{-1}\Sigma & {\mathcal C}^{-1}{\mathcal L} \\
                 -\mathbb{I} & 0 \\
               \end{array}
             \right),
\end{equation}
where $\mathbb{I}$ is a $N\times N$ identity matrix. 

Diagonalizing the Schrödinger Hamiltonian \eqref{CS}, one can obtain the frequency spectra and voltage wave functions. The time-evolution $\psi_n(t)$ is given by $\psi_n(t)=\psi_ne^{i\omega_nt}$ ($n=1,2,...,2N$), where $\omega_n$ is given by the roots of the admittance eigenvalues $j(\omega_n)=0$. For a Hermitian system, the eigenfrequencies are real and appear in pairs of $\pm \omega$, and the admittance and frequency spectra (positive part) exhibit an one-to-one correspondence.

In circuit, it is difficult to measure the admittance spectrum because one need to shift the admittance continuously by adding inductors or capacitors between each circuit node and the ground. The frequency spectrum offers an alternative method to measure the band structures, because one can adjust the driving frequency conveniently, which performs well in detecting the in-gap boundary states.

\subsection{Constructions of topolectrical circuits}
\begin{figure}
  \centering
  \includegraphics[width=0.9\textwidth]{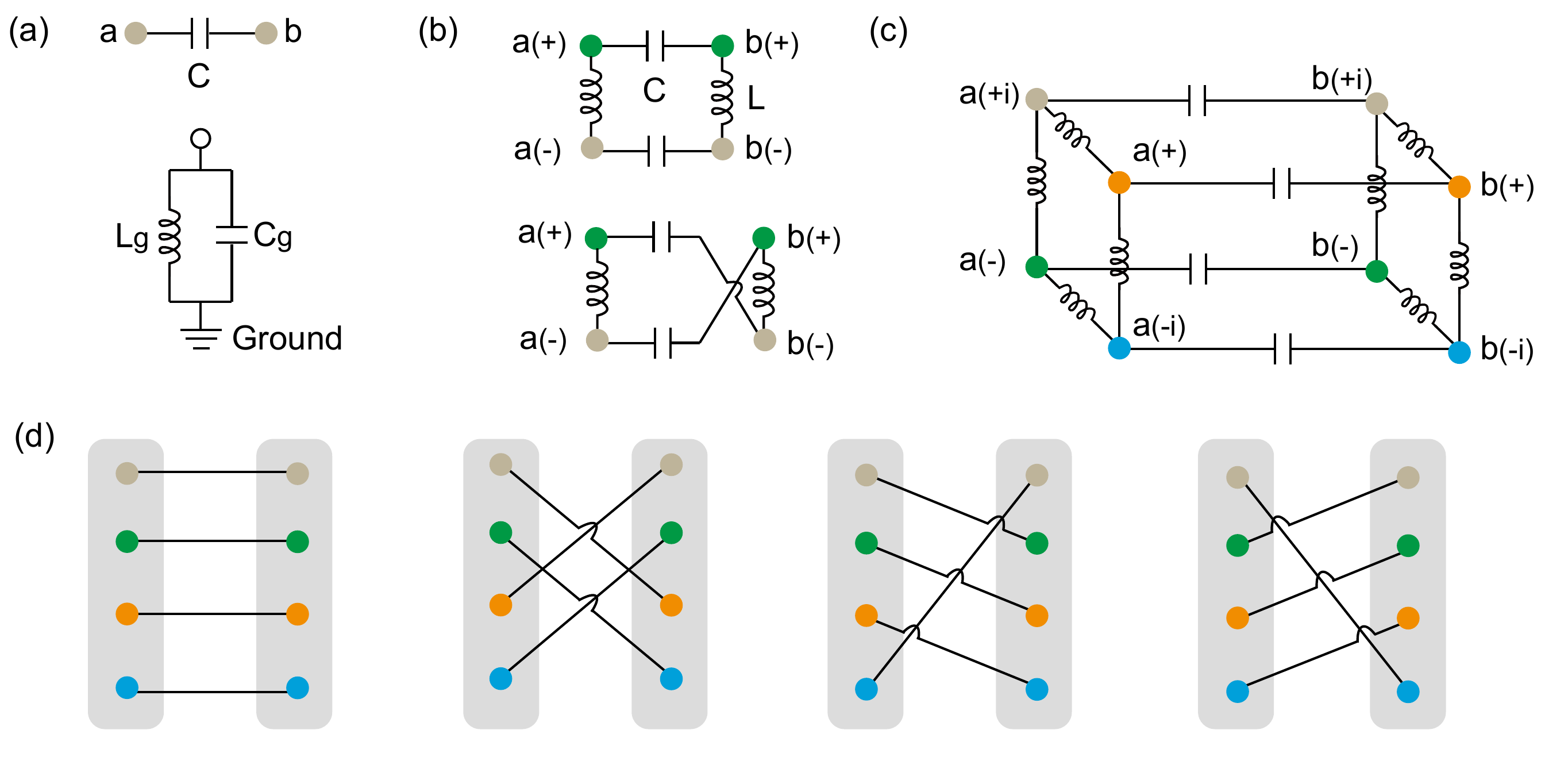}
  \caption{Construction of the topoletrical circuit with (a) one-subnode method, (b) two-subnode method, and (c) four-subnode method. Each subnode is grounded by a resonant unit, i.e., an inductor $L_g$ and a capacitor $C_g$ shown in bottom of (a). (d) Four different connection methods for four-subnode method. }\label{model}
\end{figure}

There are several methods to construct the TECs for realization different hoppings \cite{Zhao289,Helbig161114,Dong023056}. The first one is to use circuit nodes to mimic lattice nodes and capacitors to simulate the hopping terms. We dub this method as \emph{one-subnode} method, as shown in Fig. \ref{model}(a). The circuit Laplacian of the model can be written as
\begin{equation}
{\mathcal J}=D\sigma_0+i\omega C\sigma_1,
\end{equation}
with $D=i\omega [C_g-1/(\omega^2L_g)]$ and $\sigma_1$ Pauli matrix. Here, $i\omega C$ corresponds to the hopping strength and $D$ represents the on-site potential.

However, the \emph{one-subnode} method cannot simulate the case of the hoppings with phase factors, so the \emph{two-subnode} and \emph{four-subnode} structures connected by inductors are proposed, as displayed in Fig. \ref{model}(b) and Fig. \ref{model}(c). One can express the circuit Laplacian of the \emph{two-subnode} model as

\begin{equation}
{\mathcal J}=\sigma_0\otimes \hat{D}+\sigma_0\otimes \hat{C}_\alpha,
\end{equation}
where $\hat{D}=i\omega\{[C_g-1/(\omega^2L_g)]\tau_0+1/(\omega^2L)\tau_1\}$ and $\hat{C}_\alpha=i\omega C\tau_\alpha$ with $\alpha=0,1$ corresponding to the circuit in top and bottom configurations of Fig. \ref{model}(b), respectively. Here, $\sigma_0$ and $\tau_0$ represent identity matrices, and $\sigma_i$ ($i=1,2,3$) and $\tau_i$ are Pauli matrices. If we only concern the voltages on the node $a$ and $b$ regardless of the phase between two subnodes, the hopping and on-site potential can be simplified by projecting $\hat{D}$ and $\hat{C}_\alpha$ to subnode subspace spanned by $\nu_2=\frac{\sqrt{2}}{2}[1,-1]$ with the projector $\hat{P}=\nu_2\nu_2^*$. With this formula, one can obtain 
\begin{equation} \label{project}
\begin{aligned}
t_\alpha&={\rm Tr}(\hat{P}\hat{C}_\alpha\hat{P}),\\
\mu&={\rm Tr}(\hat{P}\hat{D}\hat{P}),
\end{aligned}
\end{equation}
yielding the hopping $t_\alpha=(-1)^\alpha C$ and on-site potential $\mu=i\omega[C_g-1/(\omega^2L_g)-1/(\omega^2L)]$. By repeating the two-subnode units, an arbitrary pure real hopping can be realized in TECs.

Further, the hopping with an arbitrary phase factor can be obtained by generalizing the above construction to include four subnodes instead of two subnodes on a single node. If we consider the four-subnode structures as shown in Fig. \ref{model}(c), the subnode subspace is spanned by $\nu_4=\frac{1}{2}[1,-i,-1,i]$. Similar, the hopping between node $a$ and $b$ can be computed by Eq. \eqref{project} with the projector $\hat{P}=\nu_4\nu_4^*$ and, for example, 
$\hat{C}_\alpha=\left(
                  \begin{array}{cc}
                    0 & \tau_0 \\
                    \tau_0 & 0 \\
                  \end{array}
                \right)
$ for the second connection type. It is straightforward to obtain the hopping strength for the four connection types in Fig. \ref{model}(d), i.e, $t=i\omega C$,  $t=i\omega (-C)$, $t=i\omega (iC)$, and $t=i\omega (-iC)$, respectively. The on-site potential is still the same as the two-subnode case. By combining the two foregoing connection methods, one can realize an arbitrary phase factor from $-\pi$ to $\pi$ between two supernodes, as demonstrated in Ref. \cite{Zhu115410}. 

 \begin{figure}
  \centering
  \includegraphics[width=0.9\textwidth]{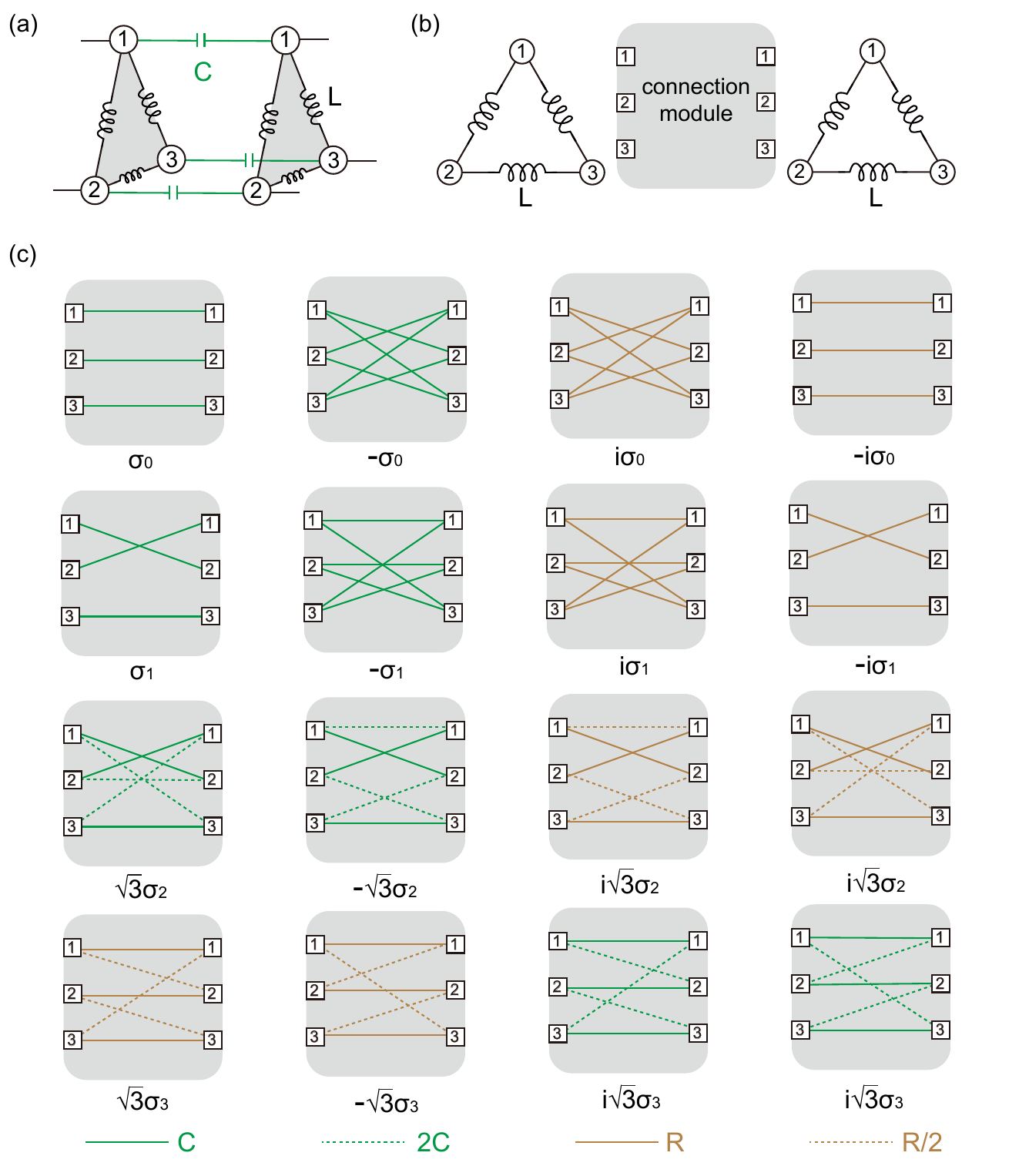}
  \caption{(a) Construction of the TECs with \emph{three-node} method. Each node is composed of three subnodes connected by inductor. (b) Sketch of the connection between two nodes. (c) Circuit realization of all Pauli matrices with different connection modules. }\label{F1Pauli}
\end{figure}

Another construction method is \emph{three-subnode} method, as shown in Fig. \ref{F1Pauli}(a). Three inductors are connected head to tail to form a triangle. For a single unit, one can write the circuit Laplacian
\begin{equation}
\mathcal{J}(\omega)=\frac{1}{i\omega L}M_L,~{\rm with}~M_L=\left(
                                 \begin{array}{ccc}
                                   -2 & 1 & 1 \\
                                   1 & -2 & 1 \\
                                   1 & 1 & -2 \\
                                 \end{array}
                               \right)
\end{equation} 
Diagonalizing this matrix by $U^\dagger M_L U=\Lambda$ where $U=\left(
                     \begin{array}{ccc}
                       1 & \epsilon & \epsilon^* \\
                       1 & \epsilon^* & \epsilon \\
                       1 & 1 & 1 \\
                     \end{array}
                   \right)
$ with $\epsilon=e^{i2\pi/3}$, the eigenvalues are 0, $-3$, $-3$, and the eigenmodes correspond to a single state $\phi_1=\frac{\sqrt{3}}{3}(1,1,1)^{\rm T}$, and twofold degenerate states with the wave functions 
\begin{equation} \label{Tspins}
\begin{aligned}
&\phi_2=\frac{\sqrt{3}}{3}(\epsilon,\epsilon^*,1)^{\rm T},\\
&\phi_3=\frac{\sqrt{3}}{3}(\epsilon^*,\epsilon,1)^{\rm T}.
\end{aligned}
\end{equation}
 The twofold degenerate eigenstates of the circuit can serve as the basis of the pseudospin space.

To realize the interaction between two units, one need to connect the units with the configurations shown in Fig. \ref{F1Pauli}. Taking $\sigma_1$ as an example and considering the interaction between node $m$ and $n$, the Ohm's Law reads
\begin{equation}
\left(
  \begin{array}{cc}
    -2I_3 & M_{\sigma_1} \\
    M_{\sigma_1} & -2I_3 \\
  \end{array}
\right)\left(
                \begin{array}{c}
                  v_m \\
                   v_n \\
                \end{array}
              \right)=\frac{1}{\omega^2LC}
\left(
  \begin{array}{cc}
    0 & M_L \\
    M_L & 0 \\
  \end{array}
\right)\left(
                \begin{array}{c}
                  v_m \\
                   v_n \\
                \end{array}
              \right)
\end{equation}
with $M_{\sigma_1}=\left(
                     \begin{array}{ccc}
                       0 & 1 & 0 \\
                       1 & 0 & 0 \\
                       0 & 0 & 1 \\
                     \end{array}
                   \right)
$. Using the matrix transformation 
\begin{equation}
\left(
  \begin{array}{cc}
    -2I_3 & U^\dagger M_{\sigma_1} U \\
    U^\dagger M_{\sigma_1} U & -2I_3 \\
  \end{array}
\right)\left(\begin{array}{c}
                  \widetilde{v}_m \\
                   \widetilde{v}_n \\
                \end{array}
              \right)=\frac{1}{\omega^2LC}
\Lambda\otimes I_2\left(
                \begin{array}{c}
                  \widetilde{v}_m \\
                   \widetilde{v}_n \\
                \end{array}
              \right)
\end{equation}
where $\widetilde{v}_{m(n)}=U^\dagger v_{m(n)}$ and 
$U^\dagger M_{\sigma_1} U=2\oplus \sigma_1$.
For the two degenerate states in Eq. \eqref{Tspins}, the hopping between cell-$n$ and cell-$m$ is characterized by matrix $\sigma_1$. Similarly, one can realize the all of the Pauli matrices with this connection methods shown in Figs. \ref{F1Pauli}(b) and (c). For further considerations, one can add the voltage follower to these units and achieve the unidirectional hopping Pauli matrices \cite{Wu635}.

With these construction methods mentioned above, one can almost build all of the models in TECs. The next step is to characterize of the topological states, and we will discuss it below.

\subsection{Observables in topolectrical circuits}
In circuit, the most convenient way is to measure the impedance between two nodes. The wave functions and the impedance of the circuit are linked by the circuit Green's function 
\begin{equation}
G=\sum_{j_n}\frac{1}{j_n}\phi_n\phi_n^\dagger,
\end{equation}
with $j_n$ the $n$-th admittance and $\phi_n$ being the wave functions of $n$-th admittance. One can characterize the topological states by the \emph{impedance measurement} \cite{Lee2018}.

\begin{figure}
  \centering
  \includegraphics[width=0.7\textwidth]{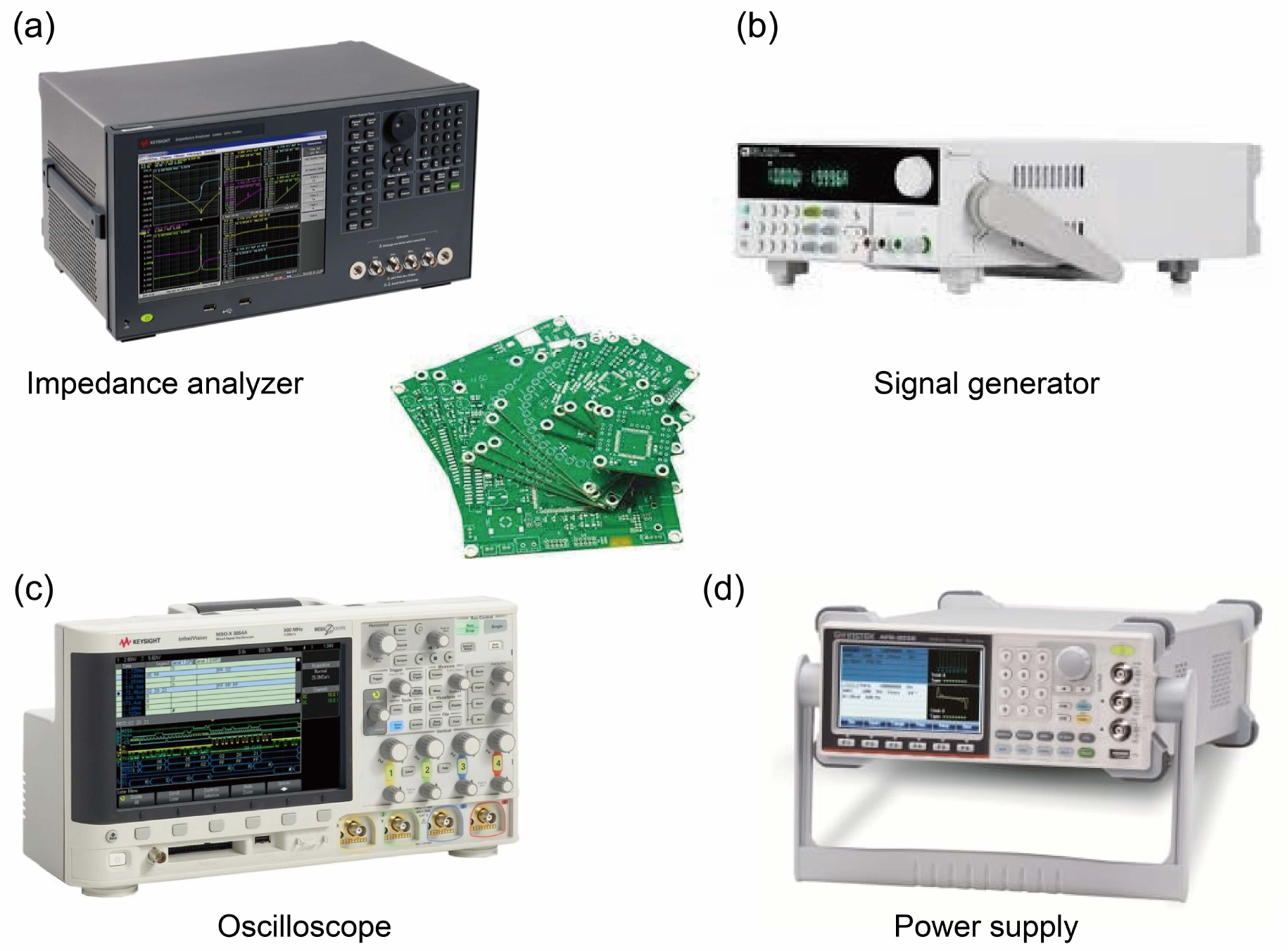}
  \caption{Instruments used in TEC experiments. (a) Impedance analyzer (measure impedance). (b) Signal generator (generate the voltage signals). (c) Oscilloscope (measure voltage response). (d) Power supply (supply power for the active circuit elements like operational amplifier). The center photo shows the printed circuit broads. \\
  {\it Source:} The figures are adapted from the instrument official websites. }\label{F1Exp}
\end{figure}

The two-point impedance of the circuit is given by the formula
\begin{equation}\label{Impedance}
Z_{ab}=\frac{V_a-V_b}{I_{ab}}=\sum_{i=a,b}\frac{G_{ai}I_i-G_{bi}I_i}{I}=
G_{aa}+G_{bb}-G_{ab}-G_{ba}=
\sum_{j_n}\frac{|\phi_{n,a}-\phi_{n,b}|^2}{j_n},
\end{equation}
where $\phi_{n,a}-\phi_{n,b}$ is the amplitude difference between $a$ and $b$ nodes of the $n$th admittance mode. 
The impedance between node $a$ and the ground is given by 
\begin{equation}
Z_{a,{\rm ground}}=\sum_{j_n}\frac{|\phi_{n,a}|^2}{j_n}.
\end{equation}
It shows that the impedance strongly reflects the wave function distribution near $j_n=0$ $\Omega^{-1}$ since it diverges there if there is no dissipation. In circuit, the loss is inevitable, caused by both the circuit elements and wires, which introduce a small imaginary part to the admittance spectrum, so the divergence of impedance can be avoided. In experiment, one can connect the circuit node and the instrument ports by wires, and the impedance can be measured directly with the impedance analyzer shown in Fig. \ref{F1Exp}(a). 

Alteratively, one can also measure the static voltage signals by adding an external current source to characterize the eigenmodes, because the voltage distributions can be expressed as
\begin{equation}
V=GI=\sum_{j_n}\frac{1}{j_n}\phi_n\phi_n^\dagger I.
\end{equation}
Similarly, the voltage distributions are determined by the eigenstates near $j_n=0$ $\Omega^{-1}$.

The static eigenmodes can be characterized by the impedance or static voltage measurement, while it is invalid to obtain the propagation of the boundary states. To see the time evolution of the signals, one can measure the \emph{time-dependent voltage propagation} in circuit. In experiment, one can inject a current signal by the wave signal generator [Fig. \ref{F1Exp}(b)] at one (sub)node and measure the voltage response at other nodes with the oscilloscopes [Fig. \ref{F1Exp}(c)], then obtain the time-dependent propagations of the voltage signals. This method is extremely important in detecting the chirality of edge states or hinge states. Figure \ref{F1Exp}(d) shows a photo of DC power supply, which is used to provide power for the active elements like operational amplifier.

So far, we have introduced the fundamentals of TECs, including the electrical elements, networks, descriptive equations, and master the knowledge of how to observe the topological states. Then, we will take advantages of these knowledge to comprehend topological states in electrical circuits. 

\section{Circuit realization of topological insulators (TIs)}  \label{S3}
The discovery of the quantum Hall effect refreshes the cognition of phase of matter, leading to a new phase classification via topological order \cite{Klitzing1980}. The quantized Hall conductance and its robustness have been successfully explained by the Thouless-Kohmoto-Nightingale-den Nijs (TKNN) topological index \cite{Thouless1982}. Later, Haldane put forward the quantum Hall effect without Landau levels in graphene by adding periodic magnetic flux \cite{Haldane2015}. Kane-Mele predicted the quantum spin Hall insulator [two-dimensional (2D) TIs] in 2005, which is the spin version of quantum Hall effect \cite{Kane2005,Kane20052}, and this interesting phase was confirmed in HgTe quantum well later \cite{Bernevig2006,Konig2007,Buttiker278}. Subsequently, the 2D TIs were generalized to three-dimensional (3D) cases \cite{FuTI3D,Moore2007,Roy2009,Fu2007}, which are divided into strong and weak TIs, hosting metallic surface states but with an odd and even number of Dirac cones, respectively. For the free-fermion systems, one can classify the topological phases by their symmetries, i.e., time-reversal, particle-hole, and chiral symmetries, called $\mathcal{AZ}$ class \cite{Chiu2016,Altland1142}. Introducing the additional spatial symmetries, such as reflection, rotation, and inversion, to the topological phase may modify the topological classifications and lead to new gapped topological phases protected by these symmetries, named topological crystalline insulators \cite{Ando361} including the higher-order topological insulators (HOTIs) \cite{BXie2021}. It can be seen that the study of TIs/HOTIs is a prosperous research field \cite{Hasan2010,Qi2011,Chiu2016}, and the abundant topological phases and their applications arouse researchers' enthusiasm.  

Bulk-boundary correspondence lies at the heart of topological physics, relating robust edge states to bulk topological invariants: non-zero bulk topological indexes indicate the boundary states one dimension lower than the system dimension. Recent works extend the conventional bulk-boundary correspondence and identify the HOTI \cite{BXie2021}, in which the codimension of boundary states is larger than one. The HOTIs were first proposed in electronic system \cite{Benalcazar2017,Bernevig2017,Song2017,Langbehn2017,Schindler2018,Ezawa2018_1,Khalaf2018,Geier2018,Queiro2019,Benalcazar2019,Peterson2018,Peterson2020}, which have attracted considerable attention by the broad community of photonics \cite{Xie2018,Noh2018,Hassan2019,Mittal2019,Chen2019,Xie2019,Ota2019,ZhangL2019,LiM2019}, acoustics \cite{Xue2019,Ni2019,Xue2019_2,He2019,ZhangX2019,ChenZ2019,ZhangZ2019}, mechanics \cite{Serra-Garcia2018,Fan2019}, and spintronics \cite{Li2019_1,Li2019_2}. Examples of HOTI states include corner states in two or higher-dimensional systems and hinge states in three or higher-dimensional systems.

The electrical circuit is found to be an excellent platform for studying topological insulating states. In this section, we follow the development venation of the (higher-order) TIs and review the results on the topological insulating states observed in electrical circuits.

\subsection{The first-order topological insulators in circuits}
First of all, we backtrack on the conventional topological states in circuit. We begin from the construction of the Su-Schrieffer-Heeger (SSH) model in circuit, and then introduce the circuit realization of Chern insulators and $\mathbb{Z}_2$ topological insulators. Beyond these standard models, we also show the circuit realization of the weak TIs and Anderson TIs and sketch other first-order topological states in circuits.

\subsubsection{Su-Schrieffer-Heeger circuits}

\begin{figure}
  \centering
  \includegraphics[width=0.8\textwidth]{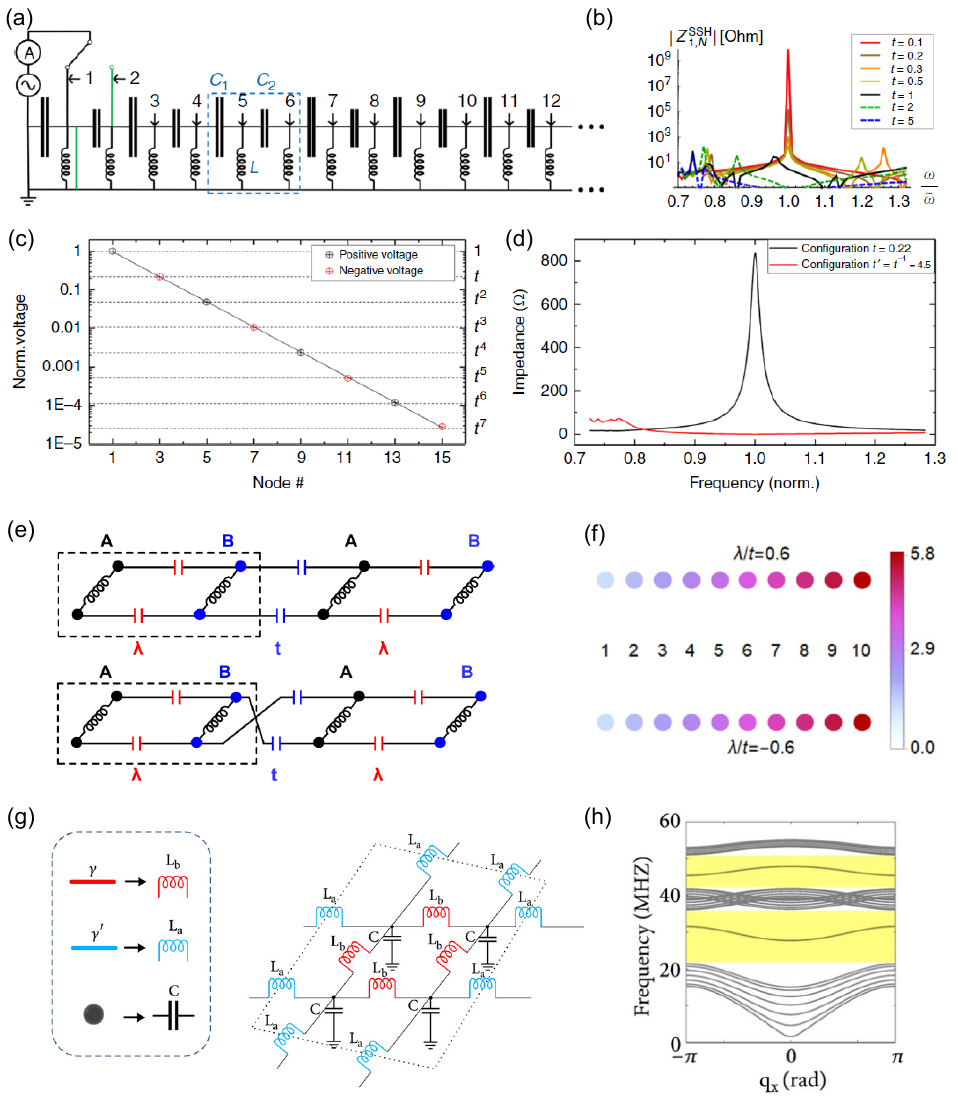}
  \caption{(a) Circuit realization of SSH model with the \emph{one-subnode} method. (b) Numerical impedance between the first and the last nodes in SSH circuit with different hopping ratios. (c) Exponentially decay of edge modes (voltage signals). (d) Measured impedance for the trivial and nontrivial case. (e) The construction of SSH circuit with \emph{two-subnode} method. (f) The impedance distribution of the SSH circuit chain with positive (top) and negative (bottom) hopping ratios. (g) 2D SSH circuit model. (h) The non-trivial band structures of 2D SSH model with the isolated edge modes located in the band gap.\\
  {\emph{Source}:} The figures are adapted from Refs. \cite{Lee2018,Dong023056,SLiu2019}.}\label{F1}
\end{figure}

The study of SSH model originates from the solitons in Polyacetylene \cite{Su1698,Heeger781}. Later, SSH model becomes an effective model to study topological states because of its nontrivial nature and concise structure. It provides valuable insights into the role of symmetry and topology in determining the electronic properties of materials and has contributed to the development of the broader fields in topology \cite{Weimann433,Liu1904784,Go134423}. The bulk Hamiltonian of SSH model is expressed as \cite{Asbothbook}
\begin{equation}\label{SSHH}
\mathcal{H}=(v+w\cos k)\sigma_x+w\sin k\sigma_y,
\end{equation}
where $\sigma_x$ and $\sigma_y$ are the Pauli matrices, $v$ and $w$ represent the hopping amplitudes within and between unit cells.

In electrical circuits, one can reproduce the SSH model merely by LC network. As shown in Fig. \ref{F1}(a), Lee et al. proposed a SSH circuit with \emph{one-subnode} method. By choosing two nodes as the unit cell [dashed blue rectangle in Fig. \ref{F1}(a)], the corresponding circuit Laplacian for the infinite SSH circuit reads 
\begin{equation} \label{SSHJ}
J_{\rm SSH}=i\omega(C_1+C_2-\frac{1}{\omega^2L})\mathbb{I}-i\omega[(C_1+C_2\cos k)\sigma_x+C_2\sin k \sigma_y],
\end{equation}
with $\mathbb{I}$ the identity matrix \cite{Lee2018}. At the resonant frequency $\omega=\sqrt{(C_1+C_2)L}$, the first term of Eq. \eqref{SSHJ} vanishes, and the circuit model is fully consistent with tight-binding model of electronic SSH model [Eq. \eqref{SSHH}]. The topological index winding number is given by 
\begin{equation}
w=\int_0^{2\pi}d\left[\tan^{-1}\frac{\sin k}{C_1/C_2+\cos k}\right]=\theta(C_2-C_1)
\end{equation}
with $\theta$ the step function, which indicates the existence of the mid-gap state for $C_1<C_2$.

Considering a finite-size SSH circuit chain, the mid-gap eigenvalue $j_0$ is
\begin{equation}
j_0=i\omega\frac{(-t)^N(1-t^2)}{1-t^{2[N/2]}},
\end{equation}
which is very close to 0 and decays with the node number $N$ exponentially. [$N/2$] indicates the down round. Therefore, one can measure the mid-gap state by impedance directly.

Figure \ref{F1}(b) shows the impedance between the first and the last nodes for different hopping ratios, where one can see a clear impedance at resonant frequency in topological regions. Further, numerical calculations show that the voltage signal of edge modes exponentially decays from the boundary nodes, as shown in Fig. \ref{F1}(c). One can distinguish the trivial and nontrivial cases by measuring impedance experimentally, as shown in Fig. \ref{F1}(d), and the impedance peak only appears in the topological phase. As we can see from this work, electrical circuit performs well in simulating the tight-binding model, and one can study topological physics in circuit readily.

Similar SSH circuits are proposed by the \emph{two-subnode} method in Ref. \cite{Dong023056}. The detailed circuits are shown in Fig. \ref{F1}(e), and the Hamiltonians reads 
\begin{equation}
\mathcal{H}=(\lambda+t\cos k)\sigma_x\pm t\sin k\sigma_y,
\end{equation}
where the $+$ and $-$ are corresponding to the top circuit describing a standard SSH model and the bottom circuit representing the SSH model with negative hopping amplitude, respectively. This system are topologically nontrivial for $|\lambda/t|<1$. Figure \ref{F1}(f) shows the impedance distribution of this SSH chain with a logarithm plot, where one can clearly observe the edge modes for both two systems.

Later, SSH model was extended to the 2D system, as shown in Fig. \ref{F1}(g), composed of inductor networks with each node grounded by a capacitor. The circuit Laplacian is
\begin{equation}
J=i\omega\left(
           \begin{array}{cccc}
             C-\frac{2}{\omega^2L_a}-\frac{2}{\omega^2L_b} & \frac{e^{-ikx}}{\omega^2L_a}+\frac{1}{\omega^2L_b} & \frac{e^{-iky}}{\omega^2L_a}+\frac{1}{\omega^2L_b} & 0 \\
             \frac{e^{ikx}}{\omega^2L_a}+\frac{1}{\omega^2L_b} & C-\frac{2}{\omega^2L_a}-\frac{2}{\omega^2L_b} & 0 & \frac{e^{-iky}}{\omega^2L_a}+\frac{1}{\omega^2L_b} \\
             \frac{e^{iky}}{\omega^2L_a}+\frac{1}{\omega^2L_b} & 0 & C-\frac{2}{\omega^2L_a}-\frac{2}{\omega^2L_b} & \frac{e^{-ikx}}{\omega^2L_a}+\frac{1}{\omega^2L_b} \\
             0 & \frac{e^{iky}}{\omega^2L_a}+\frac{1}{\omega^2L_b} & \frac{e^{ikx}}{\omega^2L_a}+\frac{1}{\omega^2L_b} & C-\frac{2}{\omega^2L_a}-\frac{2}{\omega^2L_b} \\
           \end{array}
         \right)
\end{equation}
with $k$ the wave vector. The topological phase transition is characterized by Zak phase, which takes $\pi$ for $L_a<L_b$. Considering a ribbon configuration in the topological regime, one can find two isolated bands inside the frequency band gap, corresponding to the edge states \cite{SLiu2019}, as displayed in Fig. \ref{F1}(h). Further, these edge modes are demonstrated in the finite-size circuit experiments and against the defects.

The circuit SSH model can also be used to study the effects of some symmetries on topological edge modes. Ventra et al. show that the custodial symmetry of the chiral type can be realized in a classical SSH electrical circuit with memory \cite{Ventra097701}. Memory can induce nonlinearities that breaks the chiral symmetry of SSH circuit and spread the nontrivial boundary state to bulk nodes, but the extended state is still protected against perturbations in the presence of custodial symmetry. 


\subsubsection{Chern insulators in circuits}
\begin{figure}
  \centering
  \includegraphics[width=0.9\textwidth]{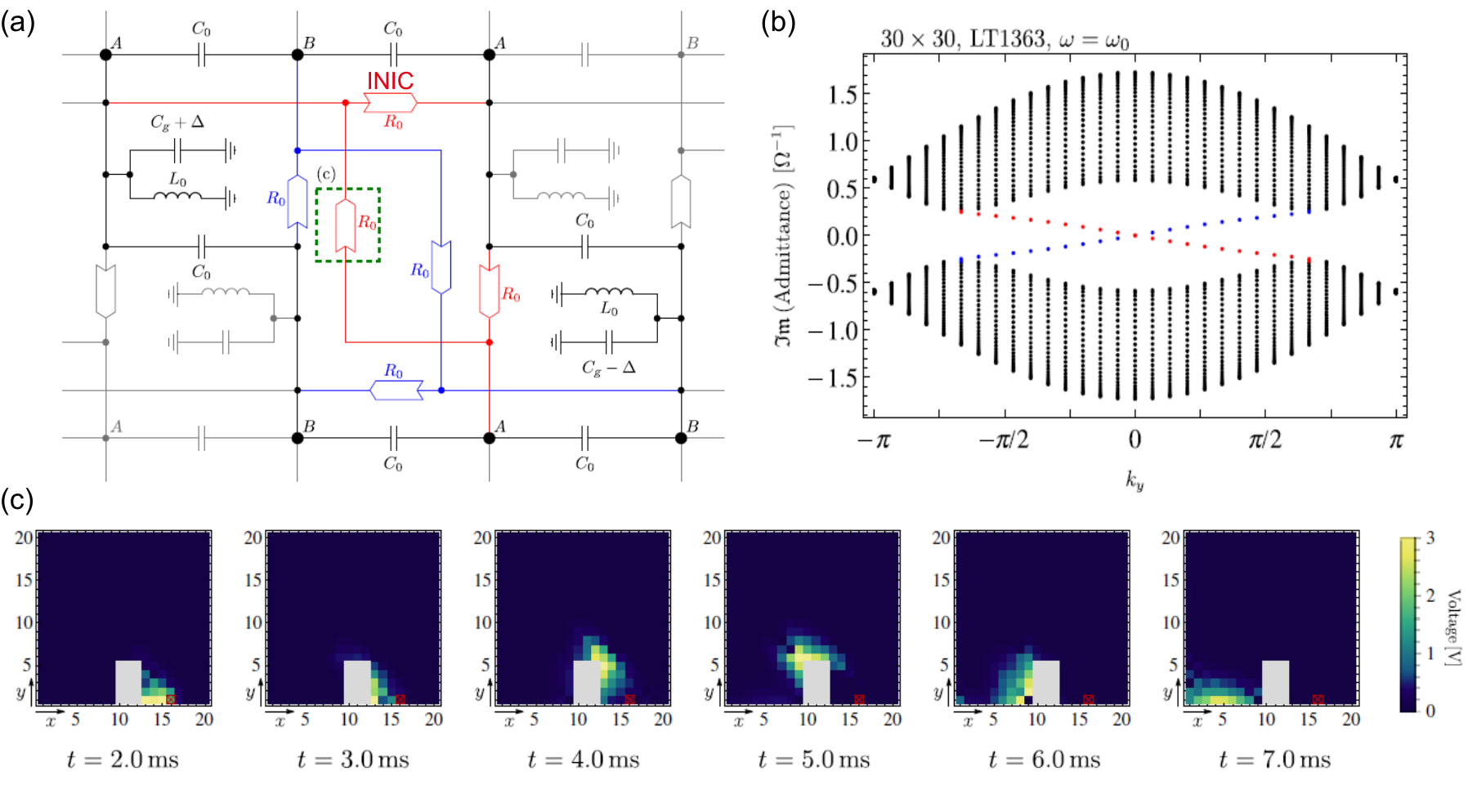}
  \caption{(a) The circuit model of the Chern insulator. (b) The admittance of Chern insulator. The red and blue spectra indicate the edge modes in two different boundaries. (c) The propagation of voltage signal undergoing a defect.\\
  {\emph{Source}:} The figures are adapted from Ref. \cite{Hofmann2019}.}\label{F2}
\end{figure}

Chern insulator, as the lattice counterpart of the quantum Hall effect, is the prototype of TIs. For realization of Chern circuit, the key point is to break time-reversal symmetry (TRS), and TRS is specifically expressed by circuit Laplacian $J({\bf k})=-J^*(-{\bf k})$. Hofmann et al. proposed a Chern circuit model with unidirectional voltage mode \cite{Hofmann2019}. This circuit consists of capacitors, inductors, and INICs, as shown in Fig. \ref{F2}(a). The INIC unit consists of three resistors, which breaks the time-reversal symmetry and the reciprocity of the circuit. The circuit Laplacian for an infinite model is written as
\begin{equation}
\begin{aligned}
J_{\rm CI}=&i\omega\big\{(3C_0+C_g-\frac{1}{\omega^2L})\mathbb{I}-
C_0[1+\cos(k_x)+\cos(k_y)]\sigma_x
-C_0[\sin(k_x)+\sin(k_y)]\sigma_y\\
&+[\Delta+\frac{2}{\omega R_0}
[\sin(k_x)-\sin(k_y)-\sin(k_x-k_y)]\sigma_z]\big\},
\end{aligned}
\end{equation}
with $\omega$ the driving frequency, $k_{x,y}$ being the wave vectors, $C_0$, $C_g$, $R$, and $\Delta$ are circuit parameters.

For a Chern insulator, one can calculate the Chern number for the lower admittance spectrum as
\begin{equation}\label{Chernnumber}
\mathcal{C}=\frac{1}{2\pi}\oint d^2k \mathcal{A}({\bf k})
\end{equation}
to characterize the bulk topology, where $\mathcal{A}({\bf k})$ is the Berry curvature. In this system, the Chern number is given by 
\begin{equation}
\mathcal{C}=\frac{1}{2}\left[{\rm sgn}\left(\Delta+\frac{3\sqrt{3}}{\omega R_0}\right)
-{\rm sgn}\left(\Delta-\frac{3\sqrt{3}}{\omega R_0}\right)\right],
\end{equation}
which is nonzero for $\omega R_0<3\sqrt{3}\Delta$.

Considering the parameters in nontrivial region, one can study an infinite-ribbon configuration to see the in-gap edge modes. The admittance spectrum for a ribbon along $\hat{y}$ direction is displayed in Fig. \ref{F2}(b). Near the admittance $j_n=0$ $\Omega^{-1}$, two isolated modes appear with opposite propagation directions labeled by the red and blue color dots inside the band gap, which correspond to the chiral edge states along $\hat{y}$ direction. Figure \ref{F2}(c) shows the time-evolution of chiral edge modes in finite-size circuit system simulated by the software LTspice. The boundary state propagates in a clockwise sense and passes the defect successfully, manifesting the robustness of edge state.

Similarly, Haenel et al. reported an electrical network that realizes Chern insulating phase for electromagnetic waves \cite{Haenel235110} with the time-reversal symmetry broken by a class of weakly dissipative Hall resistor elements. Further, Wang et al. proposed a general scheme to realize the Chern states with arbitrarily large Chern numbers in the circuit by controlling the long-range couplings precisely \cite{WangL201101}. Yang et al. reported a Haldane circuit model with modified next-nearest-neighbour interactions, and the antichiral edge states are observed in circuit \cite{Yang257011}. Thereby, circuit provides an accessible physical platform for realizing Chern insulating state, which can be used to manufacture robust unidirectional waveguides.

\subsubsection{$\mathbb{Z}_2$ TIs in circuits}

Beyond the quantum Hall effect, the spin version of the quantum Hall effect represents a large amount of TIs, i.e., 2D TIs. Kane-Mele model is a standard paradigm for studying the quantum spin Hall effect  \cite{Kane2005,Kane20052}, but it cannot be realized in the prototype graphene model due to the weak spin-orbit coupling. Fortunately, the strength of spin-orbit coupling is tunable in circuit.

\begin{figure}
  \centering
  \includegraphics[width=0.8\textwidth]{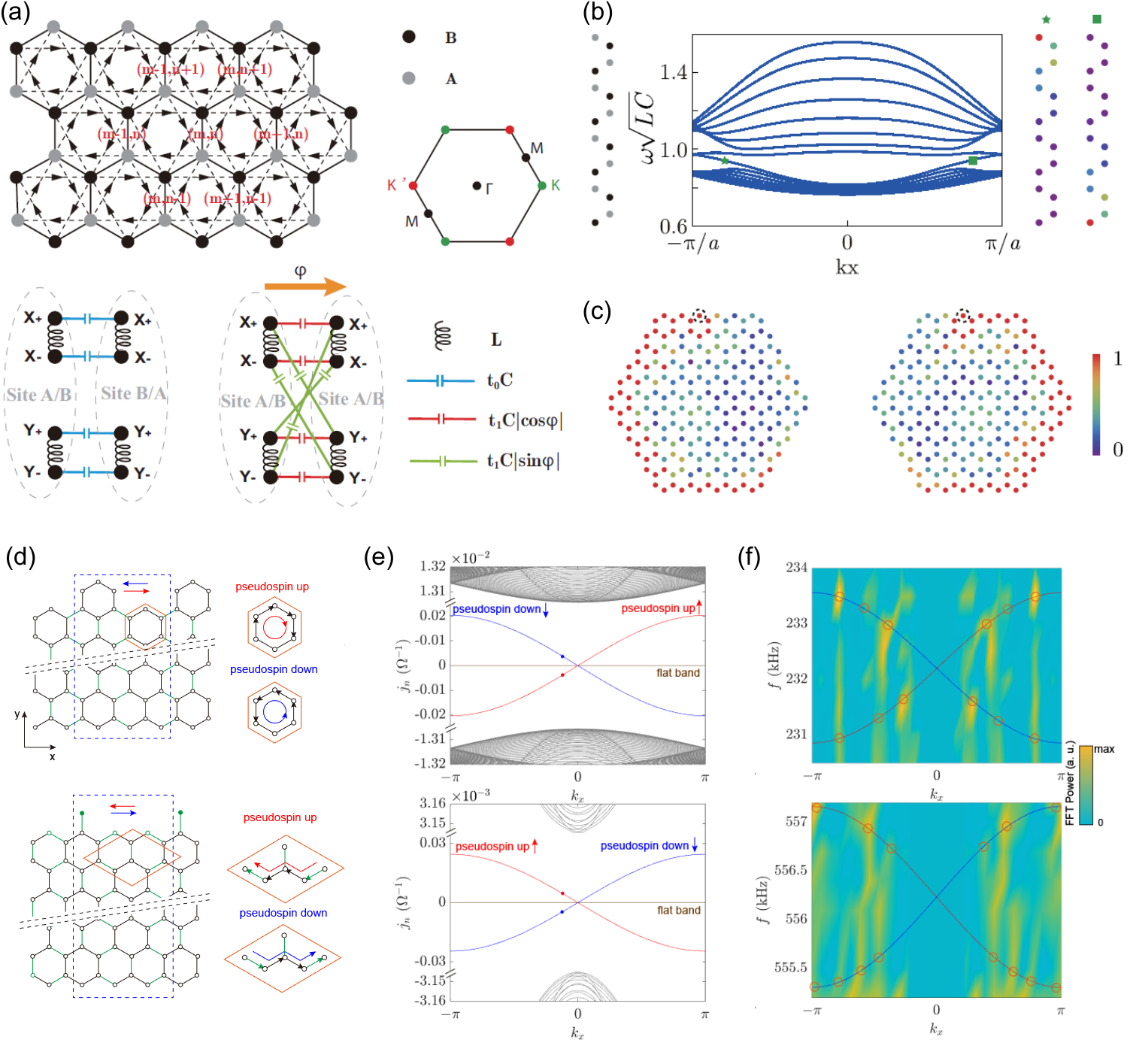}
  \caption{(a) Circuit model of $\mathbb{Z}_2$ topological insulator. (b) The frequency spectrum of $\mathbb{Z}_2$ topological insulator. The in-gap spectra indicate the helical edge modes. (c) The propagation of voltage signal. The spin-up and spin-down edge modes spread with opposite chiralities. (d) The quantum pseudospin Hall models in the two Kekul\'{e} electric circuits with molecule-zigzag and partially bearded edges. (e) Calculated admittance spectra with helical edge modes in the band gaps. (f) Measured edge spectra in frequency space.\\
  {\emph{Source}:} The figures are adapted from Refs. \cite{Zhu115410,Yang2021}.}\label{F3}
\end{figure}

As shown in Fig. \ref{F3}(a), Zhu et al. proposed a circuit model for realization of Kane-Mele Hamiltonian \cite{Zhu115410}
\begin{equation}\label{KMH}
{\mathcal H}=\sum_{i;s}u_ic_{i,s}^\dagger c_{i,s}+t_0\sum_{\langle i,j\rangle;s}c_{i,s}^\dagger c_{j,s}
+t_1e^{iv\phi}\sum_{\langle\langle i,j\rangle\rangle;s}c_{i,s}^\dagger c_{j,s},
\end{equation}
where $c_{i,s}(c_{i,s}^\dagger)$ is the annihilation (creation) operator for a particle on site $i$ with $s=\uparrow$ for spin up and $s=\downarrow$ for spin down. $\langle i,j\rangle$ and $\langle\langle i,j\rangle\rangle$ represents the sum between nearest-neighbor nodes and next-nearest-neighbor nodes, respectively. $u_i$ is the on-site potential. $t_0$ and $t_1$ represent the hopping strengths. $\phi$ denotes the phase factor.

In order to produce the Hamiltonian \eqref{KMH} in circuit, two coupled inductors ($X$ and $Y$) are used to simulate the "spin" with alternate connections to realize an arbitrary phase factor, as displayed in Fig. \ref{F3}(a). The "spins" are denoted by $U_{\uparrow,\downarrow}=U_X\pm U_Y$. The circuit model hosts the same topological regime as the Kane-Mele one \cite{Kane2005,Kane20052}. From the frequency spectra plotted in Fig. \ref{F3}(b), one can find two isolated edge modes near the resonant frequency $\omega\sqrt{LC}=1$. For a finite-size sample, the spin-up and spin-down edge modes propagate along the boundary in opposite directions, as shown in Fig. \ref{F3}(c).

Such multi-orbital TECs are also investigated in Ref. \cite{Yao021001}. Yao et al. simulated the $(p_x,p_y)$-orbital model with two sets of inductors and realized the quantum spin Hall and quantum anomalous Hall states, and found that these two states can be converted to each other by simply tuning the resistances in circuit.

In Ref. \cite{Yang2021}, Yang et al. discovered the edge-dependent quantum pseudospin Hall effects in Kekul\'{e} electric circuits with molecule-zigzag and partially bearded edges, as shown in Fig. \ref{F3}(d). The Hamiltonian of an infinite Kekul\'{e} circuit reads
\begin{equation} \label{H1}
\mathcal{H}=-\omega_0 C_A\sum_{\left<i,j\right>}c_i^\dagger c_j-\omega_0 C_B\sum_{\left<i',j'\right>}c_{i'}^\dagger c_{j'},
\end{equation}
where $\left<i,j\right>$ and $\left<i',j'\right>$ run over nearest-neighboring sites inside and between hexagonal unit cells with the hopping strengthes $\omega_0 C_A$ and $\omega_0 C_B$, respectively. The authors map the above Hamiltonian to the four-band Bernevig-Hughes-Zhang (BHZ) model and define the chirality of current inside the unit cell as pseudospin. The helical edge modes are observed by calculating and measuring the admittance and frequency spectra in admittance and frequency spaces, respectively, as shown in Figs. \ref{F3}(e) and (f). One can clearly see the helical edge modes in the band gap.

The helical edge modes can also be extended to high-frequency regimes ($\sim$GHz). Li et al. established a hexagonal LC circuit with planar microstrip arrays and found the topological interfacial electromagnetic waves propagate in opposite directions for different orbital angular momentums \cite{LiNC2018}.

In a nutshell, the spin-resolved physics can be fully explored in circuit platforms. Although the definition of pseudospins are different, the $\mathbb{Z}_2$ topological states can be established well in this emergent field.

\subsubsection{Other TIs in circuits}
Expect for the quantum (spin) Hall phase, electrical circuits can also be used to study other topological states, such as the Anderson TI and 2D weak TI. We briefly introduce these two cases below.

\begin{figure}
  \centering
  \includegraphics[width=0.85\textwidth]{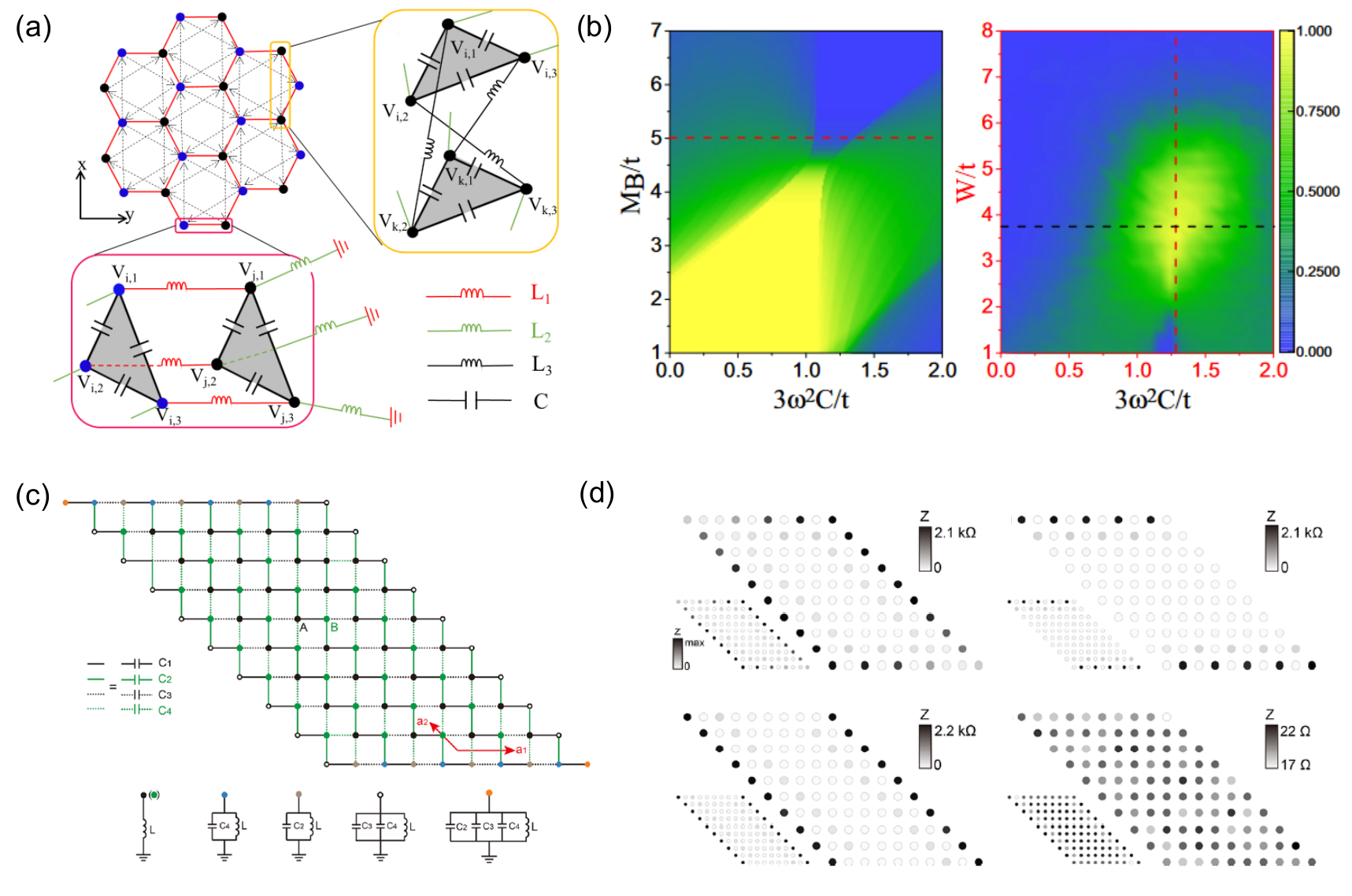}
  \caption{(a) Circuit realization of the Haldane model. (b) Disorder induced topological Anderson phase transition. (c) Circuit model of the 2D weak TIs. (d) Measured impedance distributions with edge states locating at certain sample boundaries.\\
  {\emph{Source}:} The figures are adapted from Ref. \cite{Zhang184202,Yang3125}.}\label{F4}
\end{figure}

Topological Anderson insulator is a distinct kind of TIs, which is induced by disorders with edge modes protected by the mobility gap \cite{Li136806,Groth196805}. In Ref. \cite{Zhang184202}, Zhang et al. reproduced the Haldane model in circuit as shown in disordered Fig. \ref{F4}(a). The Hamiltonian is expressed as
\begin{equation}
\mathcal{H}=-[\sum_{\left<i,j\right>}ta_i^\dagger b_j+\sum_{\ll i,j\gg}t'e^{i\frac{2\pi}{3}}(a_i^\dagger b_k
+a_i^\dagger b_k)+{\rm H.c.}]+\sum_i [(\Delta M_A+\epsilon_i)a_i^\dagger a_i+(\Delta M_B+\epsilon_i)b_i^\dagger b_i],
\end{equation}
where $a_i$, $b_i$ ($a_i^\dagger$, $b_i^\dagger$) are the annihilation (creation) operator for the AB sub-lattice. $t$ and $t'$ are the hopping strengths between nearest-neighbor and next-nearest-neighbor nodes, respectively. $\Delta M_A$ and $\Delta M_B$ are the staggering potential in the sublattice A and B, respectively, and $\epsilon_i$ is the Anderson disorder (uniformly distributed in the range [$-\frac{W}{2}$, $\frac{W}{2}$] with $W$ the disorder strength). Introducing strong disorder to topologically trivial phase of this system, one expects to find the topological edge states. It is convenient to introduce disorder to circuit systems because one can control the on-site potential for each node precisely. Figure \ref{F4}(b) (right) displays the Chern number for different disorder strengths and a topological phase appears induced by disorder.

By introducing disorders to the modified Kane-Mele model, Zhang et al. realized the Anderson TIs, as reported in Ref. \cite{Zhang195304}. The disorder recovers symmetry under statistical averaging and transforms the trivial insulating phase to the topological one. As we can see from the above two examples, circuits are convenient in studying topological phases of disordered systems.

Recently, a theoretical work predicted the existence of the 2D weak TIs \cite{Jeon121101}, as an analog of 3D weak TIs, but with the lower dimension. In electronic systems, it is difficult to detect the edge modes of weak TIs because they are fragile to disorder \cite{Noguchi2019,Zhang2021}. Yang et al. realized this proposal in circuit experiments \cite{Yang3125}, as shown in Fig. \ref{F4}(c). The circuit Hamiltonian reads
\begin{equation}
\mathcal{H}(\omega)=-iJ(\omega)=\omega\left(
              \begin{array}{cc}
                h_0 & h_1 \\
                h_1^* & h_0 \\
              \end{array}
            \right),
\end{equation}
where $h_0=C_1+C_2+C_3+C_4-1/(\omega^2 L)$ and $h_1=-C_1-C_2e^{i{\bf k}\cdot{{\bf a}_2}}-C_3e^{-i{\bf k}\cdot{{\bf a}_1}}-C_4e^{-i({\bf k}\cdot{{\bf a}_1}+{\bf k}\cdot{{\bf a}_2})}$ with {\bf a}$_1=2(1+e)\hat{x}$ and {\bf a}$_2=-(1+e)\hat{x}+(1-e)\hat{y}$ the two basic vectors and $\bf k$ being the wave vector.

Imposing some special distortions on the SSH lattice (square-to-rectangle and staggered distortions), this system exhibits different topological phases, which are characterized by the strong $\mathbb{Z}_2$ topological indexes
\begin{equation}
(-1)^{\nu_0}=\prod_{i=1}^4\delta_i,
\end{equation}
and weak topological invariants, i.e., $\nu_1\nu_2$, along {\bf a}$_1$ and {\bf a}$_2$ directions, similar to the treatment for 3D WTIs
\begin{equation}
(-1)^{\nu_k}=\prod_{n_k=1,n_{j\neq k}=0,1}\delta_{{\mathbf i}=(n_1n_2)},
\end{equation}
where $\delta_i=\xi(\Gamma_i)$ is the parity eigenvalues at four time-reversal invariant momenta $\Gamma_{i=(n_1n_2)}=\frac{1}{2}(n_1{\bf b}_1+n_2{\bf b}_2)$ ($n_{1,2}=0$ or 1), with ${\bf b}_1$ and ${\bf b}_2$ being the reciprocal-lattice vectors.

In this system, one can find the non-vanishing strong $\mathbb{Z}_2$ index which means the existence of Dirac SM phases, and the nonzero weak $\mathbb{Z}_2$ invariants that indicates the existence of dispersionless (flat-band) edge modes in certain boundaries. By constructing a finite-size sample, one can observe the edge modes by measuring the impedance distributions, as shown in Fig. \ref{F4}(d), which are consistent with the weak topological indexes ($\nu_1\nu_2$) equaling to $(11)$, $(10)$, $(01)$, and $(00)$, respectively. Zhu et al. predicted an 1D flat band also at the domain wall of 2D weak TIs \cite{Zhu041117}, and this phenomenon is also been observed in Ref. \cite{Yang3125}. In the future, circuit may be used to study more flat-band physics.

As we can see from the aforementioned works, abundant topological insulating phases have been investigated with electrical circuits, and TEC offers a powerful platform for exploring these topological phases. These topological boundary states, such as the chiral edge states in Chern insulator, provide opportunities for realizing the robust propagation of voltage signals.

\subsection{Higher-order topological insulators (HOTIs) in circuits}

In this section, we will move to another important class of topological insulating phase, i.e., the HOT states in circuit, including the circuit realization of quantized multipole insulators, HOTIs from generalized SSH models, long-range interactions induced HOTIs, and other HOTIs.

\begin{figure}
  \centering
  \includegraphics[width=0.8\textwidth]{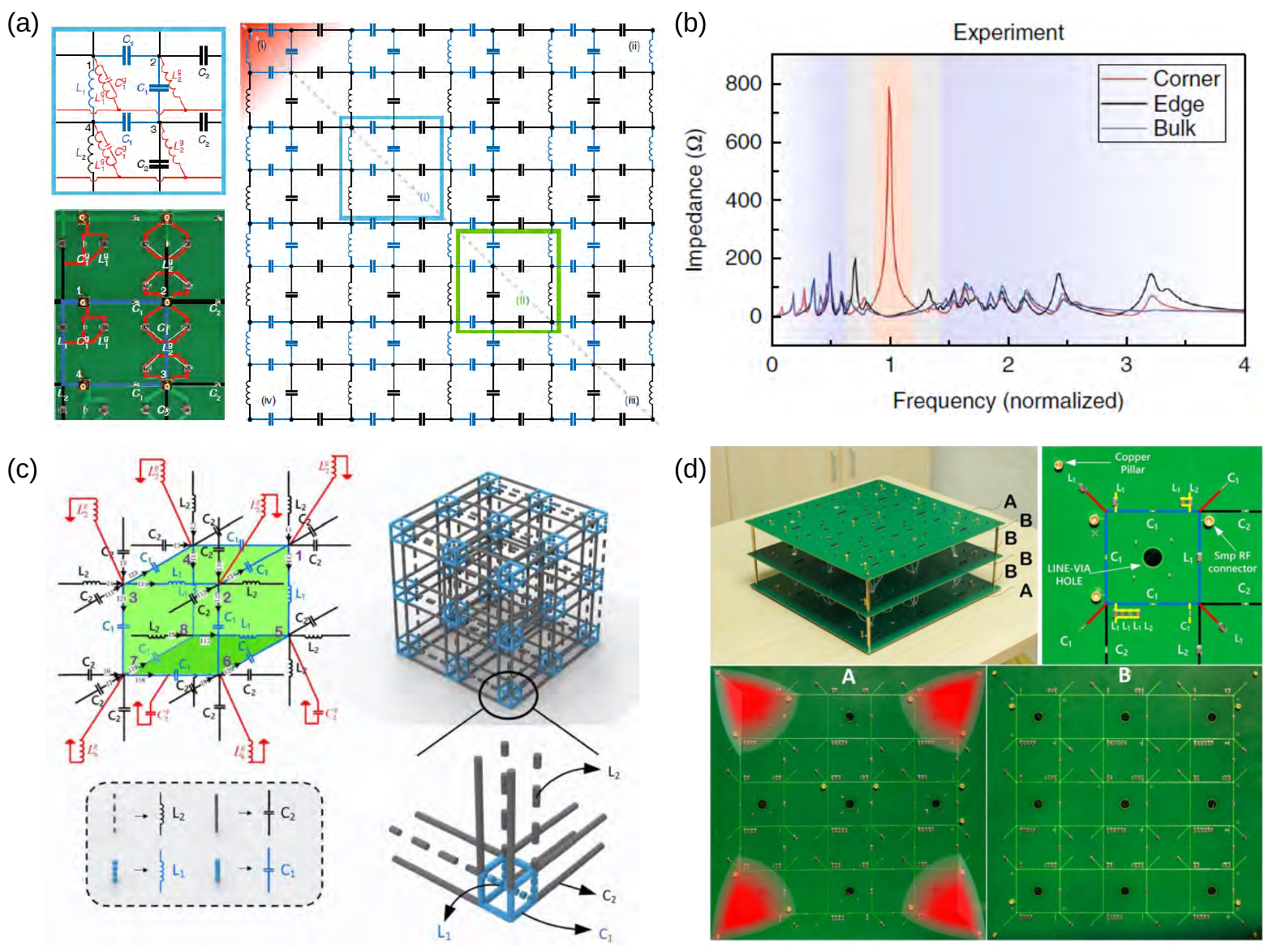}
  \caption{(a) Circuit realization of the quantized quadrupole insulator. (b) Measured impedance from corner, edge, and bulk nodes as a function of driving frequency.  (c) Circuit realization of the quantized octupole insulator. (d) Printed circuit board with corner modes labelled by the red patterns.\\
  {\emph{Source}:} The figures are adapted from Ref. \cite{Imhof2018,Bao201406}.}\label{F5}
\end{figure}

\subsubsection{Quantized multipole insulators}
The quantized multipole insulator was first proposed in electronic system, which hosts topologically protected corner states carrying fractional charge. To design the quadrupole in classical system, the most convenient method is starting from complementary protecting-symmetry perspective. For example, two noncommuting reflection symmetries $M_x$ and $M_y$ as well as the $C_4$ rotation symmetry can be used to realize the quantized quadrupole moment. Chiral symmetry is introduced to pin the boundary states to zero energy. Imhof et al. fulfilled this proposal in circuit and established a 2D circuit with quantized quadrupole moments \cite{Imhof2018}. The circuit model is shown in Fig. \ref{F5}(a), and circuit Laplacian for an infinite system is expressed as
\begin{equation}
J=i\sqrt{\frac{c}{l}}[(1+\lambda\cos k_x)\sigma_1\tau_0+(1+\lambda\cos k_y)\sigma_2\tau_2-\lambda\sin k_x\sigma_2\tau_3+\lambda \sin k_y \sigma_2\tau_1],
\end{equation}
with $\tau_0$ identity matrix, $\sigma_i$ and $\tau_i$ ($i=1,2,3$) Pauli matrices. $\lambda=C_2/C_1$ is the hopping ratio and $\sqrt{c/l}$ is a constant. The symmetry operators are defined as $M_x=\sigma_1\tau_3$, $M_y=\sigma_1\tau_1$, and $2C_4=(\sigma_1+i\sigma_2)\tau_0+(\sigma_1-i\sigma_2)(i\tau_2)$.

To characterize the topological phase here, one can use the mirror-symmetry graded winding number as topological invariant \cite{Benalcazar2017}
\begin{equation}
\nu_\pm:=\frac{i}{2\pi}\int dk~{\rm Tr}[q_\pm^\dagger(k) \partial_k q_\pm(k)]
\end{equation}
where $q_\pm(k)=\sqrt{2}(1+\lambda e^{\mp ik})$ is obtained from the circuit Laplacian. It's found that $\nu_\pm=\pm 1$ for $\lambda>1$. In experiment, $\lambda$ is set to be 3.3. Through the impedance measurement, the authors found a strong impedance response at the sample corner when the driving frequency matches the resonant condition, as shown in Fig. \ref{F5}(b), confirming the existence of "quantized quadrupole moments". It's noted that there is no Pauli principle for classical excitations in electrical circuit. The classical indication of dipole polarization is evident through the presence of midgap states (corner states here), which are inherently "polarized" at the boundary.

Later, Serra-Garcia et al. reported the circuit realization of quadrupole TI in an LC circuit as well \cite{Garcia020304}.
Their experimental results showed the existence of corner modes with tunable localization length controlled by the bias voltage. The authors also provided a technique to characterize edge modes by measuring the winding numbers.

With the extension of dimension, the third-order corner states are also reported in circuit \cite{Bao201406}. To realize this system, one can print a series of 2D circuit broads and stack them by interlayer connections to form a 3D circuit, as shown in Fig. \ref{F5}(c), and the circuit Laplacian is given by
\begin{equation}
J(\omega_0,q)=i\frac{C_1}{L_1}[\lambda\sin q_y\Gamma_1'+(1+\lambda\cos q_y)\Gamma_2'
+\lambda\sin q_x\Gamma_3'+(1+\lambda\cos q_x)\Gamma_4'
+\lambda\sin q_z\Gamma_5'+(1+\lambda\cos q_z)\Gamma_6'],
\end{equation}
where $\lambda=C_2/C_1$, $q_{x,y,z}$ are wave vectors, $\Gamma_i'=\sigma_3\otimes\Gamma_i$ ($i=0,1,2,3,4$), $\Gamma_5'=\sigma_2\otimes I_{4\times4}$, and $\Gamma_6'=i\Pi_{i=0}^{i=5}\Gamma_i'$. Here, $\Gamma_0=\tau_3\sigma_0$, $\Gamma_k=-\tau_2\sigma_k$ ($k=1,2,3$), and $\Gamma_4=\tau_1\sigma_0$.

This circuit system respects the reflection symmetry and it is an octupole circuit for $\lambda>1$ \cite{Bao201406}. To demonstrate the existence of HOT states, the authors observed the corner states by measuring the impedance distributions, indicated by the red regions in Fig. \ref{F5}(d). Zhang et al. investigated a similar system and observed the 0D corner states in a cubic lattice, which is induced by the octupole moment of the bulk circuit and protected by the anticommuting spatial symmetries of the lattice \cite{Zhang145}. 

\subsubsection{HOTIs from generalized SSH models}
Besides the quantized multipole insulators, the HOTIs generalized from SSH models belong to another important HOT classification without quantized multipole moments \cite{Xue2019,Ni2019,Liu076803,Xie205147}. Here, the higher-order topology emerges as a consequence of the filling anomaly: a discrepancy arises between the number of electrons needed to maintain charge neutrality and crystalline symmetry at the same time, resulting in the fractional corner charge \cite{Benalcazar2019}. For instance, in a four-band square lattice with a filling factor of 1/4 (where only the lowest band is filled), the Wannier center is located at the corner of the unit cell, which signifies a nontrivial phase characterized by the fractional corner charge of 1/4. In these systems, if the generalized chiral symmetry is preserved, the HOT states are pinned to the "zero" energy position.

\begin{figure}
  \centering
  \includegraphics[width=0.9\textwidth]{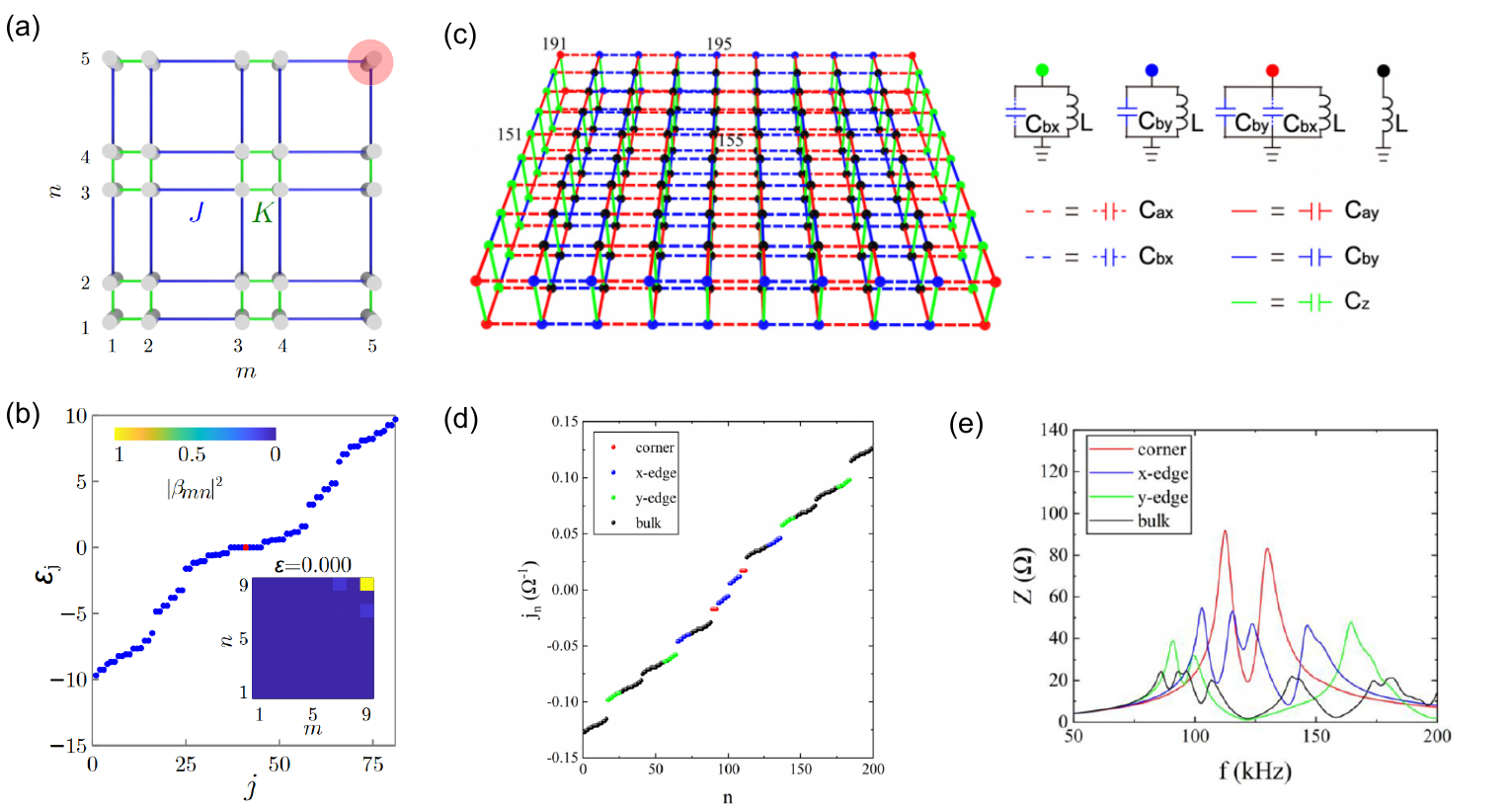}
  \caption{(a) Lattice model of generalized SSH. $J$ and $K$ are the intercell and intracell couplings. (b) The eigenvalues of this system, red dot indicates the corner state (Inset: wave functions of the corner state). (c) Circuit model of bilayer SSH model. (d) Calculated admittance spectra. Red dots show two groups of corner states. (e) Experimental impedance from different nodes.\\
  {\emph{Source}:} The figures are adapted from Ref. \cite{OlekhnoL081107,Guo073104}.}\label{HOSSH}
\end{figure}

A typical model of 2D generalized SSH model is shown in Fig. \ref{HOSSH}(a) \cite{OlekhnoL081107}. The Bloch Hamiltonian is written as 
\begin{equation}
\mathcal{H}(k)=-\left(
                           \begin{array}{cccc}
                             0 & J+K\exp(-ik_x) & 0 & J+K\exp(ik_y) \\
                             J+K\exp(ik_x) & 0 & J+K\exp(ik_y) & 0 \\
                             0 & J+K\exp(-ik_y) & 0 & J+K\exp(ik_x) \\
                             J+K\exp(-ik_y) & 0 & J+K\exp(-ik_x) & 0 \\
                           \end{array}
                         \right)
\end{equation}
with $k_x,~k_y$ being the wave vector, and $J,K$ the hopping strengthes. This Hamiltonian conserves the generalized chiral symmetry due to 
\begin{equation}
\begin{aligned} 
&\Gamma_{4}^{-1}\mathcal {H}\Gamma_{4}=\mathcal {H}_{1},\\
&\Gamma_{4}^{-1}\mathcal {H}_{1}\Gamma_{4}=\mathcal {H}_{2},\\
&\Gamma_{4}^{-1}\mathcal {H}_{2}\Gamma_{4}=\mathcal {H}_{3},\\
&\mathcal {H}+\mathcal {H}_{1}+\mathcal {H}_{2}+\mathcal {H}_{3}=0,
\end{aligned}
\end{equation}
with 
 $\Gamma_{4}=\left(
 \begin{matrix}
   1 & 0 & 0 & 0 \\
   0 & e^{\frac{ \pi i}{2}} & 0 & 0 \\
   0 & 0 & e^{\pi i} & 0\\
   0 & 0 & 0 & e^{\frac{3 \pi i}{2}} \\
  \end{matrix}\right),$
which can protect the "zero" admittance corner state for $J/K<1$. Considering a finite-size sample, the spectra can be solved by diagonalizing the corresponding circuit Hamiltonian ($J=1$ and $K=4$), and the eigenstates are plotted in Fig. \ref{HOSSH}(b). The inset displays the wave function of the red mode with highest localization, i.e., the corner states. It can be seen that this corner state mixes with the bulk mode. Further, one can distinguish the corner state by adding extra next-nearest-neighbor interactions, and the phenomena is demonstrated in circuit experiment.

Later work constructed a bilayer 2D SSH circuit model as shown in Fig. \ref{HOSSH}(c) \cite{Guo073104}. Due to the coupling between two layers, the generalized chiral symmetry is broken, and the corner states deviate the zero admittance,  as shown in Fig. \ref{HOSSH}(d). Figure \ref{HOSSH}(e) shows the experimentally measured impedance, and the red curves represent the two groups of corner states.

\begin{figure}
  \centering
  \includegraphics[width=0.9\textwidth]{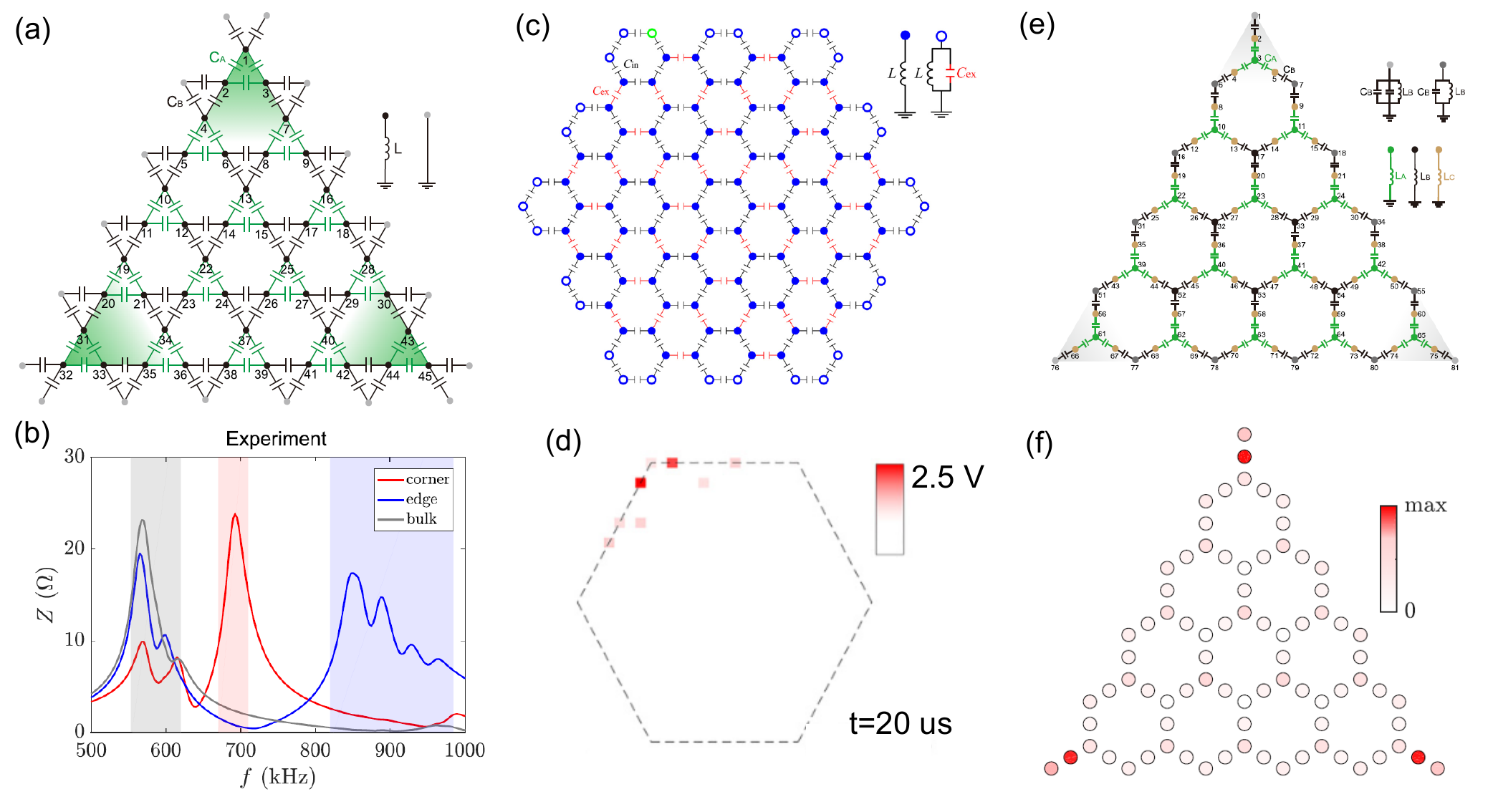}
  \caption{(a) Circuit realization of corner states in the kagome lattice. (b) Measured impedance as a function of frequency. A strong resonant peak appears at resonant frequency for corner nodes (red region). (c) The breathing honeycomb circuit model. (d) Steady-state voltage distribution of the corner state. (e) Circuit model of hybrid honeycomb-kagome lattice. (f) The distribution of impedance.\\
  {\emph{Source}:} The figures are adapted from Refs. \cite{Yang2020,Liu094903,Song2020}.}\label{F6}
\end{figure}  

Except for the standard SSH model, there are some other models can be viewed as the generalized SSH model, such as the breathing kagome lattice, honeycomb lattice, and their hybrid honeycomb-kagome lattice. These models can be constructed in circuit as well \cite{Ezawa201402,Yang2020,Liu094903,Song2020}. As shown in Fig. \ref{F6}(a), Yang et al. reported the experimental observation of corner states in the breathing kagome circuit \cite{Yang2020}. The circuit Hamiltonian reads
\begin{equation} \label{HKagome}
 \mathcal {H}=\left(
 \begin{matrix}
   Q_{0} & Q_{1} & Q_{2}\\
   Q_{1}^{*} & Q_{0} & Q_{3} \\
   Q_{2}^{*} & Q_{3}^{*} & Q_{0}
  \end{matrix}
  \right),
\end{equation}
where the matrix elements 
\begin{equation}
\begin{aligned}
&Q_0=\omega[2(C_{A}+C_{B})-\frac{1}{\omega^{2}L}], \\
&Q_1=-\omega[C_{A}+C_{B}e^{-i(k_{x}d/2+\sqrt{3}k_{y}d/2)}], \\
&Q_2=-\omega[C_{A}+C_{B}e^{-ik_{x}d}],\\
&Q_3=-\omega[C_{A}+C_{B}e^{i(-k_{x}d/2+\sqrt{3}k_{y}d/2)}]
\end{aligned}
\end{equation}
with $k_x$ and $k_y$ the wave vectors and $d$ the lattice constant. At the resonant frequency $\omega=1/\sqrt{2(C_{A}+C_{B})L}$, this Hamiltonian also satisfies the generalized chiral symmetry due to
$\Gamma_{3}^{-1}\mathcal {H}\Gamma_{3}=\mathcal {H}_{1}$, 
$\Gamma_{3}^{-1}\mathcal {H}_{1}\Gamma_{3}=\mathcal {H}_{2}$, 
$\mathcal {H}+\mathcal {H}_{1}+\mathcal {H}_{2}=0$,
with 
 $\Gamma_{3}=\left(
 \begin{matrix}
   1 & 0 & 0 \\
   0 & e^{\frac{2\pi i}{3}} & 0 \\
   0 & 0 & e^{\frac{4\pi i}{3}}
  \end{matrix}\right)$.
To characterize the HOT states, one can evaluate the $\mathbb{Z}_3$ Berry phase \cite{Zak1989,Kariyado2018,Araki2019}
 \begin{equation}
 \mathcal{\theta}=\int_{L_{1}}\text{Tr}[\textbf{A}(\textbf{k})]\cdot d\textbf{k}\ \  (\text{mod}\ 2\pi),
\end{equation}
as the topological index, where $\textbf{A}(\textbf{k})$ is the Berry connection:
$\textbf{A}(\textbf{k})=i\Psi^{\dag}(\textbf{k})\frac{\partial}{\partial\textbf{k}}\Psi(\textbf{k})$.
Here, $\Psi(\textbf{k})$ is the wave function of Eq. \eqref{HKagome} for the lowest band. $L_{1}$ is an integral path in Brillouin zone: $K^{\prime}\rightarrow \Gamma\rightarrow K^{\prime\prime}$.

It is obtained that $\theta$ is quantized to $\frac{2\pi}{3}$ for $C_A/C_B<1$, corresponding to a HOT phase. This HOT state is demonstrated by measuring the impedance of the sample. As shown in Fig. \ref{F6}(b), the authors build a finite-size breathing kagome circuit lattice with $C_A=22$ nF, $C_B=100$ nF, and $L=220$ nH, and observe a strong impedance peak at the sample corner at the resonant frequency. With the superiority of circuits, one can break the generalized chiral symmetry of the system by introducing next-nearest-neighbor interactions in adjacent unit cells between the same subnodes and verify that the corner states disappear if the generalized chiral symmetry is broken. 

The breathing honeycomb lattice can also form the HOTI as another generalized SSH model. As shown in Fig. \ref{F6}(c), Liu et al. reported the corner states in a breathing honeycomb circuit \cite{Liu094903}. They identified the corner states in the zero admittance, and, alternatively, the corner states were characterized by the attenuated voltage signal from the corner nodes, as shown in Fig. \ref{F6}(d). 

By combining the aforementioned two lattices, Song et al. reported the square-root HOT states \cite{Song2020}. The square-root operation can be traced back to the discovery of positron by squaring-root the Klein-Gordon equation for relativistic particles \cite{Dirac610}. This method leads to the discovery of a series of square-root topological states \cite{Arkinstall165109,Mizoguchi029906}. Executing a square-root operation on the breathing kagome lattice, the authors obtain the honeycomb-kagome model, as shown in Fig. \ref{F6}(e). In the topological parameter region, the authors obtained the corner states by measuring the impedance, with the impedance concentrating on the sample corners, as shown in Fig. \ref{F6}(f). 

As an extension of dimension, the third-order HOTI is confirmed in the pyrochlore TEC \cite{Guo505305}. One can establish the pyrochlore lattice in circuit and observe the zero-admittance corner states protected by generalized chiral symmetry, too.

\subsubsection{Long-range interaction induced HOTIs}
Long-range interactions (LRIs) pervade the natural world, manifesting in interactions like electromagnetism and gravity. Systems governed by LRIs often display intriguing dynamics and statistical behaviors, including temperature anomalies and negative specific heat \cite{Campa57,French1887}. These interactions are known to exert distinct influences on topological states \cite{Varney115125,Beugeling195129,Li89,Shen24045,OlekhnoL081107,Yu125113,Leykam023901}. However, it is difficult to control the strength of LRIs in electronic or other systems. Electrical circuits can help going out of the predicament, because one can realize arbitrary hoppings between two circuit nodes. Therefore, one can study how LRI dictates the topological phases carefully.

\begin{figure}
  \centering
  \includegraphics[width=0.9\textwidth]{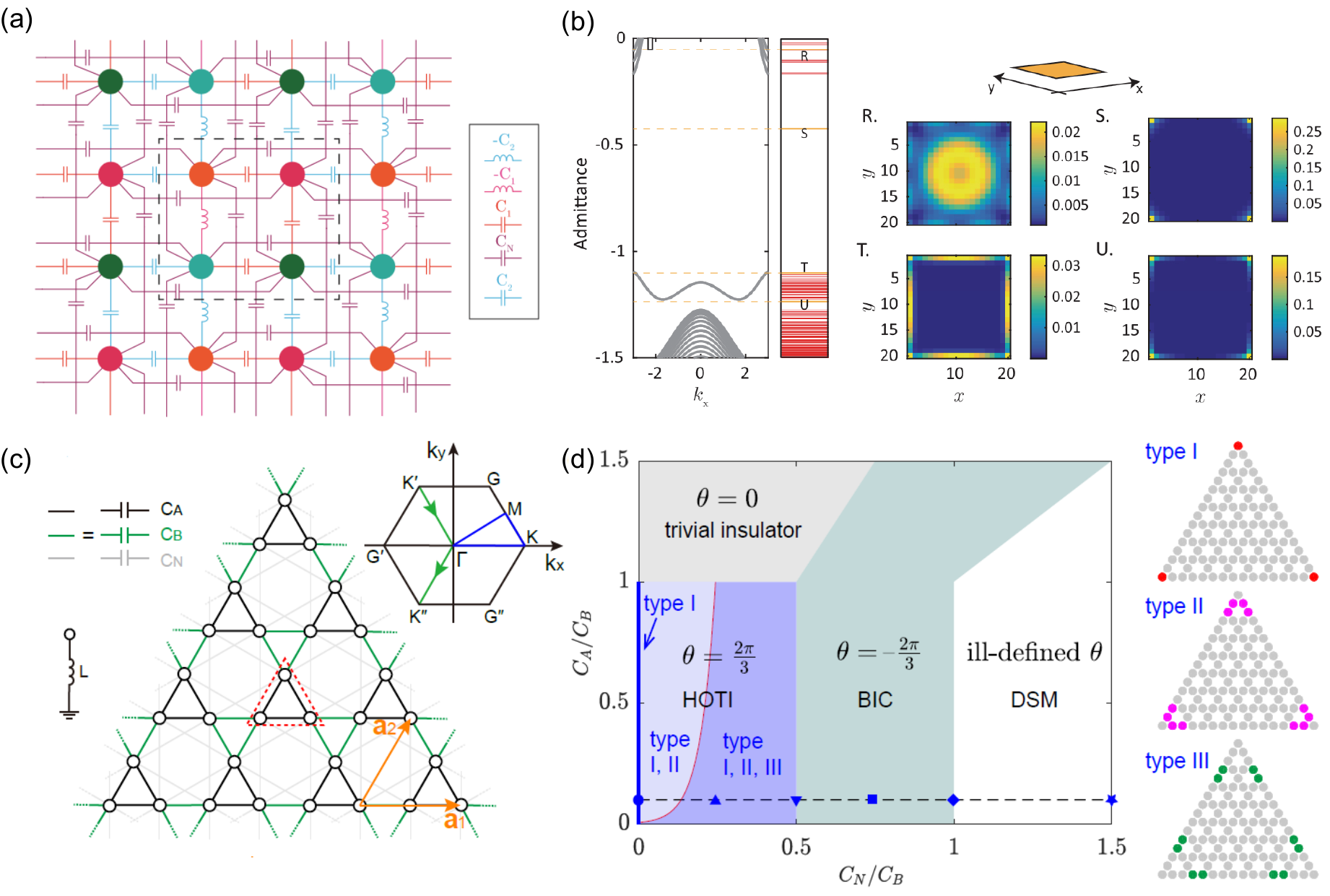}
  \caption{(a) Topolectrical circuit model that hosts HOT phases with type-II corner modes. (b) The admittance spectra and wave functions. S and U correspond to type-II corner modes, which are localized in the vicinity of the corner nodes but exponentially decay away from them. (c) The breathing kagome circuit model with LRIs. (d) The phase diagram by tuning the hopping ratios and the strength of LRIs. The right part shows the configurations of three types of corner states.\\
  {\emph{Source}:} The figures are adapted from Refs. \cite{Islam245128,Yang075427}.}\label{LRI}
\end{figure}

Islam et al. theoretically proposed a TEC based on a modified 2D SSH model with next-nearest-neighbor (NNN) couplings shown in Fig. \ref{LRI}(a) \cite{Islam245128}. The circuit Laplacian is written as
\begin{equation}
\begin{aligned}
\mathcal{J}=&-2C_N(\cos k_x+\cos k_y)\sigma_0\otimes\sigma_0-(C_1+C_2\cos k_x)\sigma_0\otimes\sigma_x-C_2\sin k_x
\sigma_z\otimes\sigma_y-(C_1+C_2\cos k_y)\sigma_y\otimes\sigma_y\\
&-C_2\sin k_y \sigma_x\otimes\sigma_y+[2(C_1+C_2+2C_N)-\frac{1}{\omega^2L}]\sigma_0\otimes\sigma_0,
\end{aligned}
\end{equation}
with $\sigma_0$ the identity matrix and $\sigma_{x,y,z}$ Pauli matrices. $C_1$, $C_2$, $C_N$, and $L$ are circuit parameters. By calculating the admittance spectra and wave functions, as displayed in Fig. \ref{LRI}(b), the authors demonstrated the emergence of "type-II" corner modes induced by the NNN interactions, which localize at sample corners and exponentially decay from the edge nodes. 

Yang et al. observed a type-III corner state in the breathing kagome circuit lattice with LRIs \cite{Yang075427}. The circuit model is shown in Fig. \ref{LRI}(c), and the Hamiltonian for an infinite lattice is written as
\begin{equation}
 \mathcal {H}=\mathcal {H}_0+\mathcal {H}_\text{NNN},
\end{equation}
where $\mathcal{H}_0$ is Eq. \eqref{HKagome} and $\mathcal{H}_{\rm NNN}$ represents the NNN coupling Hamiltonian
\begin{equation}\label{HNNN}
 \mathcal {H}_\text{NNN}=\left(
 \begin{matrix}
   Q_{N0} & Q_{N1} & Q_{N2}\\
   Q_{N1}^{*} & Q_{N0} & Q_{N3} \\
   Q_{N2}^{*} & Q_{N3}^{*} & Q_{N0}
  \end{matrix}
  \right),
\end{equation}
with the matrix elements
$Q_{N0}=4\omega C_N$, $Q_{N1}=-\omega C_N[e^{-ik_{x}}+e^{i(k_{x}/2-\sqrt{3}k_{y}/2)}]$, $Q_{N2}=-\omega C_N[e^{-i(k_{x}/2-\sqrt{3}k_{y}/2)}+e^{-i(k_{x}/2+\sqrt{3}k_{y}/2)}]$, $Q_{N3}=-\omega C_N[e^{-ik_{x}}+e^{i(k_{x}/2+\sqrt{3}k_{y}/2)}]$. By adjusting the NNN strength and hopping ratios, four
topologically distinct phases are discovered, i.e., HOT phase, the bound state in the continuum, Dirac semimetal (DSM) phase, and trivial insulator phase [see Fig. \ref{LRI}(d)]. For $C_A/C_B<1$, in the absence of NNN interactions, only the type-I corner state can be found. As the increase of NNN hoppings, the coexistence of type-I and type-II corner states appears and three types of corner states coexist for larger LRIs. The patterns of these three corner states are shown in the right panel of Fig. \ref{LRI}(d). In the BIC phase, one can find the HOT states in the bulk spectra. In the Dirac SM phase, the position of Dirac cone are dependent on the NNN interactions, which can move in momentum space. The LRIs give rise to many interesting topological phases in this system.

In principle, one can introduce interactions between two arbitrary nodes in circuit, which opens a new pathway to exploring interesting topological phenomena with LRIs.

\subsubsection{Other HOTIs}
Inspired by the realization of Anderson TIs in circuit, Zhang et al. demonstrated the appearance of the corner states induced by disorder in a modified Haldane model, characterized by the voltage measurements \cite{Zhang146802}. Li et al. observed the large-number ($N$) corner states in $\mathbb{Z}$-class HOT circuits, which host $N$ corner modes at each single corner structure \cite{Li10585}. We expect to discover more HOT states in circuit.

This ends our review of topological insulating states in circuits, and we see that TEC manifests as a reliable platform for studying both the first-order and higher-order topological insulating phases. Taking the advantages of flexibility and controllability of circuits, one can easily introduce the disorders, defects, LRIs to the circuit lattices. These unique features enable us to testify the robustness of topological states, study the disorder-induced topological states, explore the novel topological phases caused by LRIs, and investigate the protection of the symmetry on the boundary states. Next, we will move the gapless topological systems and reviw the literature of the circuit on studying TSM phases.

\section{Circuit realization of topological semimetals (TSMs)}  \label{S4}

Topological band theory also dictates the gapless topological phases, that are, TSMs. According to the degeneracy of band crossings, TSMs can be generally divided into Dirac SMs, Weyl SMs, nodal-line SMs, and nodal-surface SMs \cite{LvRMP2021}. TSMs can not only be applied to study the excitation of low-energy topological quasiparticles, such as Dirac and Weyl fermions, but also can be analogized to fundamental particles in quantum field theory and used to explore their properties. Due to the rich topological bulk and surface properties of TSMs, exotic quantum transport phenomena are induced, such as the extremely high mobility of electrons in Dirac SMs \cite{Neupane3786}, the chiral anomaly effect and negative magnetoresistance effect in Weyl SMs \cite{Huang031023}. These peculiar transport properties make TSMs appealing for potential applications in quantum control, spintronics, and other fields. The HOT phases can also be identified in TSMs, which means the coexistence of gapless band structures and HOT states. The discovery of HOT states enriches the topological phases, which has spurred fundamental interests and induced promising applications. 

In this section, we review several kinds of TSMs realized in circuit. We show how to establish the circuit models for first-order nodal-line, Dirac, and Weyl SMs, and how to stack the 2D lattice to realize the HOT SMs in higher dimensions.

\subsection{The first-order TSMs in circuits} 
According to different origins, the nodal-line SMs can be divided into type-A, type-B, and type-C, which are protected by mirror reflection symmetry, the coexistence of time-reversal and space inversion symmetry, and the nonsymmorphic space group, respectively \cite{Chen25}.

\begin{figure}
  \centering
  \includegraphics[width=0.8\textwidth]{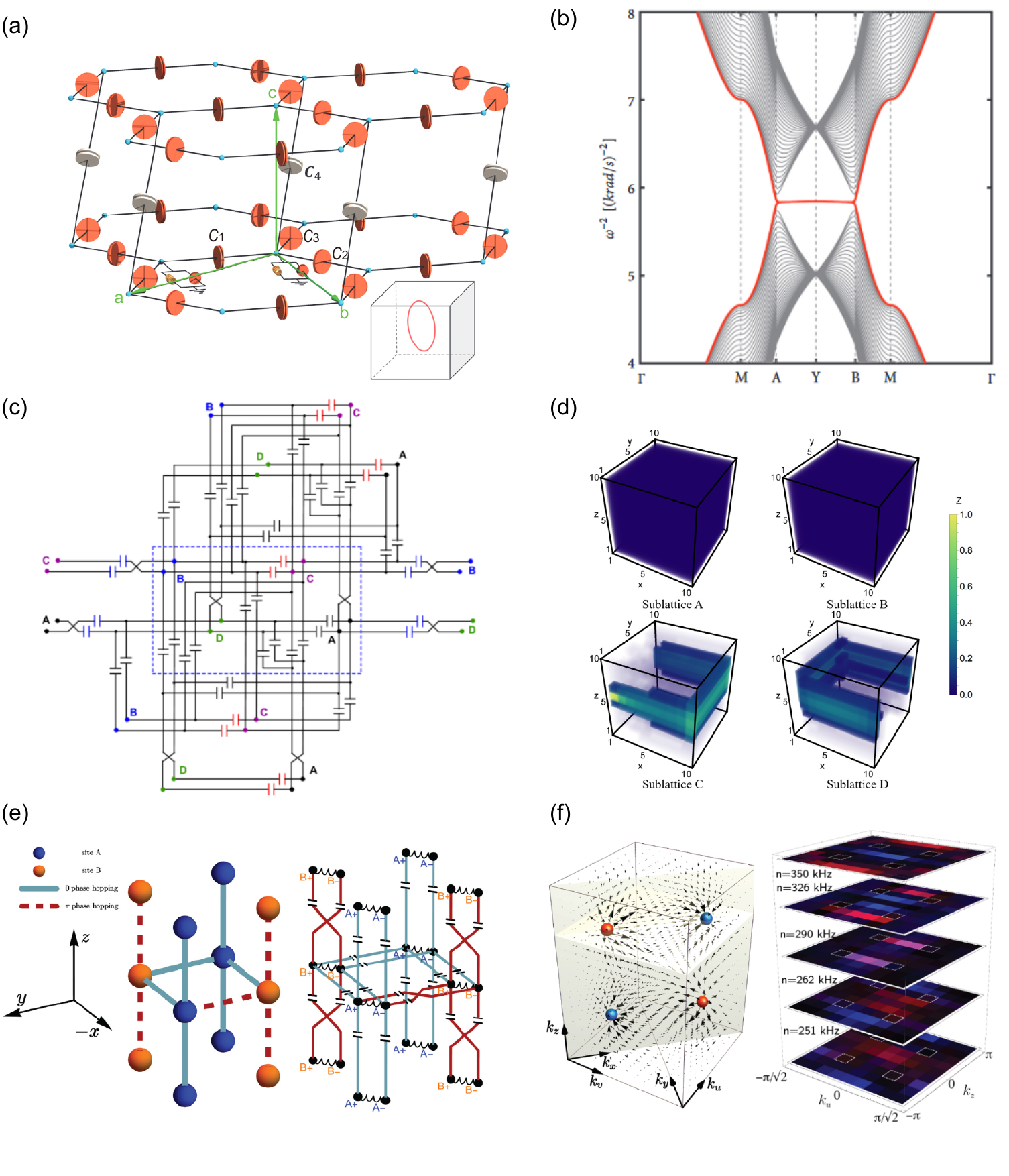}
  \caption{(a) Circuit realization of nodal-line SM. (b) The admittance spectra of nodal-line SM with the surface states (red colour lines) on the (001) surface. (c) Circuit model of Dirac SM. (d) Calculated impedance distributions indicating the existence of surface states. (e) Weyl circuit model. (f) The Weyl points in momentum space and the fermi-arc surface states in lattice space.\\
   {\emph{Source}:} The figures are adapted from Refs. \cite{Luo2018,Dong023056,Lu020302}.}\label{F7}
\end{figure}

In Ref. \cite{Luo2018}, Luo et al. proposed the circuit model for realizing type-B nodel-line SMs, as shown in Fig. \ref{F7}(a). The circuit Hamiltonian reads
\begin{equation}
\mathcal{H}=\sum_{i=0}^3 d_i({\bf k})\sigma_i,
\end{equation}
where $d_0({\bf k})=C_1+C_2+C_3+C_4+\frac{C_{GA}+C_{GB}}{2}+(C_A+C_B)(1-\cos k_c)$, $d_1({\bf k})=-C_1-C_2\cos(k_b-k_a)-C_3\cos k_a-C_4\cos(k_c-k_a)$, $d_2({\bf k})=C_2\sin(k_b+k_a)-C_3\sin k_a+C_4\sin(k_c+k_a)$, and $d_3({\bf k})=\frac{C_{GA}-C_{GB}}{2}+(C_A-C_B)(1-\cos k_c)$ with $k_a$, $k_b$, and $k_c$ the wave vectors along $a,b,c$ direction.

In this circuit, the time-reversal symmetry is conserved because only LC elements are used, and the space inversion symmetry hold by setting $C_{GA}=C_{GB}$, so this model belongs to type-B nodal-line state. By calculating the band structures of the model, the authors obtained the linear band crossing in Fig. \ref{F7}(b), the red lines indicate the drumhead-like surface states, which are the typical features of nodal-line SMs. For further considerations, the nodal-line SMs can be converted to Dirac and Weyl SMs by adding additional lattice symmetries. For example, one can break the inversion symmetry by setting $C_{GA}\neq C_{GB}$ and $C_A\neq C_B$ and obtain the Weyl SMs.

To realize the Dirac SMs, Dong et al. first constructed the quantum spin Hall insulators \cite{Dong023056}, as shown in Fig. \ref{F7}(c). The Hamiltonian is 
\begin{equation}
\mathcal{H}_{\rm QSHI}=t[\sin(k_x)\Gamma_1+\sin(k_y)\Gamma_2]
+\{m-t_0[\cos(k_x)+\cos(k_y)]\}\Gamma,
\end{equation}
with $\Gamma_1=\sigma_1\otimes\tau_2$, $\Gamma_2=\sigma_2\otimes\tau_0$, and $\Gamma_3=\sigma_1\otimes\tau_1$.
The capacitances of the black, blue, and red capacitors are $t/2$, $t$, and $m$, respectively. Then, they stacked this model in 3D space and got the Dirac SMs with the Hamiltonian
\begin{equation}
\mathcal{H}=t[\sin(k_x)\Gamma_1+\sin(k_y)\Gamma_2]+[t_z\cos(k_z)+m-t_0\sum_{j=x,y}\cos(k_j)]\Gamma_3.
\end{equation}
The features of Dirac SMs are demonstrated by computing the impedance of surface states numerically, as depicted in Fig. \ref{F7}(d).

For Weyl SMs, different intersections between band dispersions and Fermi surfaces lead to type-I, type-II, and type-III classifications \cite{Huang121110}, the feature of which can be characterized by the Weyl points in the momentum space and Fermi-arc surface states in real space. In circuits, all of the three kinds of Weyl SMs have been reported, and the transmission behaviors of heterojunctions between two different types of Weyl SMs are also studied. As shown in Fig. \ref{F7}(e), Lu et al. proposed a type-I Weyl circuit in Ref. \cite{Lu020302}, whose tight-binding model is expressed as
\begin{equation}
\mathcal{H}=\epsilon_0+{\bf h}({\bf k})\cdot{\boldsymbol \sigma},
\end{equation}
with ${\bf h}({\bf k})=2t_0[\cos(k_xa)\hat{x}-\sin(k_y a)\hat{y}+\cos(k_z a)\hat{z}]$ and $\epsilon_0$ the on-site potential. The energy spectra are
\begin{equation}
E_\pm=\epsilon_0\pm 2t_0\sqrt{\cos^2(k_xa)+\sin^2(k_ya)+\cos^2(k_za)}.
\end{equation}
There are four Weyl point located at $(k_x,k_y,k_z)=(\pm\pi/2,0,\pm\pi/2)$ with $E=\epsilon_0$. By computing the Berry curvature, two groups of Weyl points emerge in momentum space, as shown in left of Fig. \ref{F7}(f). The right panel  of Fig. \ref{F7}(f) shows the Fermi arc in $k_u-k_z$ plane, which connects the two pairs of Weyl points at the frequency $f=290$ kHz, exhibiting the most important feature of Weyl SMs. 

Further, Islam et al. designed a passive circuit to realize type-I and type-II Weyl SMs \cite{Islam023025}. These two types of Weyl SMs can be converted to each other by adjusting the hoppings, and the type-I Weyl SM shows impedance at the Weyl points, and type-II exhibits multiple secondary peaks near the Weyl points. Li et al. reported the realization of ideal type-II Weyl SMs by stacking LC resonator dimers with broken parity inversion symmetry in Ref. \cite{RLi2020}. Later, Islam et al. established a Weyl SM heterojunctions by combining two of the three different Weyl SMs and studied the transmission behaviors \cite{Islam2020}.

\begin{figure}
  \centering
  \includegraphics[width=0.9\textwidth]{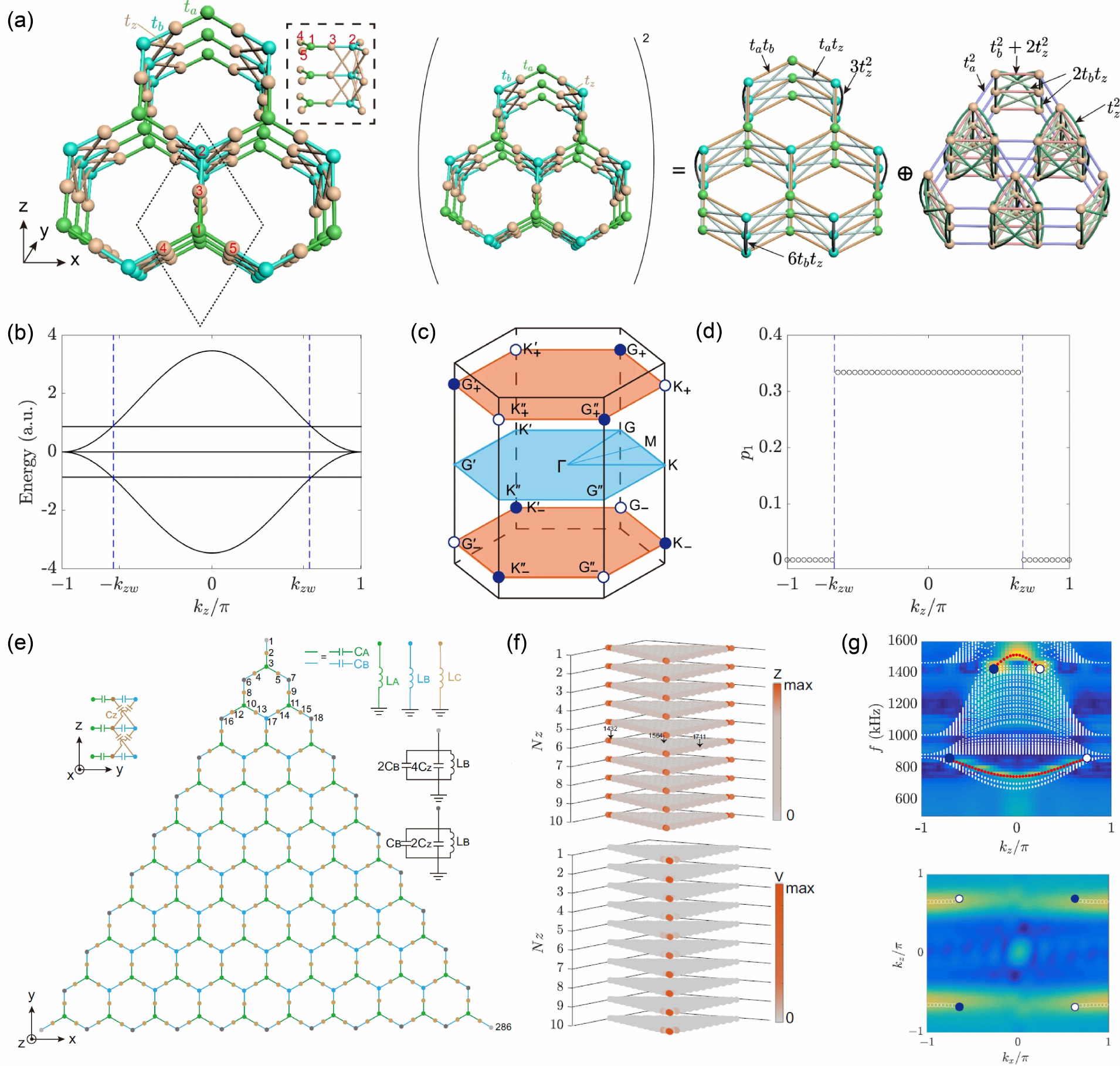}
  \caption{(a) Realization of square-root higher-order Weyl SM by stacking the 2D square-root HOTIs. (b) The energy spectrum for $t_a=0.5$, $t_b=1$, and $t_z=0.5$. (c) The distribution of the Weyl points (hollow and solid circles). (d) The bulk polarization as a function of $k_z$. (e) Top view of the circuit model with the inset showing the interlayer hoppings. (f) Impedance distributions and the steady voltage signal of hinge states. (g) The Fermi arc in frequency space and momentum space.\\
  {\emph{Source}:} The figures are adapted from Ref. \cite{Song2022}.}\label{F8}
\end{figure}

Except for the standard nodal-line, Dirac and Weyl SMs, Yang et al. established a circuit model that realizes the quantum anomalous SM \cite{Fu94,Yang211}, whose quasiparticle corresponds to the Wilson fermion. The quantum anomalous SM hosts a half-integer bulk topological index, and non-trivial bulk invariant indicates the existence of edge currents but no edge states. The edge currents are observed by forming a domain wall structures with opposite topological indexes $\pm1/2$. Besides, they mapped the structure of Wilson fermions to the magnetic solitons, which connects these two promising fields and may inspire further studies.

\subsection{Higher-order TSMs in circuits}
Through the introduction of HOT states to traditional TSM phase, the concept of higher-order TSMs are put forward \cite{Wang146401,Luo794}. In Ref. \cite{Dong023056}, the authors established a series of circuit models to realize higher-order TSMs. For example, they stacked 2D HOTIs to obtain the 3D quadrupolar Dirac SM and stacked 2D generalized HOTI to form the 3D HOT Weyl SM. In addition, Song et al. reported the realization of square-root higher-order TSMs in circuits \cite{Song2022}, by stacking the 2D square-root HOTI in a 3D space with inter layer hoppings, as shown in Fig. \ref{F8}(a), and the corresponding Hamiltonian reads
\begin{equation}
 \mathcal {H}=\left(
 \begin{matrix}
  O_{2,2} & \Phi_{\bf k}^{\dagger}\\
  \Phi_{\bf k} & O_{3,3}\\
 \end{matrix}
 \right),
 \end{equation}
where $O_{2,2}$ and $O_{3,3}$ are the $2\times2$ and $3\times3$ zero matrix, respectively, and $\Phi_{\bf k}$ is the $3\times2$ matrix
\begin{equation}
 \Phi_{\bf k}=\left(
 \begin{matrix}
  t_{a} & t_{b}+2t_{z}\text{cos}(\mathbf{k}\cdot{\mathbf{a}_3}) \\
  t_{a} & [t_{b}+2t_{z}\text{cos}(\mathbf{k}\cdot{\mathbf{a}_3})]e^{-i\mathbf{k}\cdot{\mathbf{a}_1}} \\
  t_{a} & [t_{b}+2t_{z}\text{cos}(\mathbf{k}\cdot{\mathbf{a}_3})]e^{-i\mathbf{k}\cdot{\mathbf{a}_2}}\\
 \end{matrix}
 \right).
 \end{equation}
Here $\mathbf{k}=(k_x,k_y,k_z)$ is the wave vector. $ \textbf {a}_{1}=\frac {1} {2} \hat {x} + \frac {\sqrt {3}} {2} \hat {y} $, $\textbf {a}_{2}=-\frac {1} {2} \hat {x} + \frac {\sqrt {3}} {2} \hat {y}$, and $\textbf {a}_{3}= \hat {z}$ are three basic vectors. 

Figure \ref{F8}(b) displays the energy spectrum for $(k_x,k_y)=(0,\frac{4\pi}{3})$, the gap of which is gapless at $k_z=\pm k_{zw}$. Figure \ref{F8}(c) shows the distribution of the Weyl points in the first Brillouin zone. Due to the $C_3$ symmetry of this system, six pairs of Weyl points are labeled by the hallow and solid circles. To characterize the HOT phase, one can calculate the bulk polarization as topological index, which is nonzero between two Weyl points, as shown in Fig. \ref{F8}(d). Therefore, this system belongs to the higher-order TSM because the HOT states connect the a pair of Weyl point.

In circuit, $t_a$, $t_b$, and $t_z$ are the hopping strengths, which realized by the capacitors $C_A$, $C_B$, and $C_Z$ in the circuit, respectively, as shown in Fig. \ref{F8}(e). By measuring the impedance distributions and voltage propagation, the HOT state (hinge state) was identified, as shown in Fig. \ref{F8}(f). To characterize the properties of higher-order Weyl SMs, one can calculate the Fermi-arc and measured it in a 3D printed circuit board and detect both the hinge states and the fermi-arc surfaces connecting the Weyl points, see Fig. \ref{F8}(g), demonstrating the existence of HOT SM phase.

Ni et al. found that the synthetic gauge fields can induce higher-order TSM phases \cite{Ni050802}. The authors demonstrated their principles in circuit and found the emergence of the nodal line rings and Weyl points in bulk dispersion, whose projected surfaces and hinges support surface Fermi arcs and flat hinge Fermi arcs, indicating the appearance of higher-order TSMs. Zheng et al. realized a quadrupolar surface SM, which hosts a gapped bulk but gapless nodes on the surface spectrum \cite{Zheng2022}, and they found the coexistence of states on the hinge and surface.

Till now, we have presented the results of circuit realization TSMs. In circuits, the properties of traditional TSMs can be investigated in great detail, by directly measuring their band structures and surface states. In particular, one can realize novel TSMs by stacking the 2D circuit lattice with various manners because the interlayer couplings can be freely controlled. The tunable interactions are unique features of circuits, which is arduous to be realized in other systems. In the next section, we will profile other exotic topological states with circuit features.

\section{Circuit realization of unconventional topological states} \label{S5}
Apart from the topological states mentioned in the above two sections, one can also detect other unconventional topological phase. In this section, we will review the literature on circuit realization of non-Hermitian, non-linear, non-Abelian, non-periodic, non-Euclidean, and high-dimensional topological states, to name a few.

\subsection{Non-Hermitian topological states}

Generally, the Hamiltonian in quantum mechanics should be Hermitian for closed systems \cite{Ashida249}. It ensures the conservation of probability and guarantees the real-valuedness of an expectation value of energy. For an open system, the gain and loss from environments will induce the non-Hermiticity \cite{Ashida249,Bergholtz2021}, which yields abundant physical phenomena, such as non-Hermitian skin effect and the appearance of EPs. The non-Hermitian skin effect has profoundly changed the Bloch band theory, which means the eigenstates of the system lose their scalability and localize to the boundary. Meanwhile, the conventional bulk-boundary correspondence is broken in non Hermitian-systems. These interesting properties require a re-examination of the band theory and bulk-boundary correspondence for non-Hermitian topological systems, thus sparking new researches \cite{Yao086803,Song246801}.

\begin{figure}
  \centering
  \includegraphics[width=0.9\textwidth]{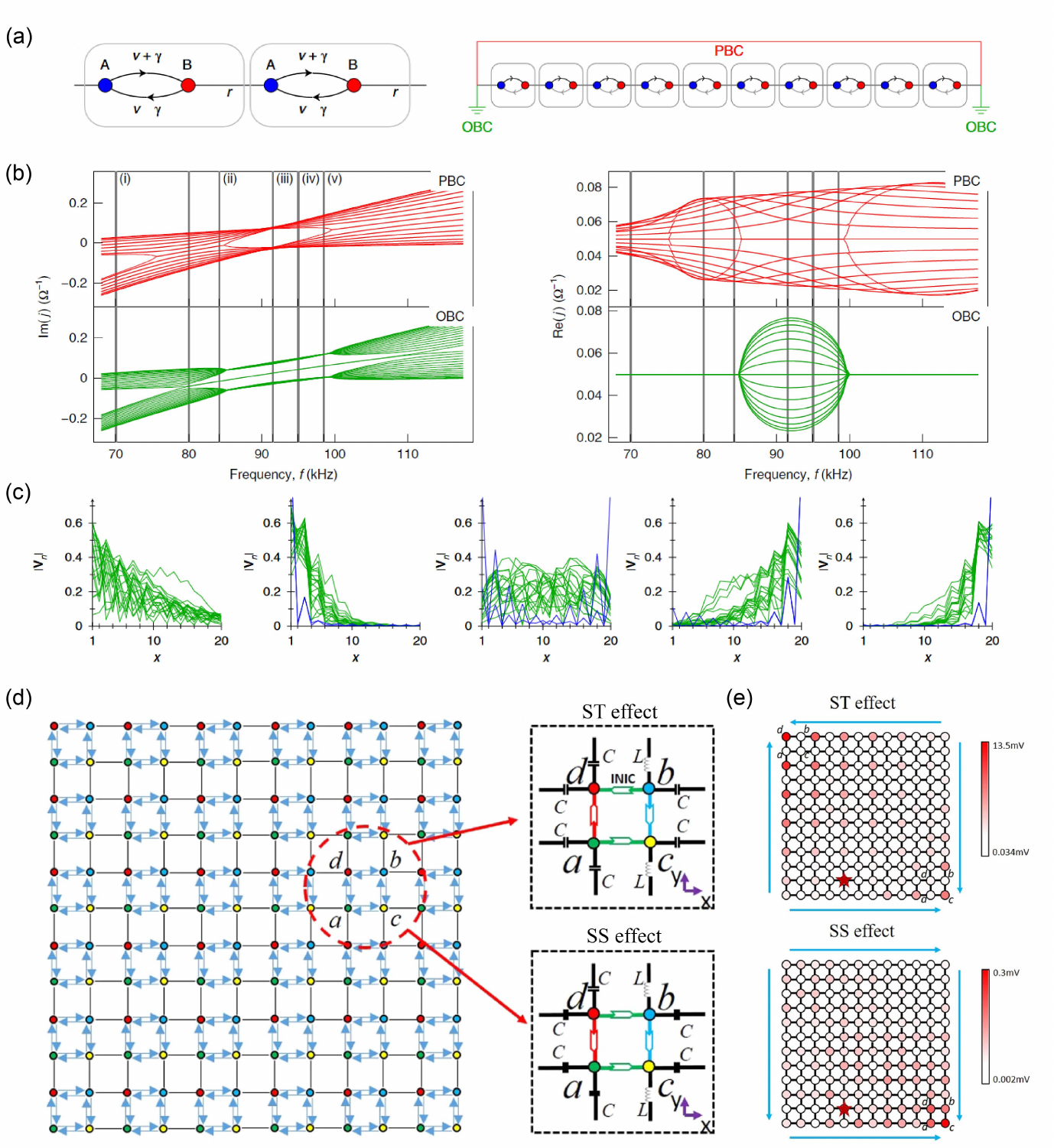}
  \caption{(a) The circuit realization of non-Hermitian SSH model with the non-reciprocal hoppings are realized by INICs. (b) The admittance spectra of non-Hermitian SSH model for the periodic and open boundary conditions. (c) Wave functions of the bulk (green) and edge (bulk) modes for the open boundary condition. (d) The circuit realization of 2D non-Hermitian SSH model. (e) Observation of the skin-topological (ST) and skin-skin (SS) modes by measuring the propagation of the voltage signals.\\
  {\emph{Source}:} The figures are adapted from Refs. \cite{Helbig2020,Zou2021}.}\label{F9}
\end{figure}

With the INIC unit introduced in Section \ref{S2}, it is convenient to induce the gain(loss) and non-reciprocal hoppings in circuit systems, which means that one can study non-Hermitian physics with circuit platform. Again, we take the SSH model in non-Hermitian lattice as the example. There are three main methods to introduce non-Hermitian terms to this system, i.e., adding imaginary mass, imaginary momentum, and imaginary anticommuting mass. As shown in Fig. \ref{F9}(a), Helbig et al. realized a non-Hermitian SSH circuit model by asymmetric intra-cell couplings $\nu\pm\gamma$ (imaginary momentum) \cite{Helbig2020}, reflected by the circuit Laplacian
\begin{equation}
(i\omega)^{-1}J(k)=\epsilon_0(\omega) \sigma_0 +[\nu(\omega)+r\cos(k)]\sigma_x+[r\sin(k)-i\gamma]\sigma_y.
\end{equation}
Here, $r$ and $\nu\pm\gamma$ are reciprocal and non-reciprocal hoppings, respectively. $\epsilon_0$ is the on-site potential. For the open and periodic boundary conditions, one can obtain totally different spectra, as shown in Fig. \ref{F9}(b). Figure \ref{F9}(c) displays the wave functions at different frequencies [(i)-(v) labeled in Fig. \ref{F9}(b)] for the open boundary condition, and one can clearly see the non-Hermitian skin effect and the modes concentrating on opposite edges for low and high frequencies, which proves the violation of bulk-boundary correspondence.

Similar results are reported in Ref. \cite{Liu5608038}, where Liu et al. demonstrated the inapplicability of the bulk-boundary correspondence in a 1D non-Hermitian circuit and the result is explained by the non-Bloch winding number. Galeano et al. realized a non-Hermitian SSH circuit model with complex hoppings \cite{Galeano217211}, and they also identified the tunable bulk-boundary correspondence and non-Hermitian skin effects. In a generalized 1D non-Hermitian model, one can discuss the topological phase and phase transition in the generalized Brillouin zone \cite{Yao086803,Yokomizo066404,Yang226402}. A more straight-forward evidence about the generalized Brillouin zone is give in Ref. \cite{Wu064307}, one can measure the non-Bloch dynamics with temporal TECs.

Ezawa investigated the non-Hermitian interface states between two domains with different topological invariants in the electric diode circuits, and found all states are localized at the interface, manifesting as the generalization of the skin-edge states \cite{Ezawa121411}. The author further extended the domain wall structures to higher dimensional systems and obtained the interface surface, line, and point states, which are conformed in latter circuit experiments \cite{Liu14987}.

In Ref. \cite{Zou2021}, Zou et al. reported the first experimental realization of hybrid higher-order skin-topological states in the 2D and 3D TEC systems. The 2D circuit model consists of LC and INICs, as shown in Fig. \ref{F9}(d), and the circuit Laplacian is written as
\begin{equation}\label{STeffect}
J(\omega)=i\omega\left(
                   \begin{array}{cccc}
                     0 & 0 & C_1-C_3+Ce^{-iq_x} & C_1\pm C_2+Ce^{-iq_y} \\
                     0 & 0 & -C_1+C_3-Ce^{iq_y} & C_1\pm C_3+Ce^{iq_x} \\
                     C_1+C_3+Ce^{iq_x} & -C_1-C_2+Ce^{-iq_y} & 0 & 0 \\
                     C_1\mp C_2+Ce^{iq_y} & C_1\mp C_3+Ce^{-iq_x} & 0 & 0 \\
                   \end{array}
                 \right),
\end{equation}
with $q_x$ and $q_y$ the wave vectors along $x$ and $y$ direction, respectively. If one chooses $-$ in $\pm$ and $+$ in $\mp$ for the circuit Laplacian \eqref{STeffect}, this system exhibits the skin-topological effect, and otherwise only the skin-skin effect can be observed. The measured static voltage distributions in Fig. \ref{F9}(e) identify the hybrid second-order skin-topological and skin-skin effects exactly. Similar phenomenon is also observed in the 3D TEC experiment.

Beyond the aforementioned works, there are many researches focusing on the skin effects that are experimentally confirmed in circuit. Hofmann et al. reported a reciprocal skin effect \cite{Hofmann023265}, and the authors found eigenmodes with opposite longitudinal momenta display opposite transverse anomalous localization. Zhang et al. proposed a topological switch that can control the on-off skin effect \cite{Zhang085426}. Yoshida et al. investigated the skin effect with mirror symmetry \cite{Yoshida2020}. Islam et al. studied the non-Hermitian skin effect with multiple asymmetrical couplings \cite{Islam043108}. Zhu et al. reported the experimental observation of rank-2 skin effect, which means the coexistence of the edge and corner localized skin modes \cite{Zhu720}. 
Ezawa et al. studied the competition between the corner modes and skin effects in the non-Hermitian 2D honeycomb lattice and 3D diamond lattice \cite{Ezawa201411}. Later, this phenomenon is demonstrated in Ref. \cite{Tang035410}. Tang et al. designed a series of circuit experiments with different non-reciprocal strengths and observed the unbalanced corner states on two acute angles of the rhombus lattice.

\begin{figure}
  \centering
  \includegraphics[width=0.9\textwidth]{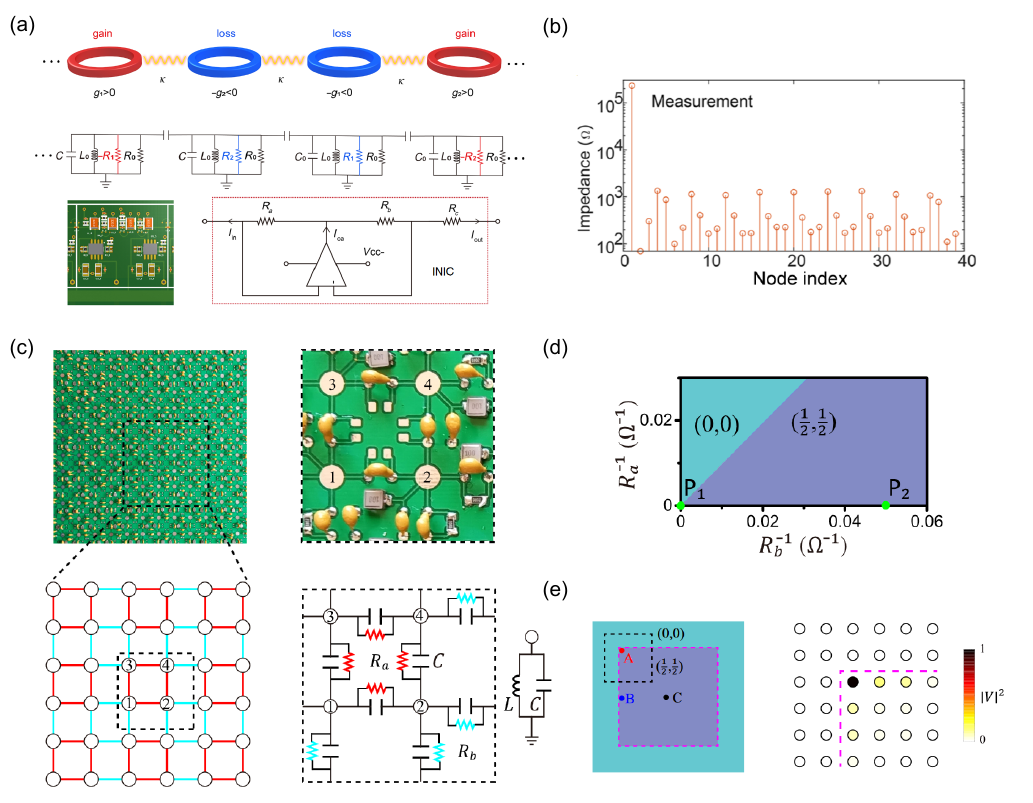}
  \caption{(a) Illustration of the non-Hermitian chain with gain and loss. Red and blue elements represent the gain and loss units, realized by negative and positive resistors in circuit, respectively. (b) Experimentally measured impedance with the edge states induced by gain and loss. (c) Circuit realization of a 2D square lattice with red and cyan segments connected by the resistor $R_a$ and $R_b$. (d) 2D Wannier centers for different $R_a^{-1}$ and $R_b^{-1}$. (e) Measured corner states formed by the interface nontrivial and trivial circuits.\\
  {\emph{Source}:} The figures are adapted from Refs. \cite{Liu2020,Wu195127}
  }\label{NonH3}
\end{figure}

Gain and loss can also induce abundant topological phenomena. In circuit, the gain and loss unit can be readily realized by the negative resistor (INIC unit) and positive resistor. Liu et al. reported the gain- and loss-induced topological insulating phase in a non-Hermitian electrical circuit \cite{Liu2020}. The circuit model is shown in Fig. \ref{NonH3}(a), an 1D single resonant rings are coupling with the strength $\kappa$, and four rings are set to be a unit cell with two gain and two loss units, described by
\begin{equation}
J(\omega_0,q)=\left(
                \begin{array}{cccc}
                  \sigma_1 & -i\omega_0C & 0 & -i\omega_0Ce^{-iq} \\
                  -i\omega_0C & -\sigma_2 & -i\omega_0C & 0 \\
                  0 & -i\omega_0C & -\sigma_1 & -i\omega_0C \\
                  -i\omega_0Ce^{iq} & 0 & -i\omega_0C & \sigma_2 \\
                \end{array}
              \right)
\end{equation}
with $q$ the wave vectors, $\sigma_1=1/R_1$, $\sigma_2=1/R_2$, and $C$ being circuit parameters.

The topological properties of this non-Hermitian system is characterized by calculated the normalized Berry phase \cite{Takata213902,Liang012118}
\begin{equation}
W=\sum_s\frac{i}{2\pi}\left[\frac{1}{2}\int_{-2\pi}^{2\pi}dq\langle\langle\psi_{B,s}|\psi_{B,s}\rangle\right]
\end{equation}
where $|\psi_{B,s}\rangle$ and $|\psi_{B,s}\rangle\rangle$ are the right and left eigenvectors of $J(\omega_0,q)$. A topological edge mode appears for $\sigma_1\sigma_2>0$ in this system. Figure \ref{NonH3}(b) shows a typical nontrivial edge state.

One can also find the nontrivial topological states induced by gain and loss in the 2D circuit lattice. Wu et al. showed a non-Hermitian second-order topological phase induced purely by loss \cite{Wu195127}. The authors introduce the resistors to the 2D SSH circuit network, as shown in Fig. \ref{NonH3}(c), and the circuit Hamiltonian is
\begin{equation}
\mathcal{H}({\bf k})=i\left(
                       \begin{array}{cc}
                         G/C & \frac{1}{L}I/C \\
                         -I & O \\
                       \end{array}
                     \right),
\end{equation}
where 
\begin{equation}
G=\left(
    \begin{array}{cccc}
      G_c & -G_a-G_be^{-ik_x} & -G_a-G_be^{-ik_x} & 0 \\
      -G_a-G_be^{-ik_x} & G_c & 0 & -G_a-G_be^{-ik_x} \\
      -G_a-G_be^{-ik_x} & 0 & G_c & -G_a-G_be^{-ik_x} \\
      0 & -G_a-G_be^{-ik_x} & -G_a-G_be^{-ik_x} & G_c \\
    \end{array}
  \right)
\end{equation}
with $G_a=1/R_a$, $G_a=1/R_a$, and $G_a=2(1/R_a+1/R_b)$. $C$ has the same form with $G$ by replace $G_a$, $G_b$, $G_c$ to $C$, $C$, and $5C$, respectively. $I$ is a $4\times4$ identity matrix.

\begin{figure}
  \centering
  \includegraphics[width=1\textwidth]{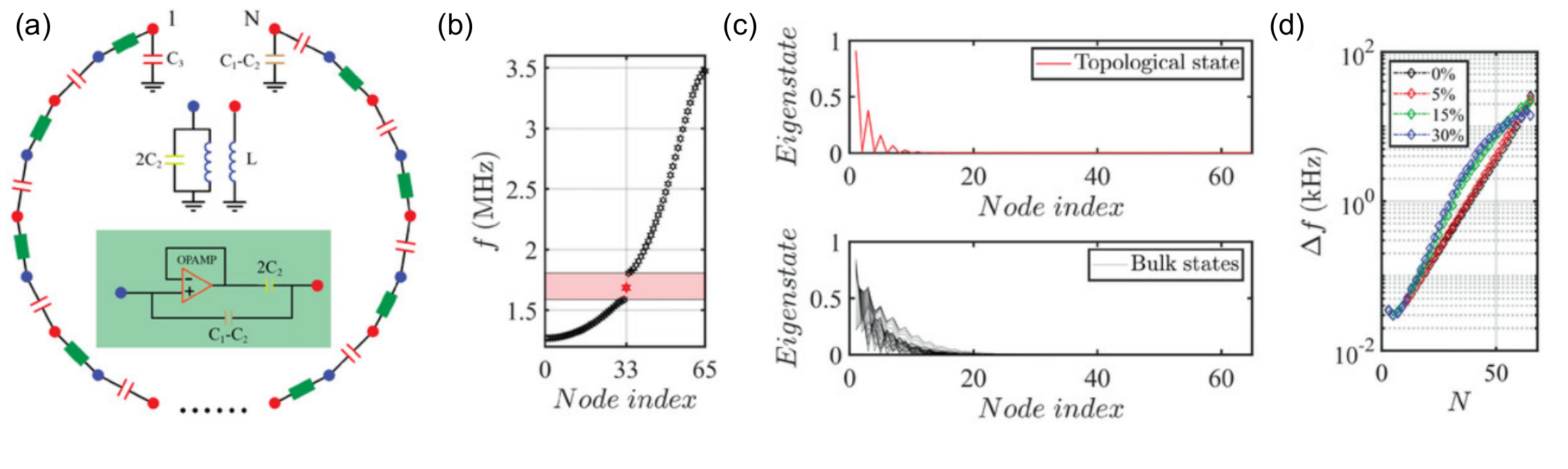}
  \caption{(a) Non-Hermitian 1D circuit model. (b) The spectra of this circuit with $N=65$. The red star indicates the edge mode. (c) The profiles of the edge (top) and bulk (bottom) states. (d) The frequency splitting dependence on the strength of disorder for different system sizes.\\
  {\emph{Source}:} The figures are adapted from Ref. \cite{Yuan13870}.
  }\label{NonH2}
\end{figure}

By evaluating the biorthogonal polarization 
\begin{equation}
W_{x,y}^{\alpha,\beta}=-\frac{1}{S}\int_{\rm BZ}A_{x,y}^{\alpha,\beta}d^2k
\end{equation} 
with $A_{x,y}^{\alpha,\beta}$ the non-Hermitian Berry connection. In some parameter regions, one can find the nonzero 2D Wannier centers, as shown in Fig. \ref{NonH3}(d). By constructing a trivial-nontrivial interface, one can observe the corner states displayed in Fig. \ref{NonH3}(e). With the controllable gain-loss parameters in circuit, both the first-order and higher-order topological have been realized. 

One can also introduce non-Hermitian elements to conventional topological systems. Based on Chern insulator model, Ezawa studied the Haldane model with non-Hermitian spin-orbit interactions and observed two non-Hermitian chiral edge modes \cite{Ezawa081401}. Later, the author explored the topological phases of the Dirac system in $2n$ dimensional space, characterized by the $n$th Chern number \cite{Ezawa075423}. After introducing the complex Dirac masses to this system, the $n$th Chern number keeps its quantization nature.

The disorder effect can also be discussed in non-Hermitian system. Su et al. simulated the non-Hermitian disordered systems in linear circuits \cite{Su184108}, and studied the competition between the skin effect and the Anderson localization behavior. Yuan et al. investigated the 1D non-Hermitian TEC with disorder \cite{Yuan13870}, as shown in Fig. \ref{NonH2}(a). In the absence of disorder, this system hosts an edge mode and multiple bulk modes, as displayed in Figs. \ref{NonH2}(b) and (c). By introducing disorder to the system, the spectra are split, as shown in Fig. \ref{NonH2}(d). One can use the phenomenon to design high-sensitive sensors. Figure \ref{NonH2}(d) also shows the influence of the system size on the splitting. The larger system size will induce a strong frequency splitting. One can also observe the size-dependent boundary effects in this non-Hermitian electric circuits.

With the non-Hermitian nature of INIC, circuits can be used to study other interesting non-Hermitian physics as well, like the non-Hermitian exceptional Landau quantization in electric circuits \cite{Zhang2020,Zhang1798}, $N$-th power root topological phases \cite{Deng033109}, and the exceptional-point splitting via non-Hermitian skin effects \cite{Xiao115427}. These works exhibit the advantages of circuit in studying non-Hermitian physics, and we will move to the nonlinear TECs below.

\subsection{Non-linear topological states}
Nonlinear systems are ubiquitous, ranging from the "butterfly effect" to "tornadoes and tsunamis". As Stanislaw Ulam said \emph{Using a term like nonlinear science is like referring to the bulk of zoology as the study of non-elephant animals}, which means most phenomena are inherently nonlinear \cite{Lapine1093,Kartashov247}. 

The study of nonlinear circuits begin from the 20th century. As we presented in Section \ref{S2}, the electrical circuits can be used to study the chaos \cite{Chenbook}, the solitons in nonlinear electric circuits governed by the Korteweg–De Vries equation \eqref{KdV} \cite{Nagashima1979,Muroya1981,Muroya1982,Kuusela1987}, and others. Combining nonlinear physics with the concept of topology, one can explore the nonlinear topological states in circuits.

\begin{figure}
  \centering
  \includegraphics[width=0.85\textwidth]{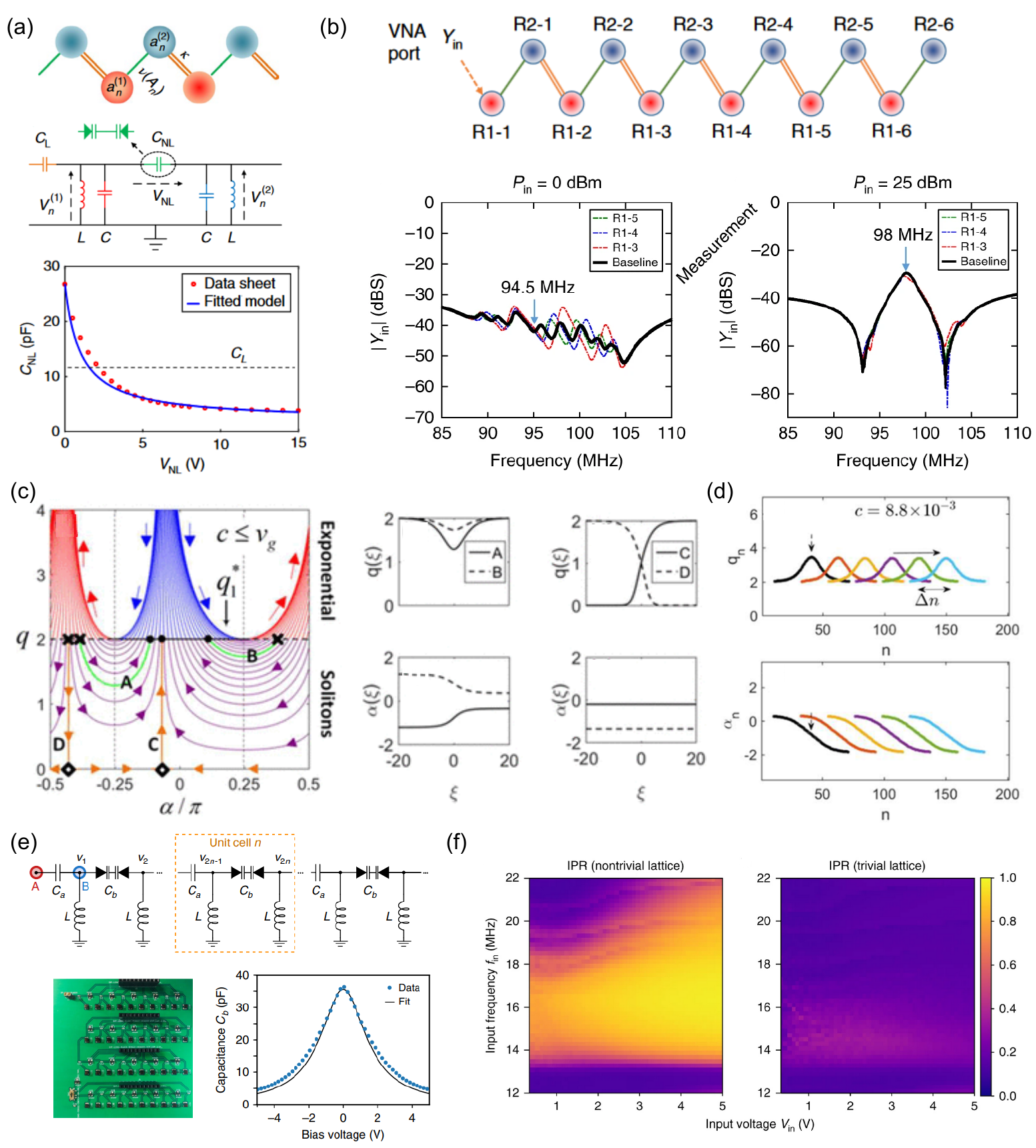}
  \caption{(a) The nonlinear circuit SSH model with nonlinear elements realized by a pair of varactors. (b) Measured input admittance versus frequency for low-intensity and high-intensity inputs, respectively. (c) Phase portrait of the nonlinear SSH model. The $A,B,C,D$ correspond to the four kinds of solitons in the right side. (d) The intensity and polarization angle evolution of the solitons. (e) Design and implementation of SSH-like left-handed transmission line. (f) Experimental observation of harmonic generation in the nonlinear circuit.\\
  {\emph{Source}:} The figures are adapted from Refs. \cite{Hadad2018,Hadad2017,Wang1102}.}\label{F10}
\end{figure}

As shown in Fig. \ref{F10}(a), Hadad et al. established a nonlinear SSH circuit \cite{Hadad2018}, and the non-linear elements are realized by a pair of varactors, the values of which depend on the voltage amplitude imposed on them, as shown in the bottom of Fig. \ref{F10}(a). The coupled non-linear circuit equations can be written as
\begin{equation}
\begin{aligned}
-j\frac{da_n^{(1)}}{dt}=\omega_0a_n^{(1)}+\nu (V_n)a_n^{(2)}+\kappa a_{n-1}^{(2)},\\
-j\frac{da_n^{(2)}}{dt}=\omega_0a_n^{(2)}+\nu (V_n)a_n^{(1)}+\kappa a_{n+1}^{(1)},\\
\end{aligned}
\end{equation}
where $a_n^{(1)}$ and $a_n^{(2)}$ are nodal amplitudes. The resonators are coupled alternately through linear and nonlinear coupling coefficients $\kappa$ and $\nu$, respectively. The nonlinear coefficient $\nu$ follows $\nu\approx(\nu_0-\nu_\infty)/(1+V_n/V_0)+\nu_\infty$ with $\nu_0=0.135\omega_0$, $\nu_\infty=0.02\omega_0$, $V_0=1$ V, and $V_n$ the imposed voltage.

For a trivial SSH model, there is no edge state, as shown in the bottom left of Fig. \ref{F10}(b). However, the edge modes appear at a strong nonlinear pumping, see the bottom right of Fig. \ref{F10} (b). Further, the authors demonstrate that edge modes can evolve into solitons in this special system, which are sustained by propagating domain walls induced by the local intensity distribution \cite{Hadad2017}. As shown in Fig. \ref{F10}(c), the phase portrait and four types of solitons are displayed, marked as $A$, $B$, $C$, and $D$. To show the stability of the soliton, the evolution of the soliton is examined from the initial black configuration to blue one, as displayed in Fig. \ref{F10}(d), and the shape of solitons are unchanged. The solitons can also be realized in higher-order topological systems. Tao et al. constructed a 3D electric network to simulate the second-order TIs with hinge states \cite{Tao103058}. By introducing the nonlinear inductors to this system, the authors observed the propagation of the solitons at hinges formed by the voltage distributions. 

Except for the solitons, the cnoidal wave localization is another important feature of nonlinear circuit. Hohmann et al. observed the localization behavior in a nonlinear TEC, followed the Korteweg-de Vries equation \cite{Hohmann2023}. Nonlinear topological systems can be used to explore other interesting phenomena. In Ref. \cite{Wang1102}, Wang et al. showed experimentally that in a left-handed nonlinear transmission lines analog of the SSH lattice, see Fig. \ref{F10}(e). One can realize a strong propagating higher-harmonic signals in the presence of topological edge mode, as shown in Fig. \ref{F10}(f). Tang et al. observed strongly nonlinear topological phases of cascaded TECs \cite{Tang33311}. The authors found that the nonlinear topological interface modes arise on domain walls of the circuit lattices, whose topological phases are controlled by the amplitudes of nonlinear voltages.

\subsection{Non-Abelian topological states}
The gauge field stands as a pivotal concept in contemporary physics, holding significance across various domains such as electromagnetism, high-energy physics, and condensed matter physics. These fields can be categorized into Abelian and non-Abelian based on the commutativity of their associated symmetric groups. Many fascinating physical phenomena are related to non-Abelian physics, such as the non-Abelian TIs \cite{Jiang235428,Jiang6471}, non-Abelian monopoles \cite{Auzzi207}, non-Abelian Aharonov–Bohm effect \cite{Horvathy407}. 

\begin{figure}
  \centering
  \includegraphics[width=0.9\textwidth]{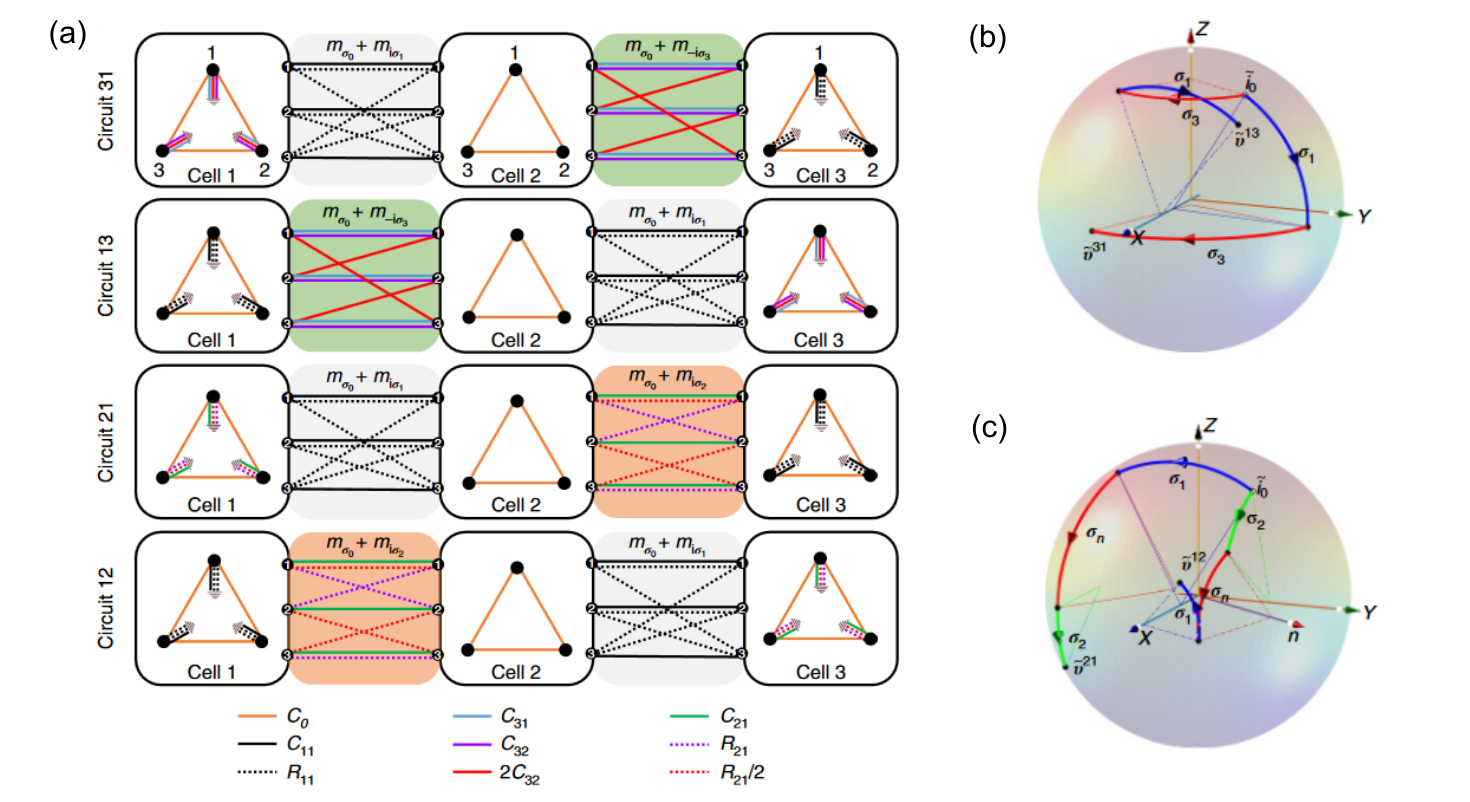}
  \caption{(a) Non-reciprocal circuit that generates a non-Abelian phase factor.  The same initial state leads to different final states under the non-Abelian gauge field: (b) circuit 31 and 13 and (c) circuit 21 and 12.\\
  {\emph{Source}:} The figures are adapted from Ref. \cite{Wu635}.}\label{F11}
\end{figure}

Recently, it was shown that electrical circuits can be employed to study non-Abelian effects. Wu et al. reported that non-Abelian gauge fields can be synthesized in circuits created from building blocks that consist of capacitors, inductors, and resistors ($three-subnode$ method) \cite{Wu635}. With these building blocks shown in Fig. \ref{F11}(a), one can create circuit networks for realizing spin-orbit interaction and the topological Chern state, representing non-Abelian gauge fields in momentum space. Besides, the non-reciprocal circuits were designed to implement the non-Abelian Aharonov-Bohm effect in real space. To show the direct evidence of the non-Abelian phase factor, the authors proved that the same initial state leads to different final states under the non-Abelian gauge field between circuits 31 and 13, as well as circuits 21 and 12, as displayed in Fig. \ref{F11}(b) and (c).

Another important branch of non-Abelian physics is the Majorana fermion, which are promising building blocks for developing fault-tolerant topological quantum computer \cite{Nayak1083}. Due to the presence of linear algebra structures in the Kirchhoff Laws, recent research has demonstrated that the electric-circuit approach has the capability to replicate the Kitaev $p$-wave topological superconductor model hosting the Majorana fermions \cite{Ezawa2075424}. One can simulate qubits, unitary transformation, superposition, and entanglement in circuit platform, which are beneficial to the development of quantum computations.

\subsection{Non-periodic and non-Euclidean topological states}

\begin{figure}
  \centering
  \includegraphics[width=0.8\textwidth]{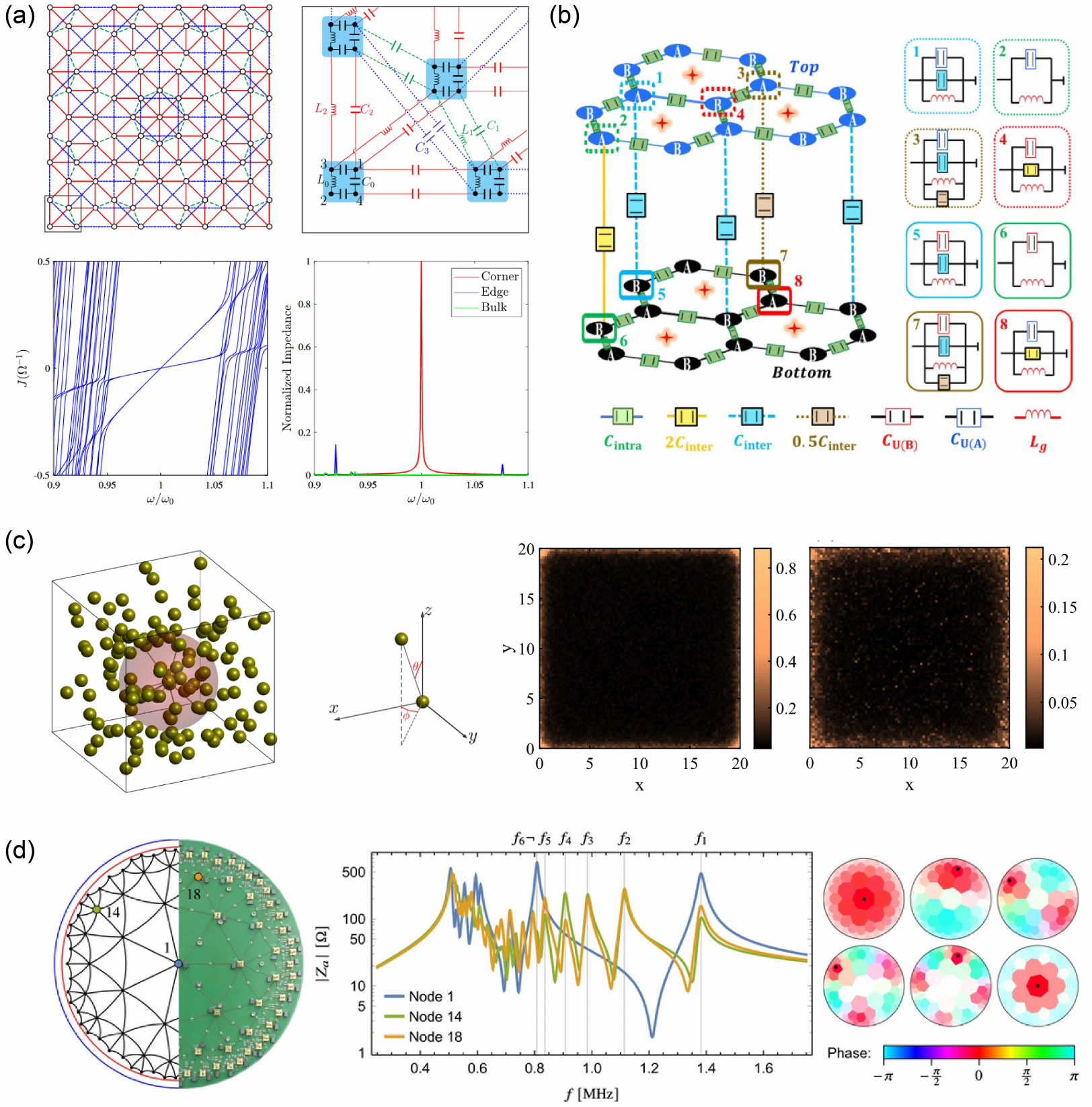}
  \caption{(a) The corner state in a quasicrystal. (b) Moir\'{e} circuit. (c) Model of topological amorphous metal (left) and its edge modes for $m_z=2$ (middle) and $m_z=6$ (right). (d) Circuit model in hyperbolic space and its edge mode.\\
  {\emph{Source}:} The figures are adapted from Refs. \cite{Chen036803,ZhangL201408,YBYang2019,Lenggenhager4373}. }\label{F13}
\end{figure}

Except for the aforementioned topological states in periodic lattices, other exotic topological states can be realized in non-periodic lattice structures (quasicrystal, Moiré, and amorphous lattices) and non-Euclidean spaces (hyperbolic space). All of these special lattice structures are generated by the flexible connection methods between circuit nodes, which are hard to achieve in other systems. Chen et al. proposed a quasicrystal model with HOT states and designed a Ammann-Beenker tiling quasicrystal with four sites in a cell \cite{Chen036803}. The Hamiltonian is written as
\begin{equation}
\mathcal{H}_{\rm QI}=\gamma\sum_jc_j^\dagger(\Gamma_2+\Gamma_4)c_j+\frac{\lambda}{2}\sum_{j\neq k}f(r_{jk})c_j^\dagger T(\phi_{jk})c_k,
\end{equation}
with $T(\phi_{jk})=|\cos\phi_{jk}|\Gamma_4-i\cos\phi_{jk}\Gamma_3+|\sin\phi_{jk}|\Gamma_2-i\sin\phi_{jk}\Gamma_1$. Here, $c_j^\dagger=(c_{j1}^\dagger,c_{j2}^\dagger,c_{j3}^\dagger,c_{j4}^\dagger)$ is the creation operator in cell $j$. $\gamma$ and $\lambda$ are the intracell and intercell hoppings. $\Gamma_4=\tau_1\tau_0$ and $\Gamma_\nu=-\tau_2\tau_\nu$ with $\nu=1,2,3$ with $\tau_i$ the Pauli matrices. They mapped the Hamiltonian to a circuit model and predicted a second-order corner state, as shown in Fig. \ref{F13}(a). Later, Lv et al. realized this quasicrystalline quadrupole TIs in circuit experiment, as reported in Ref. \cite{Lv108}, and the corner states are identified by impedance measurements. 

Zhang et al. engineered the Moir\'{e} physics in electric circuits \cite{ZhangL201408}, as shown in Fig. \ref{F13}(b). By designing the interlayer coupling and biasing of one sublattice for the twisted bilayer circuit, the low-energy flat bands with large band gaps away from other states can be realized at various twist angles. Pei et al. proposed a bilayer circuit network to engineer novel valley topologies. By tuning the interlayer coupling pattern and the distribution of node grounding, the valley-dependent chiral edge states located at two kinds of domain walls, AB-AB and AB-BA interfaces, are fulfilled in the designed bilayer circuit \cite{Pei128242}.

Further, Zhang et al. realized the topological amorphous metal in 3D system with completely random sites \cite{YBYang2019}, as shown in left of Fig. \ref{F13}(c). The Hamiltonian reads
\begin{equation}
\mathcal{H}=\sum_{\bf x}\big[\sum_{\bf R}t(R)\hat{c}_{\bf x}^\dagger H_0(\theta,\phi)\hat{c}_{{\bf x+R}(\theta,\phi)}
+m_z\hat{c}_{\bf x}^\dagger\sigma_z\hat{c}_{\bf x}\big],
\end{equation}
where $\hat{c}_{\bf x}^\dagger=(\hat{c}_{{\bf x},\uparrow}^\dagger,\hat{c}_{{\bf x},\downarrow})$ is the creation operator, which is a random vector uniformly distributed in the cubic lattice. $H_0(\theta,\phi)=\sigma_z+i\sin\theta\cos\phi\sigma_x+i\sin\theta\sin\phi\sigma_y$ is the hopping matrix for the neighboring sites. $t(R)=-\exp[\lambda(1-R)]/2$ is the exponentially decaying hopping strength and $m_z$ is the mass term. Due to the non-periodic nature of this system, their topological features can be characterized by the Bott index and the Hall conductivity, with the boundary states being displayed in Fig. \ref{F13}(c).

Furthermore, electrical circuits can be used to realize topological states in negatively curved (hyperbolic) space. In Ref. \cite{Lenggenhager4373}, Lenggenhager et al. discussed and experimentally demonstrated that the spectral ordering of Laplacian eigenstates for hyperbolic and flat 2D spaces has a universally different structure, as shown in Fig. \ref{F13}(d). The authors use a lattice regularization of hyperbolic space in an electric-circuit network to measure the eigenstates of a "hyperbolic drum" in a time-resolved experiment, where they verify signal propagation along the curved geodesics. Later, Chen et al. realized hyperbolic matter as a paradigm for topological states through TEC networks utilizing a complex-phase circuit element \cite{Chen622}. Zhang et al. fabricated the hyperbolic circuits to demonstrate hyperbolic band topological states protected by second Chern numbers \cite{Zhang1083}. It can be seen that electrical circuits offer the opportunities to explore topological states in all of these non-periodic lattice structures.

\subsection{High-dimensional topological states}
In the classical topology tenfold classes of fermionic Hamiltonians, one may expect the topological states in different dimensions, from one to ten dimensions. However, due to the limitation of real dimensions, the classification of TIs contains hypothetical high-dimensional phases that cannot be realized with real materials. Nevertheless, the circuit systems can extend the dimension readily.

\begin{figure}
  \centering
  \includegraphics[width=0.72\textwidth]{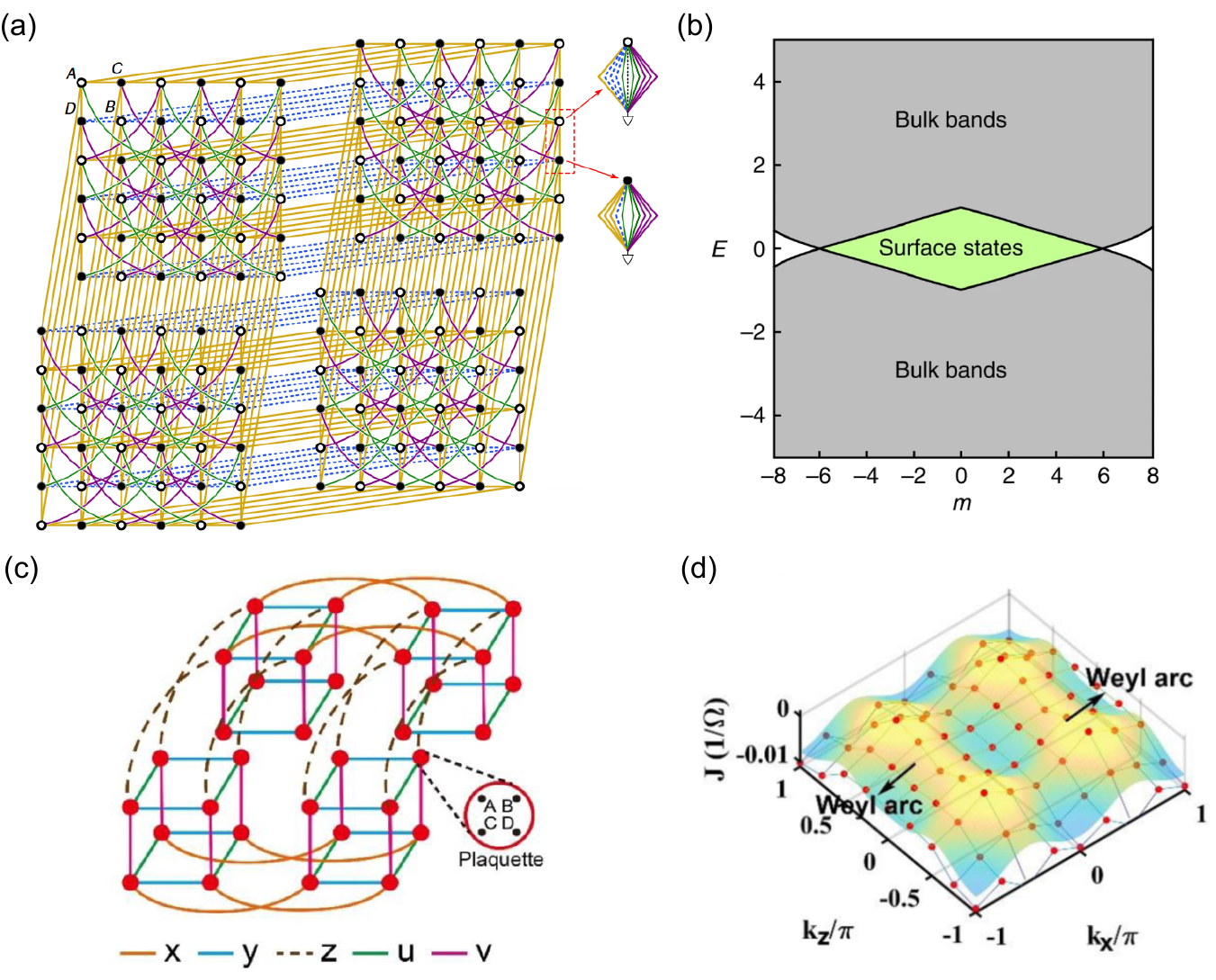}
  \caption{(a) Model of the 4D quantum Hall lattice with circuit implementation. (b) Calculated band diagram for different Dirac mass $m$. The bulk bands are shown in grey. For $|m|<6$, there is a bandgap associated with nontrivial second Chern number, accompanied by topological surface states (shaded green). For $|m|>6$, the bandgap is trivial. (c) Five-dimensional Weyl circuit. (d) The Weyl arc in admittance space.\\
  {\emph{Source}:} The figures are adapted from Refs. \cite{Wang2356,Zheng033203}.}\label{F12}
\end{figure}

As shown in Fig. \ref{F12}(a), Wang et al. established a four-dimensional (4D) circuit belonging to the AI class of topological tenfold classes. The Hamiltonian reads \cite{Wang2356}
\begin{equation}
{\mathcal H}=(2\cos k_x+\cos k_y)\Gamma_1+\sin k_y \Gamma_2+(2\cos k_z+\cos k_w)\Gamma_3
+\sin k_w\Gamma_4+[m+4\cos(2k_x+2k_z)-4\cos(2k_x-2k_z)]\Gamma_5,
\end{equation}
with $\Gamma_1=\sigma_x\otimes\tau_z$, $\Gamma_2=\sigma_y\otimes\tau_0$, $\Gamma_3=\sigma_x\otimes\tau_x$,
$\Gamma_4=\sigma_x\otimes\tau_y$, and $\Gamma_5=\sigma_z\otimes\tau_0$. $m$ is the Dirac mass. One can observe topological surface states in 4D circuits that are associated with the nonzero second Chern number but vanishing first Chern number, see Fig. \ref{F12}(b). Similar results were found in Ref. \cite{Yu1288}, and the 4D spinless TI was reported in the circuit system with pairs of 3D Weyl boundary states.

With the same strategy, the exotic TSMs in high-dimensional systems can also be realized. Zheng et al. constructed five-dimensional electric circuit platforms in real space and experimentally observed topological phase transitions in this system \cite{Zheng033203}, as shown in Fig. \ref{F12}(c), described by the following Hamiltonian 
\begin{equation}
{\mathcal H}=2i\sqrt{\frac{C_1}{L_1}}\Big\{\sum_{i=1}^5\zeta_i({\bf k})\Gamma^i+ia\frac{[\Gamma^4,\Gamma^5]}{2}\Big\},
\end{equation}
with $\Gamma^{1,2,3,4,5}=\{\sigma_3\tau_1,\sigma_3\tau_2,\sigma_3\tau_3,\sigma_1\tau_0,\sigma_2\tau_0\}$. Not only are Yang monopoles and linked Weyl surfaces observed experimentally [Fig. \ref{F12}(d)], but various phase transitions in five dimensions are also proved, such as from a normal insulator to a Hopf link of two Weyl surfaces and then to a five-dimensional TI. In addition, Li et al. realized a full 3D-imaging of nodal boundary Seifert surfaces in a 4D topological circuit \cite{Li2019}.

The HOT states can also be realized in higher-dimensional systems with distinctive features. Zhang et al. theoretically propose and experimentally demonstrate the realization of a classical analog of 4D hexadecapole TI based on the electric circuits \cite{Zhang100102}. 

Due to the freedom of connection methods in circuit, one can establish TECs in principle with an arbitrary dimension. It is noted that connections between circuit nodes are irrelevant to their positions, so high-dimensional circuits can be realized merely by 2D printed circuit boards.

\begin{figure}
  \centering
  \includegraphics[width=0.9\textwidth]{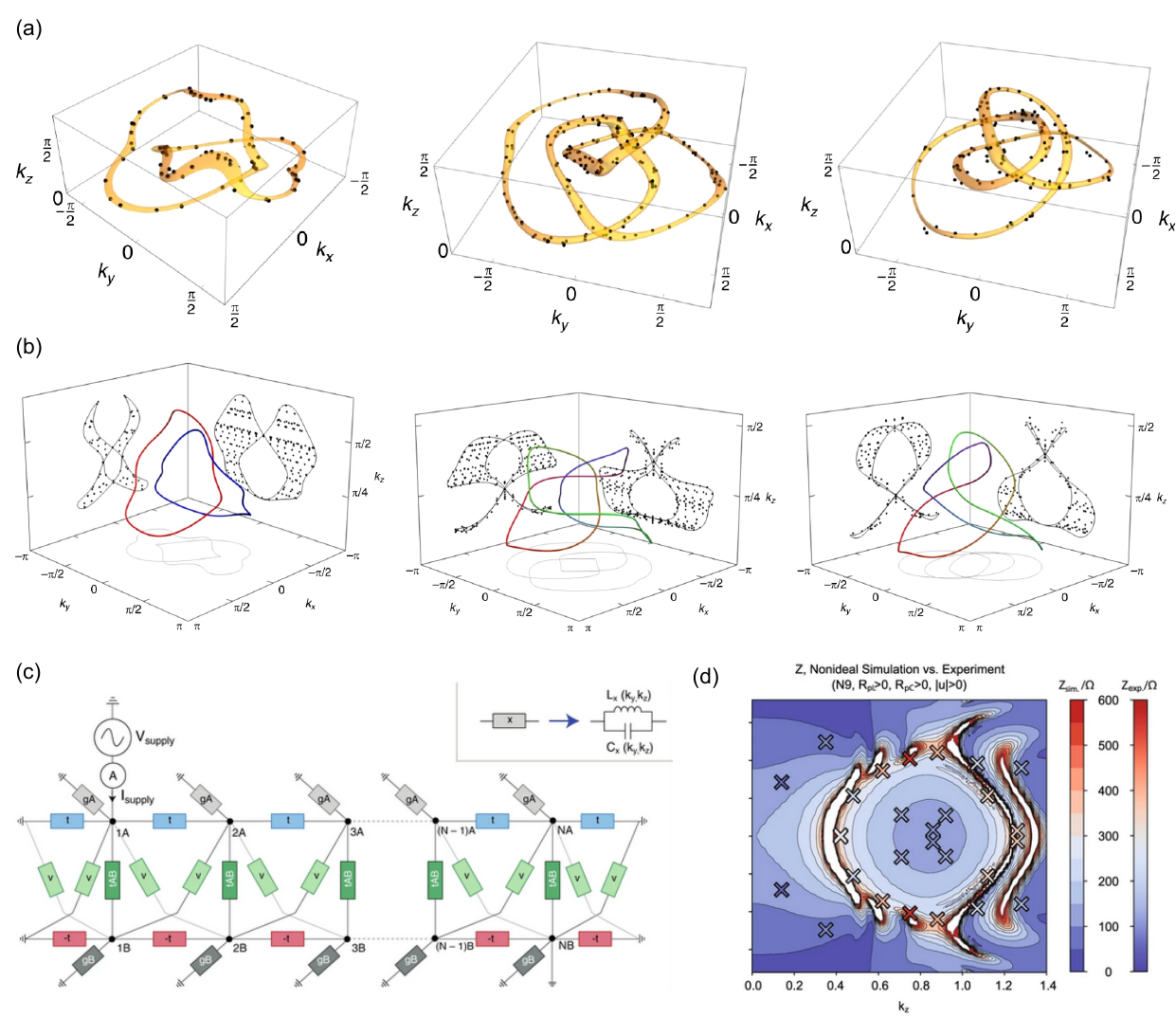}
  \caption{(a) Nodal knots for Hopf-link, trefoil knot, and Figure-8 knot from left to right. (b) Simulated drumhead states for open boundary condition. (c) Circuit model for realizing Hopf-link. (d) Experimentally measured drumhead state.\\
  {\emph{Source}:} The figures are adapted from Ref. \cite{Lee4385}.}\label{F13Knots}
\end{figure}

\subsection{Other novel topological states} 
Momentum space knots, in particular, have been challenging to study because they require precise adjustments to their long-range hoppings, as well as thorough investigation of their intricate connections and the detection of topological drumhead surface states. Lee et al. established the nodal knots in the RLC circuit \cite{Lee4385}. 
The nodal knots are built from a momentum space graph Laplacian
\begin{equation}
J({\bf k})=l_0+{\rm Re}f({\bf k})\tau_x+{\rm Im}f({\bf k})\tau_z
\end{equation}
with $l_0$ a uniform offset and $f({\bf k})$ an even function. By choosing different $f({\bf k})$, one can realize the many kinds of nontrivial linked nodal structure, such as Hopf-link, trefoil knot, and Figure-8 knot, as shown in Fig. \ref{F13Knots}(a). These different nodal knots manifest as drumhead states under open boundary condition, displayed in Fig. \ref{F13Knots}(b). To prove the drumhead state in circuit, a ladder LC circuit is designed with Hopf link, as shown in Fig. \ref{F13Knots}(c), and the topological "drumhead" surface states are measured by characterizing impedance in Fig. \ref{F13Knots}(d).

In Ref. \cite{Wang057201}, Wang et al. reported the realization of Hopf insulators in circuit, whose realization requires special forms of long-range spin-orbit coupling. The authors designed a circuit network and calculated the Hopf invariant, then found that the circuit realizes a Hopf insulator with a Hopf invariant of 4 and demonstrated the bulk-boundary correspondence.

Due to the presence of active elements in circuit (operational amplifier), one can study the pumping effects. Yatsugi et al. observed the bulk-boundary correspondance in topological pumping based on a tunable electric circuit \cite{Yatsugi180}. Katwal et al. designed principles for active TECs that can self-excite topologically protected global signal patterns \cite{Kotwal2021}, which can be used to protect information transmission and wave guidance. Stegmaier et al. realized a topological adiabatic pump in an electrical circuit network that supports remarkably stable and long-lasting pumping of a voltage signal \cite{Stegmaier023010}.

Up to now, we have shown the capabilities of TECs in studying peculiar topological physics. With the flexibility of TECs, one may envision other topological phases in the future.

\section{Circuit simulation of other systems}\label{S6}

As a powerful platform, electrical circuit can be used to simulate the novel phenomena in other systems, such as the photonic, magnetic, and quantum systems. In this section, we will discuss the versatility for the circuit simulation of other systems.

\subsection{Circuit simulation of photonic system}

\begin{figure}
  \centering
  \includegraphics[width=0.9\textwidth]{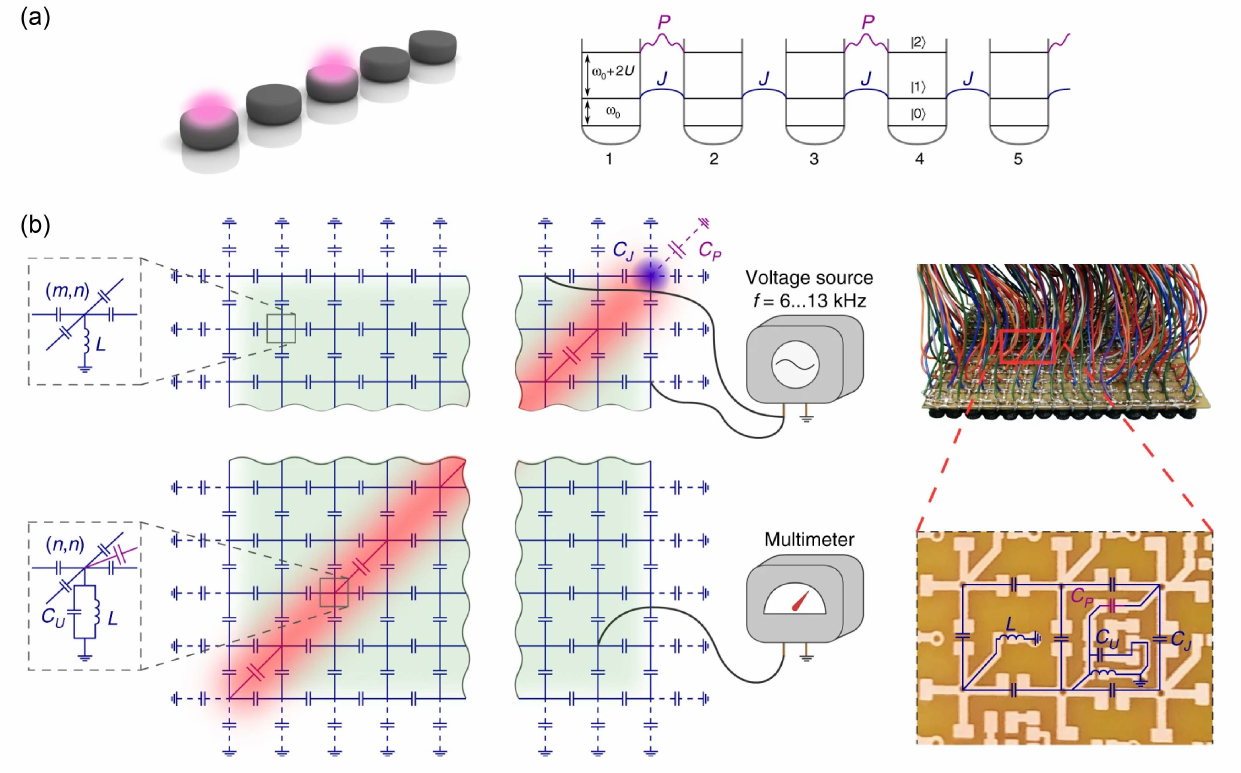}
  \caption{(a) Two interacting photons in an 1D chain. (b) Topolectrical circuit model for the extended Bose-Hubbard model, describing the two-interaction photons.\\
  {\emph{Source}:} The figures are adapted from Ref. \cite{Olekhno2020}.}\label{F13Hybrid}
\end{figure}

Olekhno et al. designed a 2D TEC to simulate an 1D quantum-optical system with two-interaction photons, as shown in Fig. \ref{F13Hybrid} (a) \cite{Olekhno2020}. The system is described by the extended Bose-Hubbard Hamiltonian
\begin{equation}
\mathcal{H}=\omega_0\sum_m\hat{n}_m-J\sum_m(\hat{a}_m^\dagger\hat{a}_{m+1}+\hat{a}_{m+1}^\dagger\hat{a}_{m})
+U\sum_m\hat{n}_m(\hat{n}_m-1)+\frac{P}{2}\sum_m(\hat{a}_{2m-1}^\dagger\hat{a}_{2m-1}^\dagger\hat{a}_{2m}\hat{a}_{2m}
+\rm{H.c.})
\end{equation}
with $\hat{a}_m^\dagger$ and $\hat{a}_m$ the creation and annihilation operators of photon at the $m$th resonator, $n=\hat{a}_m^\dagger\hat{a}_m$ the photon number. $\omega_0$, $U$, $J$, and $P$ are corresponding to the resonant frequency, on-site potential, tunneling coupling between the nearest neighbors, and direct two-photon tunnelings, respectively.

This Hamiltonian can be realized by the circuit model in Fig. \ref{F13Hybrid}(b), and one can observe the characteristic voltage patterns for two-photon scattering states (green), doublons (red), and doublon edge state (blue). It is proved that TEC can simulate the effective photon-photon interactions successfully and observe the doublon edge state. This approach can similarly be applied to mimic the behavior of $N$ interacting particles in the one-dimensional system by $N$-dimension circuits.

\subsection{Circuit simulation of magnetic system}
In spintronics community, skyrmion, a topologically protected spin texture, has aroused much attention. Magnetic skyrmion has many advantages, such as small size ($\sim$nm), high stability, and low driving current density, and one can generate, manipulate, and detect skyrmion in the room temperature. Therefore, skyrmion may be a good candidate for the next-generation information industry as the information carrier \cite{Yu9051,Chen2000857,Tokura2857,Fert17031}. Recent works found the electrical circuit can be used to simulate the behaviors of skyrmion, and we will discuss it in the section.

\subsubsection{Skyrmion in momentum space}
In Ref. \cite{Yang211}, Yang et al. reported the circuit realization of several skyrmion spin textures in momentum space. The construction of spin texture starts from a massive Dirac Hamiltonian with Wilson mass
\begin{equation}
\mathcal{H}=\sum^d_{i=1}\frac{\hbar v}{a}\sin(k_ia)\alpha_i+m v^2 \beta+\frac{4b}{a^2}\sin^2\frac{k_ia}{2}\beta
\end{equation}
where $k_i$ is the wave vector, $\alpha,\beta$ are the Dirac matrices, $a$ is the lattice constant, $\hbar v$ the hopping strength, and $d$ is the space dimension.

\begin{figure}
  \centering
  \includegraphics[width=0.95\textwidth]{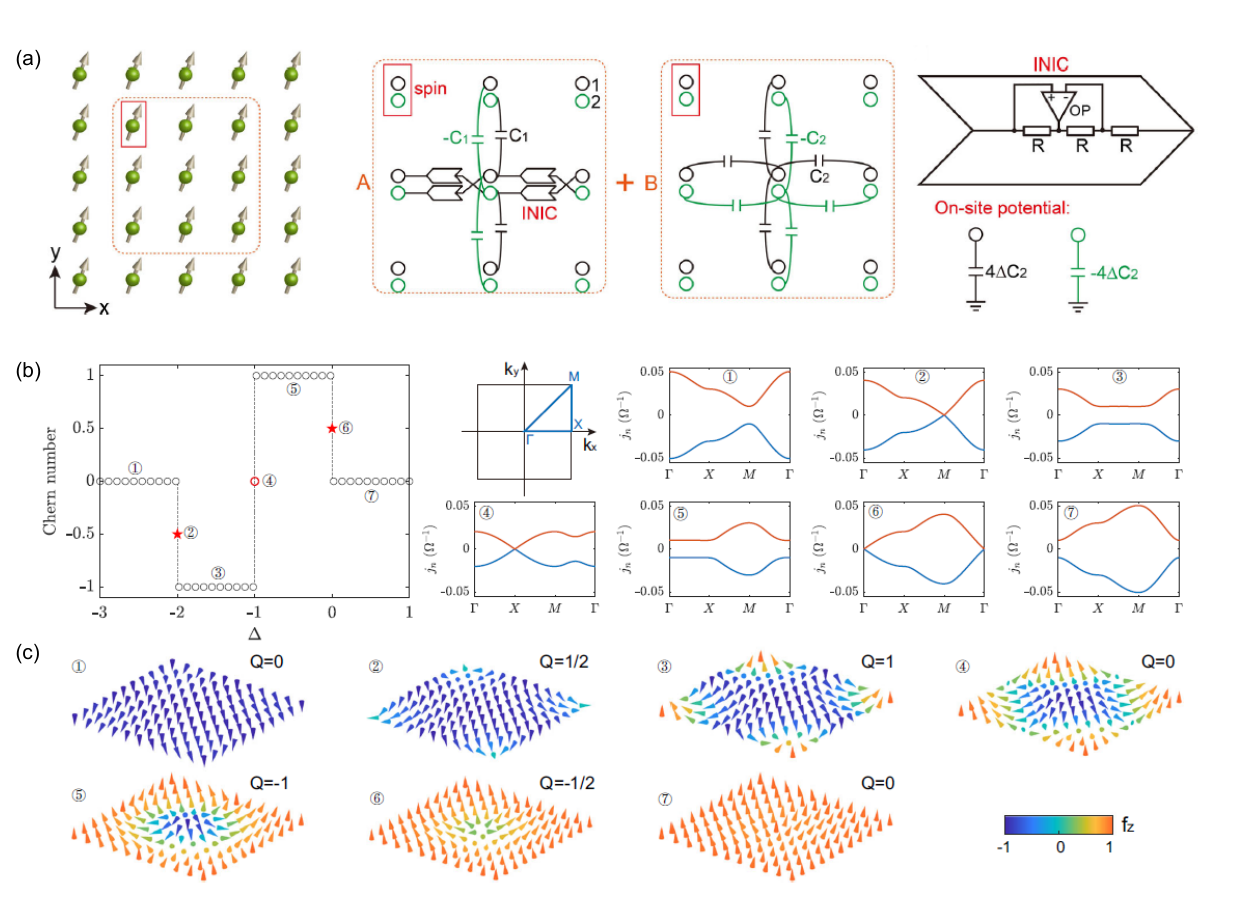}
  \caption{(a) Lattice and circuit models for the realization of Dirac Hamiltonian with Wilson mass. (b) The Chern numbers and the band structures for different Dirac mass $\Delta$. (c) The spin texture for different Chern numbers (topological chargers).\\
  {\emph{Source}:} The figures are adapted from Ref. \cite{Yang211}.}\label{F13skyrmion}
\end{figure}

One can realize this Hamiltonian for $d=2$ with the circuit diagram shown in Fig. \ref{F13skyrmion}(a), and the circuit Hamiltonian is expressed as 
\begin{equation}\label{WilsonH}
\mathcal{H(\omega)}=2G\sin k_x\sigma_x+2 \omega C_1\sin k_y\sigma_y+4 \omega \Delta C_2\sigma_z+
4\omega C_2(\sin^2{\frac{k_x}{2}}+\sin^2{\frac{k_y}{2}})\sigma_z,
\end{equation}
where $\sigma_x$, $\sigma_y$, and $\sigma_z$ are Pauli matrices, and $G$, $C_1$, and $C_2$ are circuit parameters.

The Hamiltonian \eqref{WilsonH} can be rewritten as $\mathcal{H}={\bf f}({\bf k})\cdot{\boldsymbol\sigma}$. One can define a spin vector $\hat{\bf f}({\bf k})$ as  
\begin{equation}
\hat{{\bf f}}({\bf k})=\frac{\bf f ({\bf k})}{\vert{\bf f ({\bf k})}\vert}
\end{equation}
with ${\bf f}({\bf k})=(f_x,f_y,f_z)$ is the coefficient of Pauli matrices and $\vert{\bf f ({\bf k})}\vert=\sqrt{f_x^2+f_y^2+f_z^2}$. 

The spin textures are close related the Berry curvature and the topological Chern number $\mathcal{C}$ with the formula in Eq. \eqref{Chernnumber}. Figure \ref{F13skyrmion}(b) shows the Chern number for different Dirac mass. Three groups of values, 0, $\pm1/2$, and $\pm1$, are found in this system, which are corresponding to four types of spin textures in Fig. \ref{F13skyrmion}(c).

Figure \ref{F13skyrmion}(c) shows the spin vectors for different $\Delta$. One can see the uniform ferromagnetic spin texture and a half skyrmion pair for $\mathcal{C}=0$, half skyrmion for $\mathcal{C}=\pm1/2$, and skyrmion for $\mathcal{C}=\pm1$. The topological charges of the spin textures are labelled at the top right corner, computed by 
\begin{equation}
Q=\frac{1}{4\pi}\int_{\rm BZ} \hat{\bf f}\cdot\left(\frac{\partial \hat{\bf f}}{\partial k_x}\times \frac{\partial \hat{\bf f}}{\partial k_y}\right) dk_xdk_y.
\end{equation}
One can see that $Q=-\mathcal{C}$. These results construct a close connection between circuit and magnetic systems.

For application, such as the racetrack memory, the manipulation of skyrmion motion is necessary. One can control the circuit skyrmion motion in momentum space by continuously changing the circuit parameters, and the details are discussed in Ref. \cite{Yang211}. 

\subsubsection{Skyrmion motion in lattice space}

In spintronics field, the skyrmion and other spin textures emerge in the lattice space. In order to describe the motions of the magnetic solitons, such as magnetic bubbles, vortex, skyrmions, one can use Thiele equation for interpretation, which is based on the approximation of collective coordinate for magnetization vector. 

\begin{figure}
  \centering
  \includegraphics[width=0.75\textwidth]{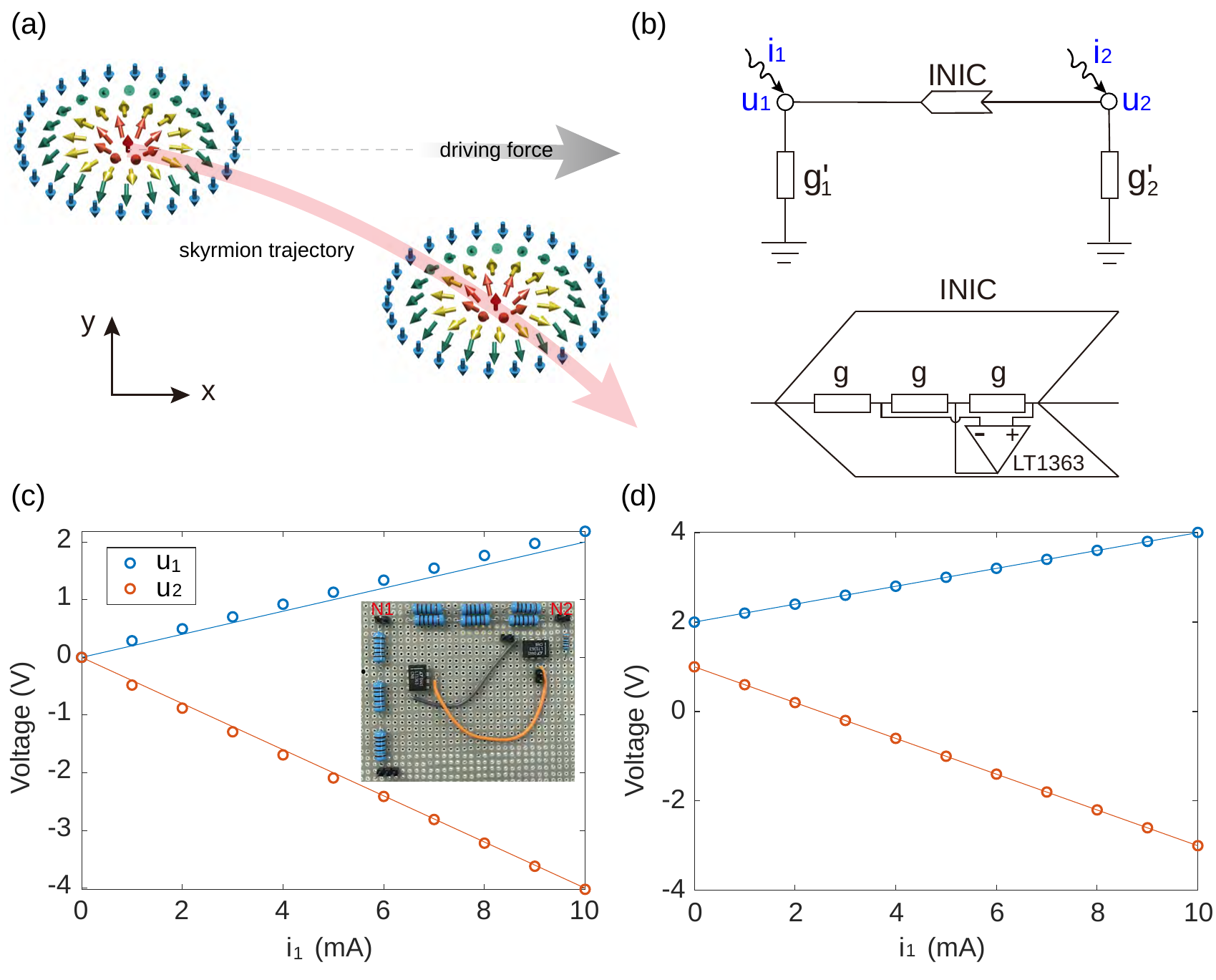}
  \caption{(a) Illustration of the skyrmion motion under a driving force {\bf F}, exhibiting the skyrmion Hall effect. (b) Two-node circuit model connected by a INIC unit. Node 1 and Node 2 are grounded by two resistors $g_1'=1/R_1$ and $g_2'=1/R_2$, respectively. $i_1$ and $i_2$ are the current flowing into the two nodes. (c) Voltage response $u_1$ and $u_2$ for different current $i_1$ with $i_2=0$ mA. The inset shows the real circuit board used in experiment. The dots are measured in experiment and lines are theoretical formula. (d) Voltage response for various $i_1$ and a fixed current $i_2=5$ mA.}\label{F13Thiele}
\end{figure}

Thiele's equation was firstly proposed by A. Thiele in 1973 for describing the motion of magnetic domains \cite{Thiele1973}, which reads
\begin{equation}
{\bf G}\times {\bf v}-\mathcal{D}\alpha {\bf v}+{\bf F}=0
\end{equation}
where ${\bf G}=4\pi Q\hat{z}$ with $Q$ being the topological charge of solitons, $\mathcal{D}$ represents the dissipative force, and $\alpha$ is the Gilbert damping constant. ${\bf v}=(v_x,v_y)^T$ is the velocity of magnetic solitons and ${\bf F}=(F_x,F_y)^T$ is the driving force. In this section, we will show how to map electrical equation to Thiele's equation.

One can express the Thiele equation in matrix form
\begin{equation}
\left(
  \begin{array}{c}
    v_x \\
    v_y \\
  \end{array}
\right)=\frac{1}{(4\pi Q)^2+(\alpha \mathcal{D})^2}\left(
          \begin{array}{cc}
            \alpha \mathcal{D} & -4\pi Q \\
            4\pi Q  & \alpha \mathcal{D}\\
          \end{array}
        \right)\left(
                 \begin{array}{c}
                   F_x \\
                   F_y \\
                 \end{array}
               \right),
\end{equation}
which describe the velocity of magnetic soliton motion under driving force. Figure \ref{F13skyrmion}(a) shows a typical motion trajectory of magnetic skyrmion with $Q=-1$ and force imposed on $\hat{x}$ direction, exhibiting the skyrmion Hall effect along $\hat{y}$ direction.

For describing it conveniently in circuit, we convert the above equation to
\begin{equation}
\left(
  \begin{array}{c}
    F_x \\
    F_y \\
  \end{array}
\right)=\left(
          \begin{array}{cc}
            \alpha \mathcal{D} & 4\pi Q \\
            -4\pi Q & \alpha \mathcal{D}\\
          \end{array}
        \right)\left(
                 \begin{array}{c}
                   v_x \\
                   v_y \\
                 \end{array}
               \right)
\end{equation}

We then consider the following mapping: the current $\left(
  \begin{array}{c}
    i_1 \\
    i_2 \\
  \end{array}
\right)$ is used to represent the driving force $\left(
  \begin{array}{c}
    F_x \\
    F_y \\
  \end{array}
\right)$ and the voltage response $\left(
  \begin{array}{c}
    u_1 \\
    u_2 \\
  \end{array}
\right)$ is utilized to display the skyrmion velocity $\left(
                 \begin{array}{c}
                   v_x \\
                   v_y \\
                 \end{array}
               \right)$.

Considering the motion of a skyrmion with $Q=-1$, one can construct a circuit model to realize
\begin{equation} \label{CE}
\left(
  \begin{array}{c}
    i_1 \\
    i_2 \\
  \end{array}
\right)=\left(
          \begin{array}{cc}
            g' & -g \\
            g  & g'\\
          \end{array}
        \right)\left(
                 \begin{array}{c}
                   u_1 \\
                   u_2 \\
                 \end{array}
               \right)
\end{equation}
This formula can be produced by a circuit model with two nodes shown in Fig. \ref{F13skyrmion}(b). One can write the circuit equations by $node-analysis$ method 
\begin{equation}
\left(
  \begin{array}{c}
    i_1 \\
    i_2 \\
  \end{array}
\right)=\left(
          \begin{array}{cc}
            g_1'+g & -g \\
            g  & g_2'-g\\
          \end{array}
        \right)\left(
                 \begin{array}{c}
                   u_1 \\
                   u_2 \\
                 \end{array}
               \right)
\end{equation}

If we choose some special parameters, such as $g_1'=-1$ mS and $g_2'=3$ mS, we can obtain $g=2$ mS and $g'=1$ mS in Eq. \eqref{CE}. Therefore, the Thiele equation is fully realized in circuit. For examination, we have established the circuit board as shown in the inset of Fig. \ref{F13skyrmion}(c). By adding a current $i_1$ ranging from $0$ to $10$ mA, we measured both the voltage $u_1$ and $u_2$, shown by the dots in Fig. \ref{F13skyrmion}(c). The lines are solved from Eq. \eqref{CE}, which compares well with the experimental results. Here, the nonzero voltage $u_2$ at node 2 is similar to the Hall motion in Thiele equation. We also measure the voltage response for $i_2=5$ mA, and display the results in Fig. \ref{F13skyrmion}(d). One can find a compensation current $i_2$ leading to $u_2=0$.

According to these results, it is possible for simulating the Thiele equation in circuit. With this platform, one can study the skyrmion Hall effect in circuit. In future studies, one can study the mass effect and thermal effect in circuit with the modified Thiele's equations.

\subsection{Circuit simulation of quantum behaviors}
\begin{figure}
  \centering
  \includegraphics[width=0.9\textwidth]{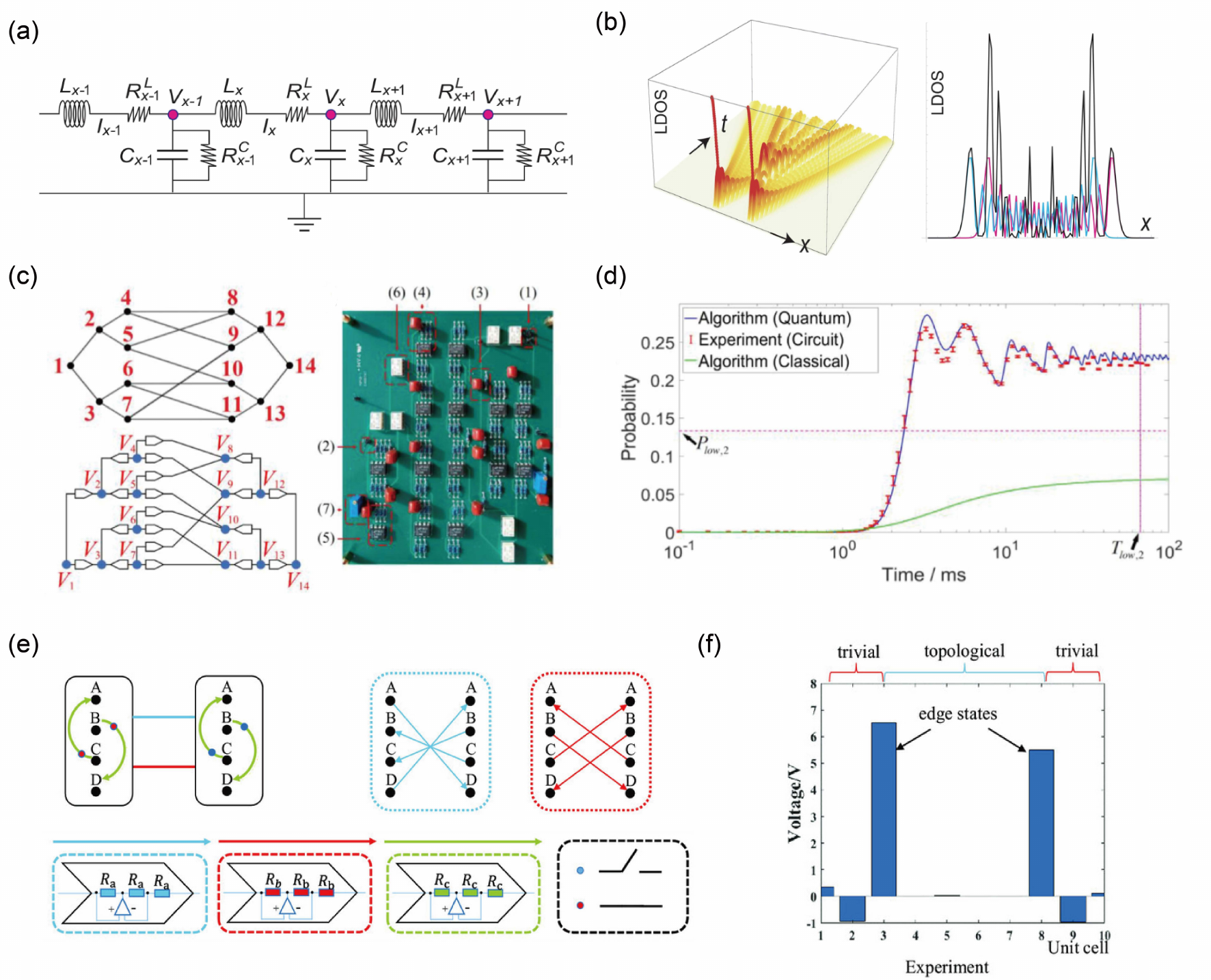}
  \caption{(a) Illustration of an electric circuit realizing an inhomogeneous telegrapher equation. (b)Time evolution of two quantum walkers starting from two points. The absolute value of the LDOS at a fixed time. (c) The framework and experimental realization of quantum search. (d) The average probability of the electric signal arriving at the exits at a given time $T$. (e) Circuit model for realizing the Kitaev chain. (f) The Majorana-Like edge states of circuit Kitaev model.\\
  {\emph{Source}:} The figures are adapted from Refs. \cite{Ezawa165419,Zhang2100143,Zou2300354}.}\label{Quantum}
\end{figure}

Recent works found that electrical circuits can be employed to simulate quantum behaviors. As shown in Fig. \ref{Quantum}(a), Ezawa simulated the Schrödinger equation in electric circuits and investigated the quantum walks in LC circuits \cite{Ezawa165419}. The author designed an electric-circuit simulation of the interference experiment, as displayed in Fig. \ref{Quantum}(b).

The fast quantum search can also be realized in electrical circuits. Pan et al. found quantum search algorithm can also be realized using classical electric circuits \cite{Pan9793071}. Zhang et al. provided the
quantum walkers evolving on unbalanced trees in circuit, demonstrating faster hitting performances than classical random walk \cite{Zhang2100143}, as shown in Fig. \ref{Quantum}(c) and (d). Later, the fast quantum search of multiple vertices is experimentally realized in electric circuits \cite{Ji172}. These works suggest an outset for quantum search with circuit platforms. 

Besides, the quantum computations and logics are also realized in circuit \cite{Zhang2200232,Zou2300354,Tong251}. In Ref. \cite{Zou2300354}, the Kitaev model is simulated by the electrical circuit with the circuit model in Fig. \ref{Quantum}(e), and the Laplacian is 
\begin{equation}
J=\left(
    \begin{array}{cc}
      {\rm Im}J_K(k) & -{\rm Re}J_K(k) \\
      {\rm Re}J_K(k) & {\rm Im}J_K(k) \\
    \end{array}
  \right),
\end{equation}
with
\begin{equation} 
J_K(k)=\frac{1}{2C}\left(
                     \begin{array}{cc}
                         -4R_b^{-1}\cos k-2R_c^{-1} & 4iR_a^{-1}e^{-i\phi}\sin k \\
                        -4iR_a^{-1}e^{i\phi}\sin k & -4R_b^{-1}\cos k-2R_c^{-1} \\
                          \end{array}
                          \right).
\end{equation}
By choosing ${R_aC}^{-1}=\Delta/4$, ${R_bC}^{-1}=t/4$, and ${R_cC}^{-1}=\mu/2$, this circuit can be fully mapped to Kitaev model with $\phi=0$ superconductor phase. With this circuit model, one can realize the Majorana-like edge mode, as shown in Fig. \ref{Quantum}(f). In order to operate the braiding of the Majorana-like edge mode, the T junction is realized in circuit. Based on these results, one can perform quantum computing with classical circuits, such as the one- and two qubit unitary operations.

In Ref. \cite{Zhang2200232}, the quantum NAND-Tree is mapped onto a classical circuit network, leading to the design of a new type of classical logic circuit incorporating NAND gates capable of executing quantum algorithms. These newly designed classical logic gates offer exponential speedup functions compared to conventional logic gates, which has great potential in information processing.

In summary, we show the utilities of the circuit for exploring other physical systems, including the photonic, magnetic, and quantum systems. With the transformation of the basic circuit equations, it is believed that many kinds of equations can be explored in circuit. In the next section, we will present a summary of this review and outlooks for the future research directions.

\section{Conclusions and Outlooks} \label{S7}

\begin{figure}
  \centering
  \includegraphics[width=0.95\textwidth]{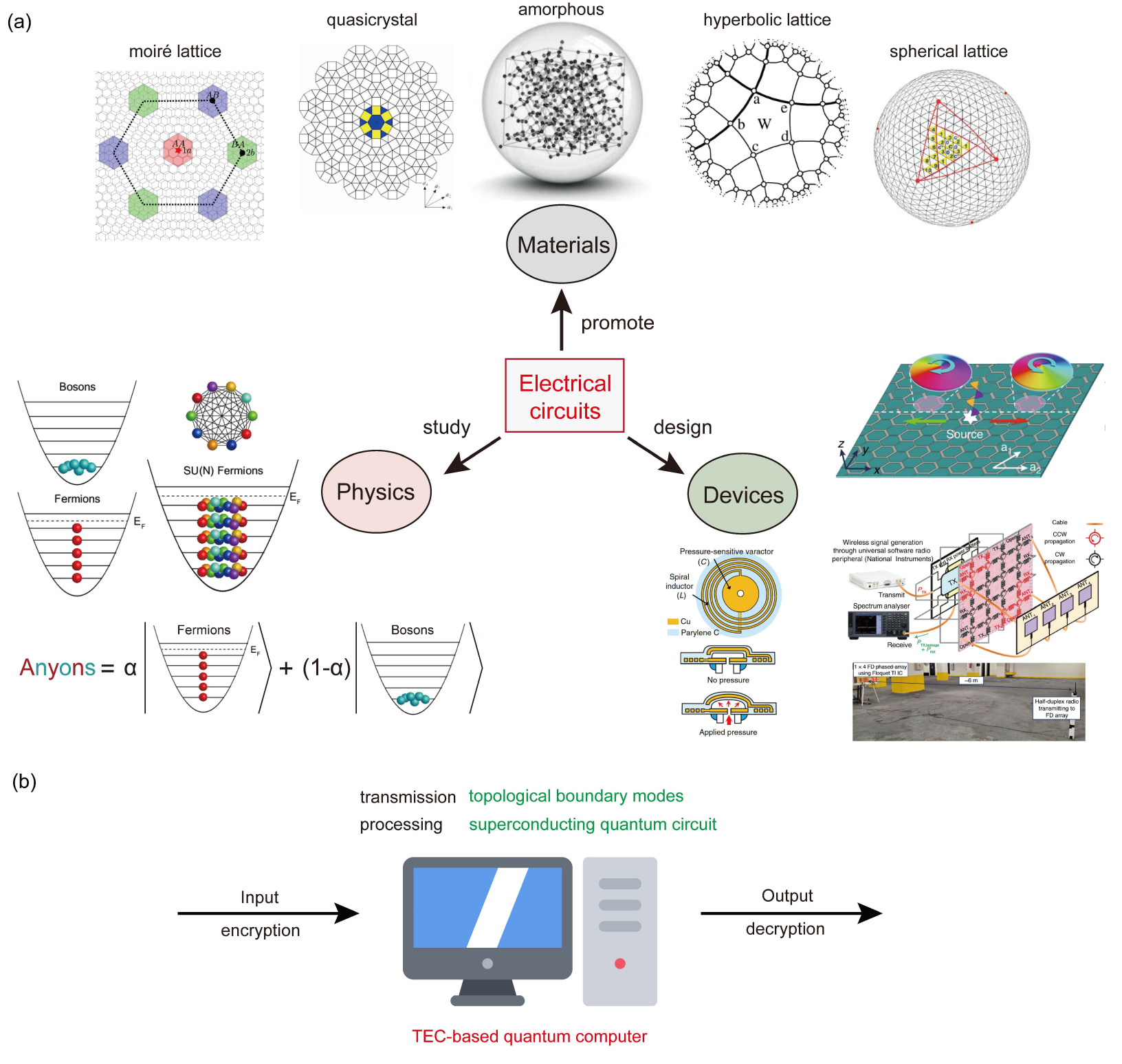}
  \caption{(a) The interplays linked by electrical circuits between materials, physics, and devices. (b) Set up of the multifunctional integrated circuit computer. \\
  {\emph{Source}:} The figures are adapted from Ref. \cite{Herzog081540,Liaobook,Ueda2007,Bowick2007,LiNC2018,Chen297,Nagulu300,Sonderhouse2020}.}\label{MPC}
\end{figure}

In conclusion, electrical circuits manifest as a powerful platform in studying topological physics. For conventional (higher-order) TIs and TSMs, one can directly map the circuit Laplacians to their tight-binding models. By measuring the impedance distributions or voltage propagations, the typical features of the topological states can be examined. Further, due to the diversity of electrical elements and their connection types between circuit nodes, the nonlinear, non-Hermitian, non-Abelian, non-periodic, non-Euclidean, high-dimensional, and other novel topological phenomena have been reported so far. By mapping the circuit equations to other systems, one can also explore their physical phenomena, such as photon-photon interaction and magnetic skyrmions. The study of TECs thus paves the way to enriching the topological phases, deepening the understanding of topological physics, setting paradigms for other metamaterial systems, and providing new opportunities for device applications. TECs build the connections between the materials \cite{Herzog081540,Liaobook,Ueda2007,Bowick2007}, physics \cite{Sonderhouse2020}, and devices, as we summarized in Fig. \ref{MPC}(a).

In electronic systems, topological states and conventional integrated circuits are two independent systems, and how to combine their advantages is still an open issue. However, electrical circuits can form a complete system. One may integrate the TECs with CMOS technique and envision a TEC-based quantum computer with the signal input, transmission, processing, output, even including the encryption and decryption processes, as we displayed in Fig. \ref{MPC}(b), which paves a new way for the development of the modern electronic technology.

\begin{figure}
  \centering
  \includegraphics[width=0.85\textwidth]{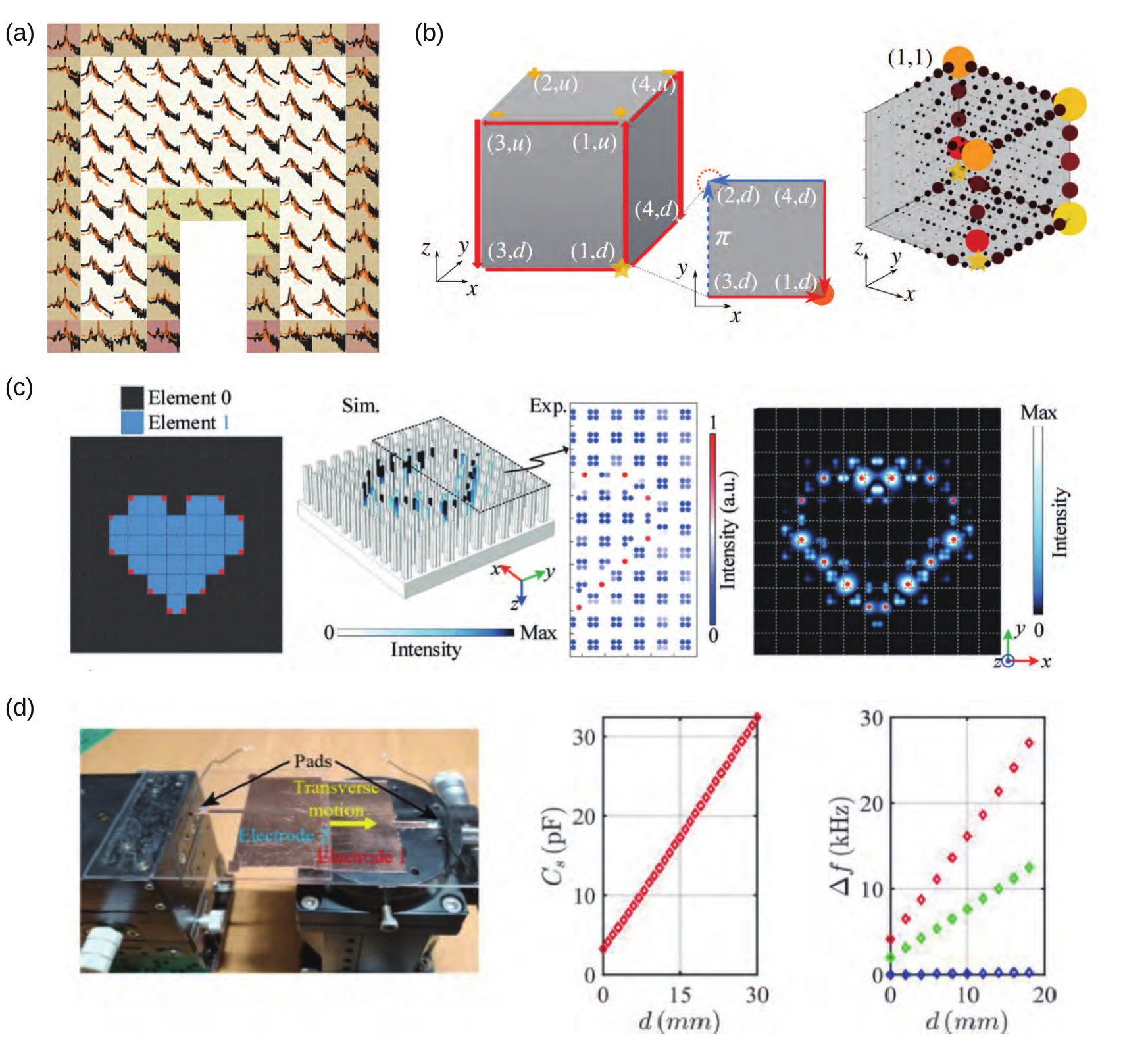}
  \caption{(a) Self-organized, self-sustained nonlinear oscillations in 2D active topolectrical circuits recapitulate topological edge mode phenomena . (b) Robust multiplexing with hinge modes. (c) Imaging with higher-order topological states. (d) The ultra-sensitive sensor of displacement based on the non-Hermitian topolectrical circuit. \\
  {\emph{Source}:} The figures are adapted from Refs. \cite{Kotwal2021,Ni064031,Yuan13870}.}\label{Applications}
\end{figure}

Utilizing the topological states in circuit, one may expect many applications. For the robust edge states or hinge states, one can use them to transport signals in the edge/hinge channels, as shown in Fig. \ref{Applications}(a). Especially in Ref. \cite{Ni064031}, Ni et al. realized a robust multiplexing with topolectrical higher-order Chern insulators, as displayed in Fig. \ref{Applications}(b). Besides, according to the positions of the localized state, such as the corner modes, one can design some imaging devices in Fig. \ref{Applications}(c), as reported in Ref. \cite{Zhang04682}. Due to the splitting of spectra in non-Hermitian, one can design high-sensitivity sensors to detect the distance, rotation angle, and liquid level with the designed capacitive devices, as shown in Fig. \ref{Applications}(d). Based on the chaotic circuits, one can design a series of crytography by encoding the information in chaotic phase diagram with different initial conditions \cite{Kocarev6}.

\begin{figure}
  \centering
  \includegraphics[width=0.85\textwidth]{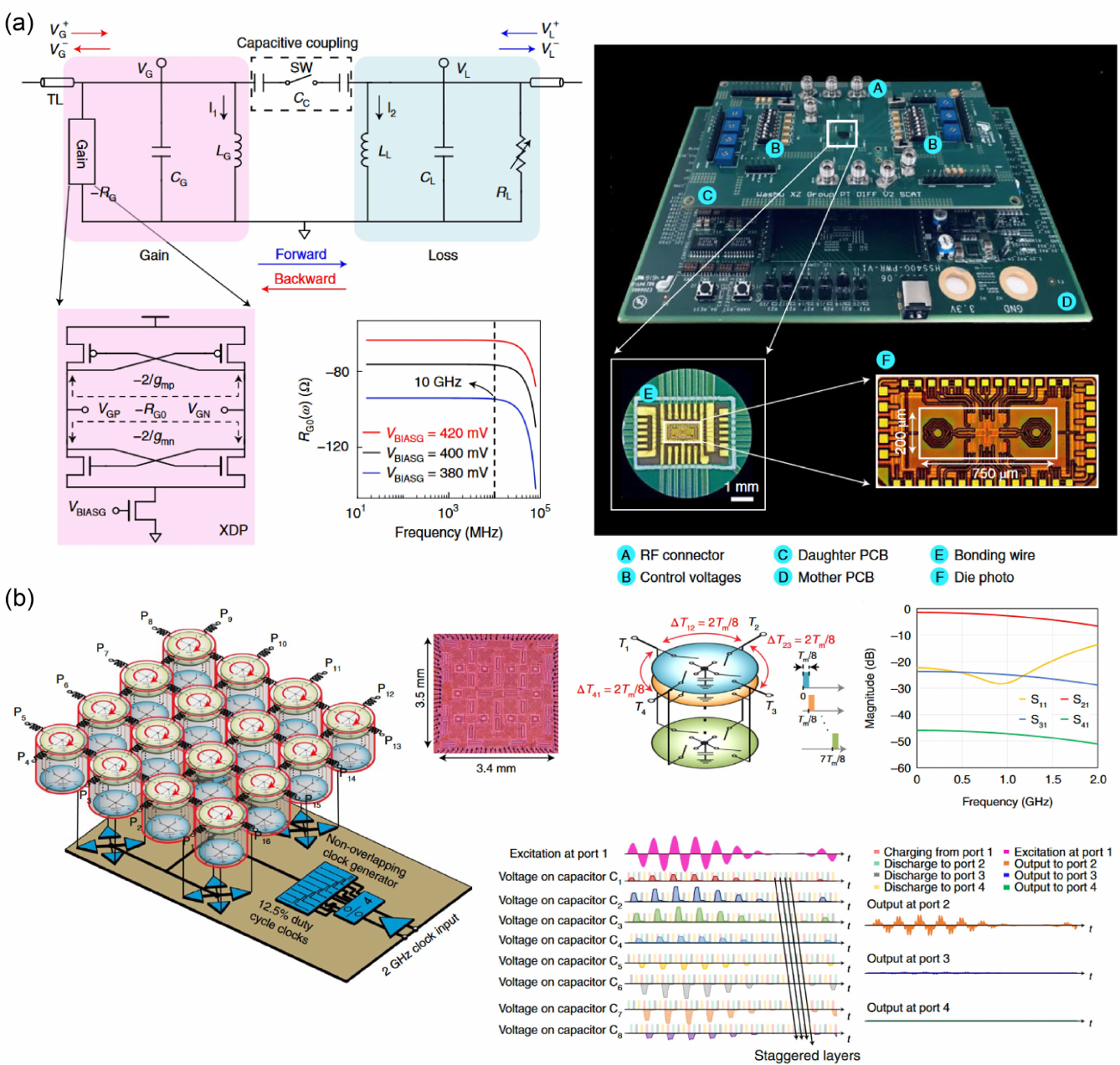}
  \caption{(a) Fully integrated PT-symmetric electronic system. (b) Conceptual diagram and chip microphotograph of the Floquet TIs implemented in a 45 nm silicon-on-insulator CMOS process. \\
  {\emph{Source}:} The figures are adapted from Refs. \cite{Cao262,Nagulu300}.}\label{F14}
\end{figure}


Due to the feasibility and promise of studying TIs with integrated circuit technology, we use two examples to show how to integrate TEC into a complementary metal-oxide-semiconductor chip. In 2022, Cao et al. presented a fully integrated PT-symmetric electronic system, as shown in Fig. \ref{F14}(a) \cite{Cao262}. The system can generate wide-band microwaves and design high-sensitivity sensors near exceptional points. As shown in Fig. \ref{F14}(b), Nagulu et al. propose the realization of Floquet TIs implemented in a 45 nm silicon-on-insulator CMOS process \cite{Nagulu300}. Based on the SSH circuit model, Liu et al. designed a fully integrated TEC chain using multiple capacitively-coupled LC resonators \cite{Liu13410}. The authors provided a detailed layout of this proposal and observed the robust edge modes. Therefore, we are optimistic to see the integration of TECs with small size and multifunctions.

As a highly effective platform for studying topological physics, TECs are promising in many further researches, and we will discuss them below.

\emph{(Meta)material design}. TECs can be viewed as a new kind of metamaterial. On the one hand, most of the structures of TECs originate from electronic materials, and the properties of materials can be explored in this platform, especially for the ones of high-dimensional lattice, Moiré lattice, quasicrystalline lattice, and amorphous lattice. On the other hand, one can identify brand-new materials with the TEC platform, which can promote the discovery and design of materials. Analogy to the electrical circuits, other metamaterials may be developed, ranging from photonic crystals, sonic crystals, magnetic soliton arrays, mechanics, transmission line networks, to surface plasmon polaritons. These metamaterials host rich and varied physical properties, which will lead to significant scientific interest and broad application prospects for future technological development.

\emph{Fundamental physics}. In the field of topology, symmetry is of vital importance, such as the time-reversal, chiral, and particle-hole symmetries in $\mathcal{AZ}$ 10-fold classification and the translation, inversion, rotation, and mirror symmetries in topological crystalline insulators. TECs can contribute to investigating the protection of symmetries on topological states because it is convenient to introduce the symmetry-broken perturbations by adding various interactions between any node \cite{Yang2020}. Defects and disorders are frequently discussed in topological physics. In circuits, one can create defects and introduce disorder to circuit lattice either arbitrarily or controllably. Not only the robustness of the boundary states against the defect and disorder can be examined directly, but also the disorder-induced localization \cite{Pereira036801,Baboux066402}, topological Anderson insulator \cite{Li136806,Groth196805}, and topological defect effects \cite{Baboux211} can be investigated.

\emph{Non-Hermitian physics}. Non-Hermitian systems exhibit many fascinating phenomena, like non-Hermitian exceptional degeneracies, skin effects, non-Bloch band structures, and non-Hermitian bulk-boundary correspondence \cite{Bergholtz2021}. Although the study of non-Hermitian system has made some progresses recently, there are still many open issues that are in demand to explore. With regard to EPs, higher-order EPs and EPs in higher-dimensional systems are rarely explored \cite{Luo065102}, and the encircling of EPs is also an interesting topic. A comprehensive theory addressing non-Hermitian skin effect exceeding one dimension has yet to be formulated. The intricate relationship between band topology and EP topology within multi-band systems needs to be clarified. Additionally, the interplay between non-Hermitian skin effect and topological edge modes may yield new topological phenomena, thus paving an exciting avenue for exploration. Circuit should be a suitable platform to study these physics with its controllable INIC unit.

\emph{Nonlinear physics}. In classical systems, most of the nonlinear phenomena are hard to harness precisely, which prevents the development of nonlinear physics. Taking advantage of nonlinear elements in circuit, it is possible to explore the phenomena in strongly nonlinear domains. With this platform, one can study other interesting nonlinear effects, such as self-localized topological edge solitons \cite{Leykam143901,Hadad1974}, self-induced topological states \cite{Hadad155112,Lumer243905}, nonlinear Chern insulator \cite{Sone2024}, and bulk boundary correspondence of nonlinear systems \cite{Isobe126601}.

\emph{Non-periodic physics}. Traditional topological band physics is built on the periodic lattices. Based on the flexibility of circuit connections, a large amount of artificial circuit lattices can be constructed. Although some pioneer works of TECs have focused on these non-periodic lattices, many nontrivial topological physics are in demand for further exploration, such as the topological states in quasiperiodic \cite{Kraus106402,Kraus226401,Huang126401}, amorphous \cite{Agarwala236402,Mitchell380,Sahlberg013053,Agarwala012067}, and fractal lattices \cite{Bandres011016,Pai155135,Biesenthal1114,Li023189}. 

\emph{Non-Euclidean physics}. In addition, electrical circuits can be fabricated to simulate the topological states in the curved space for further studies \cite{Lenggenhager4373,Chen622,Zhang1083}. In non-Euclidean spaces, namely the spherical and hyperbolic ones \cite{Lee196804,Krioukov036106}, there are many interesting topological phenomena, for example, the nontrivial quantum spin connection and active helical surface states in spherical spaces \cite{Parente075424,Imura235119} and larger families of strong and weak topological band insulators in hyperbolic spaces \cite{Urwyler246402}. With the TEC platform, these non-Euclidean phenomena can be fully explored in experiments.

\emph{Higher-dimensional topology}. In condensed matter systems, the topological states are limited to three spatial dimensions, but synthetic dimension offers a possibility for studying topological states in higher dimensional space. Electrical circuits pave a new way to exploring higher dimensional topological phases, such as the 4D spinless topological insulator, 5D topological Weyl state \cite{Zheng033203}, and the 6D chiral topological superconductors.

\emph{Topological braids}. Braids and knots are common in nature. Recent works found rich topological braiding structures, ranging from the multimode braiding in real space \cite{Chen179,Zhang390}, band node entanglements in 2D and 3D TSMs \cite{Wu1273,Bouhon1137,Lee4385}, and the complex energy braiding in non-Hermitian systems \cite{Wang59,Shen146402}, described by braiding group \cite{Iadecola073901,Pan355}. With the abundant topological phases of TECs, all of the aforementioned braids can be explored, which may open an avenue for devising and producing robust non-conservative systems exploiting the topological properties linked to knot invariants.

\emph{Fermions and bosons}. Electrical circuits can also be applied to study the behaviors of elementary particles, including fermions and bosons, and their interactions. Earlier, three types of fermions have been reproduced in circuits, such as the massive Dirac \cite{Dong023056}, massless Weyl \cite{Lu020302,Islam023025,RLi2020,Islam2020}, and Majorana fermions \cite{Ezawa2075424,Yao043032}. In circuit experiments, both the band structures and boundary states were measured directly. In other aspects, bosons are simulated in circuit systems as well. Zhou et al. imitated the Bose-Hubbard model and studied the interactions between bosons, observing the interaction-induced two-boson flat-band localizations and topological edge states in circuits \cite{Zhou035152}. It can be seen that electrical circuits are suitable to study the properties of these (quasi)particles and these studies may deepen the understanding of elementary particles.

\emph{Anyons}. Anyons, like Majorana zero modes, have attracted substantial attention because of their potential in building a fault-tolerant quantum computer \cite{Google264}. Recent works reported the realization of anyons in circuits as well, the statistical behaviors of which are investigated \cite{Olekhno205113,Zhang139}. In recent years, researchers have been devoted to finding Majorana fermions \cite{Fu107001}, including Majorana quasiparticle, Majorana zero mode, Majorana-like bound states \cite{HSu196801,Pahomi032068}, and Majorana neutrino. TECs may be an effective platform to simulate the behaviors of Majorana fermions. 
 
\emph{Dynamical phenomena}. The non-trivial dynamic evolutions of states can be studied in topological phases, such as Bloch oscillation, Zitterbewegung, Klein tunneling, and charge pumping. Zhang et al. realized an extended 1D Bose-Hubbard in circuit, with two strongly correlated bosons exhibiting Zitterbewegung and Klein tunneling effects in relativistic quantum mechanics \cite{Zhang250}. This work may inspire the further studies on the topological dynamics with TECs.

\emph{Floquet engineering}. Topological states are not limited to static systems, and the discovery of Floquet TIs paves a new way to realize the nontrivial state, which offers more controllability and loosens the strict requirement on material structures \cite{Kitagawa235114}. For a Floquet system, one needs a new topological invariant to characterize it, and several gaps emerge for the edge modes, the localization of which depends on the strength of drive. In circuit, one can realize Floquet dimension by an extra spatial dimension, as reported in Ref. \cite{Dabiri084303}. Floquet TIs have been realized in an integrated circuit platform \cite{Nagulu300}, and further research may focus on other interesting Floquet topological phases. 

\emph{Multidimensional topological physics}. Due to the discovery of HOT states, different dimensional topological boundary modes can coexist in the same system. Zhang et al. mimic the quantum spin Hall effect in a sonic crystal and break the space symmetry with quantized Zak phase, leading to the zero-dimensional topological corner states \cite{Zhang582}. Similar multidimensional topological phases are reported in acoustic lattices \cite{Xiong125144,Meng151903}. One can envision a more diverse multidimensional topology in circuit because of the unrestraint of space dimension, which offers more freedom for multidimensional physics  \cite{Yang165406,Islam085430}.

\emph{Superconducting circuit}. Another burgeoning field of electrical circuits is the superconducting quantum circuit \cite{Devoret1169,Wendin106001}, which consists of simple circuit elements, e.g., paralleling a capacitor $C$, an inductor $L$, and a Josephson junction (cross) with inductance $L_J$. One may expect the combinations of the superconducting quantum circuit with the topology, and find peculiar topological phenomena in superconducting quantum circuit \cite{Xue0017}.

\emph{Challenges}. In TEC, a technical challenge originates from the circuit system itself. The loss is inevitable in circuit, both for the circuit elements and wires. Although the boundary states of TIs or TSMs are robust, the propagation of boundary states will decay very fast in practical experiments. This phenomenon will hinder the application of edge modes in signal transmission. Additionally, the design of TEC board is a complex issue, but only the simplified models of circuit elements are considered in the present TECs, ignoring all of the parasitic parameters. For practical applications, one needs to choose and optimize the elements and the layout of printed circuit boards.

\addcontentsline{toc}{section}{Declaration of competing interest}
\section*{Declaration of competing interest}
The authors declare no competing financial interests that could have appeared to influence the work reported in this paper.
\addcontentsline{toc}{section}{Acknowledgments}
\section*{Acknowledgments}
The authors would like to thank Zhixiong Li, Chan Tang, Yuanyuan Liu, Xiansi Wang, and Xiangrong Wang for collaborations and Xingen Zheng for helpful discussions. This work was supported by the National Key Research and Development Program under Contract No. 2022YFA1402802 and the National Natural Science Foundation of China (Grants No. 12374103, No. 12074057, No. 11604041, and No. 11704060).

\end{document}